\documentclass[sigconf]{acmart}

\AtBeginDocument{%
  \providecommand\BibTeX{{%
    \normalfont B\kern-0.5em{\scshape i\kern-0.25em b}\kern-0.8em\TeX}}}

\setcopyright{acmcopyright}
\copyrightyear{2022}
\acmYear{2022}
\acmDOI{10.1145/1122445.1122456}

\acmConference[CHI '23]{CHI '23: ACM Conference on Human Factors in Computing Systems}{ April 23--28, 2023}{Hamburg, Germany}
\acmBooktitle{Hamburg '23: ACM Conference on Human Factors in Computing Systems,
April 23--28, 2023, Hamburg, Germany}

\acmSubmissionID{9251}

\PassOptionsToPackage{table}{xcolor}
\newcommand{\SYSTEMNAME}{{\texttt{AutoML Trace}}}
\usepackage{soul}
\usepackage[figuresleft]{rotating}

\colorlet{lightblue}{gray!95!}
\colorlet{lightred}{gray!30!}
\colorlet{lightyellow}{gray!20!}

\definecolor{anagreen}{HTML}{CCD5AE}
\definecolor{anagreen2}{HTML}{3b9459}

\newcommand{\CatText}[1]{\textcolor{anagreen2}{#1}}

\newcommand{\Dim}[1]{\colorbox{lightblue}{\textbf{\textcolor{white}{#1}}}}

\newcommand{\Cat}[1]{\colorbox{lightred}{\textbf{#1}}}

\newcommand{\Char}[1]{\colorbox{lightyellow}{\textit{#1}}}

\newcommand{\artificatCat}[1]{\colorbox{anagreen}{\textbf{#1}}}

\newcommand{\suppTaxSpread}{{\href{https://osf.io/h5ung?view_only=19962103d58b45d289b5c83421f48b36}{\textcolor{black}{\texttt{{[SM1]}}}}}}

\newcommand{\suppTaxSpreadFinal}{{\href{https://osf.io/94mhe?view_only=19962103d58b45d289b5c83421f48b36}{\textcolor{black}{\texttt{{[SM1-F]}}}}}}

\newcommand{\suppTaxIter}{{\href{https://osf.io/qtrs6?view_only=19962103d58b45d289b5c83421f48b36}{\textcolor{black}{\texttt{{[SM2]}}}}}}
\newcommand{\suppDesignSpec}{{\href{https://osf.io/hnvub?view_only=19962103d58b45d289b5c83421f48b36}{\textcolor{black}{\texttt{{[SM3]}}}}}}
\newcommand{\suppDesignExamples}{{\href{https://osf.io/exs3q?view_only=19962103d58b45d289b5c83421f48b36}{\textcolor{black}{\texttt{{[SM4]}}}}}}

\newcommand{\revMR}[1]{\textcolor{black}{#1}}

\newcommand{\revCHIADD}[1]{\textcolor{black}{#1}}
\newcommand{\revCHISUB}[1]{}

\begin{document}


\title[AutoML Trace]{\revMR{Tracing and Visualizing Human-ML/AI Collaborative Processes through Artifacts of Data Work}}




\author{Jen Rogers}
\affiliation{%
  \institution{University of Utah}
  \city{Salt Lake City, UT}
  \country{USA}}
\email{jen@sci.utah.edu}

\author{Anamaria Crisan}
\affiliation{%
  \institution{Tableau Research}
  \city{Seattle,WA}
  \country{USA}}
\email{acrisan@tableau.com}

\renewcommand{\shortauthors}{Rogers, et al.}

\begin{abstract}
    Automated Machine Learning (AutoML) technology can lower barriers in data work yet still requires human intervention to be functional. However, the complex and collaborative process resulting from humans and machines trading off work makes it difficult to trace what was done, by whom (or what), and when. In this research, we construct a taxonomy of data work artifacts that captures AutoML and human processes. We present a rigorous methodology for its creation and discuss its transferability to the visual design process. We operationalize the taxonomy through the development of~\SYSTEMNAME\, a visual interactive sketch showing both the context and temporality of human-ML/AI collaboration in data work. Finally, we demonstrate the utility of our approach via a usage scenario with an enterprise software development team. Collectively, our research process and findings explore challenges and fruitful avenues for developing data visualization tools that interrogate the sociotechnical relationships in automated data work. 

\noindent\textbf{Availability of Supplemental Materials:} \url{https://osf.io/3nmyj/?view_only=19962103d58b45d289b5c83421f48b36}

\end{abstract}



\begin{teaserfigure}
    \includegraphics[width=\textwidth]{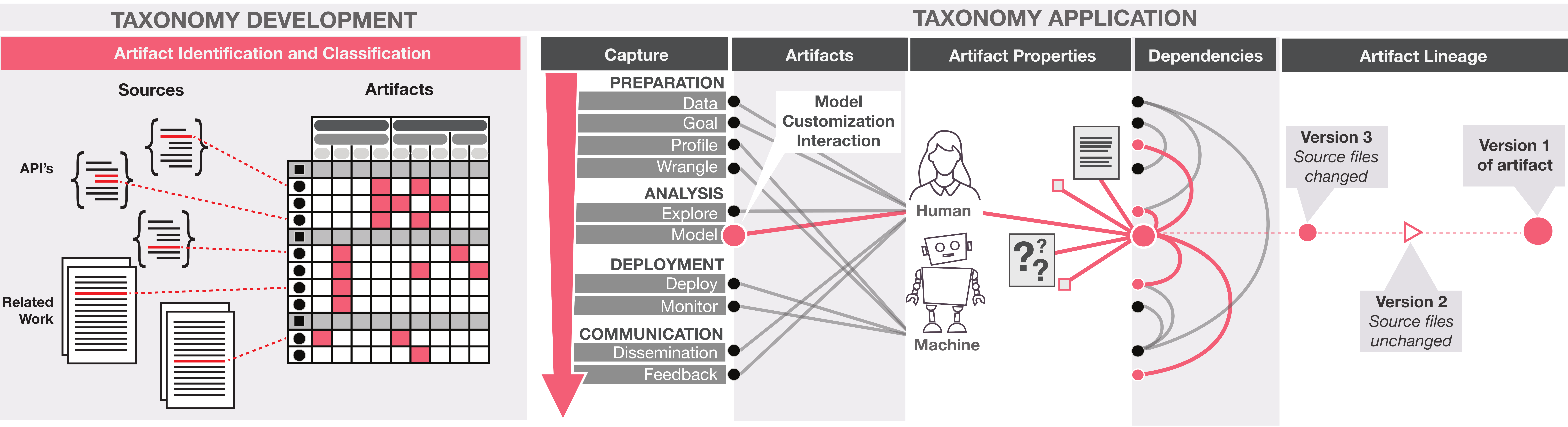}
     \caption{\revCHIADD{We developed an  artifact taxonomy that captures both human and ML/AI processes in automated data work. We assess our taxonomy's utility through a collaboration with an enterprise software development team creating an AutoML system.} \revCHISUB{We first developed a taxonomy from a broad literature review, synthesizing a set of human and machine-generated artifacts across data science work. We next conducted a case study to assess the utility of our taxonomy}.}
     \Description{A visual diagram of our study processes. We use simple and stylized icons to describe our approach and contributions. The first panel in the figure shows the taxonomy development from APIs and literature sources. The subsequent panels show the process of getting artifacts from across a data science process, classifying whether they had human or machine origins, delineating their properties, and finally, show the history of that object over time.}
    \label{fig:artifacts}
\end{teaserfigure}

\maketitle

\section{Introduction}
Data work comprises multiple interrelated phases that leverage statistical and computational techniques for data preparation, analysis, deployment, and communication~\cite{Crisan:Batton:202}. The skills required to conduct data work remain sufficiently complex, making it inaccessible to many experts with the relevant domain context but needing more technical acumen~\cite{sant:automl:2021}. \revCHIADD{Recent innovations have developed techniques that automatically carry out data work, for example, model selection or certain data preparation steps, thereby lowering barriers of use to non-technical experts~\cite{Bie:AutoDS:2022}. Initially, this so-called automated machine learning technology (AutoML)  focused primarily on the analysis phase. However, recent research is pushing the boundaries of AutoML to encompass a more end-to-end data  workflow~\cite{sant:automl:2021,Crisan:AutoML:2021, He:AutoML_Stota:2021,Zoller:BenchmarkAS:2021}. The expanded scope of AutoML can now involve automating wrangling in the data preparation phase, hyperparameter tuning and algorithm selection in the analysis phase alerts to drift within a deployed model, and even auto-generated reports for communication and dissemination.} However, in practice, AutoML still requires considerable human labor and coordination to be functional~\cite{Crisan:AutoML:2021,Drozdal:trust:2020,Xin:wither_automl:2021}. Moreover, these prior studies point to friction amongst data teams when AutoML remains a `black box'~\revCHIADD{and it is difficult to interrogate how people collaborate with AutoML technology to complete data work}. Even if full automation were possible, human oversight and intervention are still desired~\cite{wang:DS_automation:2021,sant:automl:2021,Lee2019AHP,Crisan:AutoML:2021}. \revCHIADD{Unfortunately, existing AutoML tools rarely consider this human element, resulting in many unaddressed needs even as this technology advances. To bridge this gap, we explored how visual analysis could help technical and non-technical experts trace AutoML-assisted data work.}

\revCHIADD{To acknowledge the shared labor of AutoML and humans, we treat the challenge of traceability in automated data work as one of human-ML/AI collaboration. Recent research from the HCI community~\cite{Yang:HAI_hard:2020} highlights several difficulties in human-ML/AI collaboration, and two issues were especially motivating in our research. }
The first challenge is that human-ML/AI collaboration adds uncertainty to the capabilities and outputs of an ML/AI system. They assert that this uncertainty is difficult to address with existing design methodologies. 
\revCHIADD{Though techniques and visual systems do exist to make AutoML processes more transparent through provenance tracking and auditing (e.g., ~\cite{Renan:ml_prov:2019,schelter:MLProv:2017,pipelineprofiler2020,wang:atmseer:2018}), these approaches often focus on the analysis phase of the pipeline, prioritizing machine learning engineers and data scientists over data workers with less technical expertise. This gap is significant for the development of AutoML systems, as it does not account for the diversity of teams involved in human-ML/AI collaboration~\cite{Parasuraman2000,Horvitz:tradoff:2019,Wang:hai_collab:2019,Hong:interpret:2020,Heer:agency:2019,Crisan:AutoML:2021,Liu:Paths:2020,Xin:wither_automl:2021}. Developing systems to support transparency for the full spectrum of data workers is essential, albeit challenging
~\cite{Nithya:datawork:2021}.
This reflects the second challenge Yang~\textit{et al.} identify in their work: close collaboration between user-oriented researchers and ML/AI engineers is important, but there are barriers to this collaboration stemming from a lack of mediation for such an interdisciplinary dialogue, such as \textit{``shared workflow, boundary objects, or a common language for scaffolding''}~\cite{Yang:HAI_hard:2020}.}

\revCHIADD{We encountered these challenges of human-ML/AI collaboration in our work with the enterprise team. Our initial goal was to develop a solution for visually tracing human and AutoML processes across a data workflow facilitated by their software. 
However, we realized the collaboration was missing a common language for shared discourse for their developing AutoML system. Through the lens of the broader HCI literature, and Yang~\textit{et. al.} in particular, what we lacked were boundary objects -- abstract or concrete information that established a shared understanding for collaboration with the software team~\cite{john2004identifying}. We established that we needed first to characterize what could be captured in an AutoML pipeline from both humans and automated processes before we could develop a visual analysis tool.}
\revCHISUB{However, we quickly encountered a chicken-and-egg problem: while seeking to develop a visualization tool for end-users, we simultaneously needed a visualization tool as a mediator to ideate with our collaborators. Although the visual design process is inherently iterative, a lack of specific scaffolds or common language for collaboration impeded our progress -- echoing issues raised.}

Research in human-human collaboration and knowledge sharing has highlighted the importance of capturing artifacts ~\cite{Kreiner:tacit_artifact:2002,Lee:BO:artifacts:2007} for tracing~\cite{Tiwana:kx_system:2001,Fischer:kx_mgmt_og:2001,Mariano:kwx_mgmt_survey:2016} complex collaborative processes. 
Grounding our research in this prior work and Yang~\textit{et al.}~\cite{Yang:AutoML:2020}, we present our approach for making automated data work traceable through developing an AutoML artifact taxonomy and the~\SYSTEMNAME\ visualization tool.  Our taxonomy is drawn from examining both existing and theoretical AutoML and human-ML/AI interactive systems. It defines the broad scope of both human and AutoML-derived artifacts. The precise meaning of an artifact is dependent on its context. In our research, artifacts represent tangible and abstract items generated by humans (i.e., goals, tasks, documentation, datasets, source code, etc.) or AutoML processes (e.g., feature sets, the choice of model, automated insights, etc.) within data work. \revCHIADD{The taxonomy served as a boundary object that scaffolded for dialogue with the software team, allowing us to ideate around their existing system and outstanding end-user needs.} We further operationalize this taxonomy through~\SYSTEMNAME\-- a high-fidelity visual interactive sketch that, in the words of Greenberg and Buxton~\cite{greenberg:usability_eval:2008}, aims to  \textit{``make vague ideas concrete, reflect on possible problems and uses, discover alternate new ideas, and refine current ones.''}. \revCHIADD{Developing an interactive sketch, instead of a high-fidelity and more full-featured prototype, allowed to engage more flexibly in a co-design process~\cite{Sanders:sketch:2014} with the software team.}~\SYSTEMNAME\ identifies, captures, and contextualizes artifacts defined by our taxonomy and shows their dependencies and evolution over time. Although we apply~\SYSTEMNAME\ to our collaborators' AutoML software, it can be applied to others.

\revCHIADD{Collectively, our research presents three contributions. \textbf{The first contribution of this work is the AutoML Artifact Taxonomy} that accounts for the nuances of human-ML/AI collaboration across the continuum of automation in data work. Two additional contributions of this work emerged as a natural progression from the development of the taxonomy and our engagement with the enterprise team. \textbf{The second contribution is the ~\SYSTEMNAME\ interactive sketch that operationalizes our taxonomy} and serves as a medium for engagement with AutoML systems. \textbf{Finally, the third contribution is a definition of traceability} that characterizes what an understandable and observable data workflow involves.}
While we focus primarily on the challenges of automating data work, we also reflect on using taxonomies and visual sketches to broadly develop frameworks and systems for designing human-ML/AI collaboration.

\section{Related Work}\label{related_work}
We review related work concerning taxonomies' utility for creating a common language within and between complex systems, existing taxonomies for AutoML and data visualization, and existing visualization systems for AutoML that can surface these artifacts.

\subsection{Taxonomies, Ontologies, and Schemas}\label{rel:prior-tax}
Taxonomies provide structure to knowledge and enable comparison and identification of relationships between items~\cite{Nickerson:taxonomy:2012}. The Vis, HCI, and ML communities use taxonomies to inform the development of systems, define requirements, and provide a common language for communication \cite{Lam:GoalstoTasks:2018,domova2022model,lee2021deconstructing}. We intended the same utility for our taxonomy. 
\revCHIADD{However, we sought to develop our AutoML artifact taxonomy in a rigorous manner to ensure our work is seated on a solid theoretical foundation.
Our taxonomy development is informed by the work of Nickerson~\textit{et al.} for rigorous taxonomy development in information systems \cite{Nickerson:taxonomy:2012}, which was motivated by the often ad hoc methods for constructing taxonomies identified in their own community.} We reviewed existing taxonomies in AutoML and data visualization to understand their respective conceptual characterization, utility, and granularity in relation to our taxonomy. We group existing taxonomies and similar works in three groups: ML processes, human-in-the-loop automation, and visual analysis. 

\subsubsection{Provenance, Tractability, and Reproducibility in ML Processes}
We are not the first to formalize ML and AI processes as a taxonomy. Tatemen~\textit{et al.}~\cite{tateman:repo_taxonomy:2018} proposed a taxonomy for the reproducibility of ML research. Their research identifies low to high reproducibility examples based on the artifacts their research process produces. With a similar aim of reproducibility, Publio~\textit{et al.}~\cite{Publio:MLSchemaET:2018} proposes ML-Schema, an ontology for representing and interchanging artifacts of ML processes, which includes code, data, and experimental documentation. They aim to automatically produce ML model meta-data descriptors to improve the interpretability of ML processes. 
Souza~\textit{et al.}~\cite{Renan:ml_prov:2019} builds on the ML Schema along with PROV-DM to create a specific schema for provenance tracking of multiple ML workflows. 
While these taxonomies and schema for provenance in ML are important, they do not sufficiently account for the ways that human processes and interventions at various stages, as our research attempts to do. However, in developing our taxonomy, we also considered how existing taxonomies connect to ours to add more granular details to a specific data science process. 

\subsubsection{Human-in-the-loop and Hybrid Automation} 
\revCHIADD{In their characterization of provenance in visual analytics, Ragen~\textit{et al.} illustrate the heterogeneity of a given workflow as well as the importance of \textit{``capturing user thoughts, analytical reasoning, and insights,''}~\cite{ragan2015characterizing}}.
More recent work \cite{von_Rueden:informedml:2021,dellermann:future_collab:2021} generated taxonomies that begin to explicitly account for a variety of  human-generated artifacts in ML processes. Dellerman~\textit{et al.}~\cite{dellermann:future_collab:2021} focuses on human intervention in AutoML technology; their work most closely approximates ours in spirit and uses the same methods that we do to develop a taxonomy. However, these taxonomies primarily focus on the model optimization phase, whereas ours is considered an end-to-end data science process, from preparation to communication. Taxonomies from the Human-Computer Interaction (HCI) and Computer Supported Cooperative Work (CSCW) communities\cite{sant:automl:2021,wang:DS_automation:2021} propose ways for marrying different levels of automation, across an end-to-end data science process, with human collaboration. Karamaker~\textit{et al.}~\cite{sant:automl:2021} propose six automation levels depending on the extent of successfully automated tasks. Their appendix provides a detailed view of different ML approaches, the scope of automation, and the role of human interventions. Wang \textit{et al.}.~\cite{wang:DS_automation:2021} suggest similar levels of human-directed and system-directed automation, which they describe within a larger human-in-the-loop AutoML framework. 
\subsubsection{Visualization of ML Provenance, Traceability, and Models} As our approach explores how artifacts can be surfaced via data visualization, we consider prior research in the visualization community. Sacha~\textit{et al.}~\cite{Sacha:Vis2ML_ontology:2019} formulate an ontology for visualization-assisted ML, which fits into the paradigm of human-in-the-loop ML/AI. It represents artifacts as input and output entities that constitute data, models, or knowledge; however, they do not provide more granular information on the properties of these entities. Spinner~\textit{et al.}~\cite{Spinner:explainer:2020} presents a framework for explainability in visual and interactive ML whose processes align with those of automated data science processes driven by AutoML technology. They also primarily view artifacts as input/output entities but do not further define what those entities are.

\subsubsection{Bridging the Gap} These different taxonomies, ontologies, and frameworks share the goal of defining a set of entities and actions across automated data science work. However, they lack a consistent description of entities generated or shared across data work.  We propose artifacts to be this entity. By developing our taxonomy, we argue that our research can help bridge these prior works.


\subsection{AutoML Visualization Systems}\label{rel:prior-vis}
\revCHIADD{Interaction and visualization of machine learning pipelines both facilitate user engagement and intervention and build trust in the results of an ML process \cite{boukhelifa2020challenges}.}
Many visualization tools for AutoML have emerged in recent years. ATMSeer~\cite{wang:atmseer:2018} performs an automated search for machine learning models and visualizes the summary statistics from the search space for end-users with an automatically generated dashboard of linked views. 
ModelLineUpper~\cite{Narkar:Model_lineupper:2021} also uses multi-linked views of different visual encodings to compare ML models generated by AutoML processes. AutoVizAI~\cite{Weidele:autovizai:2020} similarly explores the narrow scope of model configurations but uses conditional parallel coordinate plots to visualize the model generation across possible configurations. Lastly, Visus~\cite{Santos:Visus:2019} targets how domain experts specifically can tackle model building using AutoML. 

Other systems view AutoML processes more broadly, beyond the modeling phase. PipelineProfiler~\cite{pipelineprofiler2020} integrates with Jupyter notebooks and provides an overview of the results using a matrix juxtaposed with aligned views to indicate the different components and outputs of the AutoML pipeline in each step. AutoDS~\cite{wang:autods:2021} uses a network diagram to show different possible ways to configure an end-to-end AutoML pipeline. AutoDS exists as a stand-alone tool or embedded with a Jupyter notebook. The Boba~\cite{Liu:boba:2021} system and its underlying DSL use a similar visual design to AutoDS for visualizing the stages and results of different data science processes. The design inspiration for Boba builds off of earlier user studies conducted by Liu~\cite{liu:mlvis_analysis:2017} that visualized the analysis patterns of data workers via a network diagram. Swatai~\textit{et al.} similarly found that network diagrams effectively capture varied user paths through interactive analytic flows~\cite{Swati:design_nonexperts:2021}. Xin~\textit{et al.}~\cite{Xin:Analysis_DAG:2018} have leveraged this graph structure to develop techniques for inserting humans into automated machine-learning processes. Research is also oriented toward capturing user interactions with visual analytics systems; Knowledge Pearls~\cite{stitz:kx_pearls:2020} and Trrack~\cite{Cutler:Trrack:2020} are two examples that also use an underlying graph to manage and visualize analysis paths.

Through our taxonomy, we aim to broaden what artifacts are visualized with additional context about the artifact's origin, dependencies, and history. We draw inspiration from the visual encoding choices of these prior systems in the implementation of AutoML Trace~(Section~\ref{automl-trace}). 

\section{Traceability for Human-Machine Collaboration}\label{trace-define}
Tracing the collaborative relationship between humans and ML/AI processes is essential for ensuring the \textit{entire process of data work is transparent and scrutinizable, not just the end product (i.e., the model or result)}~\cite{Winfield:ethics_transparency:2018}. The traceability of artifacts has been explored in software and design engineering contexts~\cite{rogers:trrrace:2021,Sundaram:traceaing:2010}, the social sciences~\cite{Kreiner:tacit_artifact:2002,Lee:BO:artifacts:2007}, and knowledge management communities ~\cite{Tiwana:kx_system:2001,Fischer:kx_mgmt_og:2001,Mariano:kwx_mgmt_survey:2016}  for some time and has more recently been explored for machine learning~\cite{mora:trace_and_trust:2021,schelter:MLProv:2017,Cambronero:Software_AutoML:2021}.
~However, the definitions of traceability vary widely. Here, \textbf{we define traceability for ML/AI as encompassing provenance, transparency, and context}. \textbf{\textit{Provenance}} is the process of recording individual artifacts and their origins; what generated the artifact and other artifacts dependent upon it. \textbf{\textit{Transparency}} concerns the ability to understand how the model arrived at its conclusions. Finally, \textbf{\textit{context}} indicates where the artifact exists with the analysis. \revMR{Here, we propose tracing artifacts within data work, from preparation to communication phases, resulting from human-ML/AI collaboration  across these phases over time. } We consider an artifact to be traceable if there is a clear definition of what it is, how and when it was generated, and there exists a lineage of how it has changed. 
\section{Motivation and Methodology for an AutoML Artifact Taxonomy}\label{taxonomy-method}

\begin{figure*}[h!]
    \centering
    \includegraphics[width=0.85\textwidth]{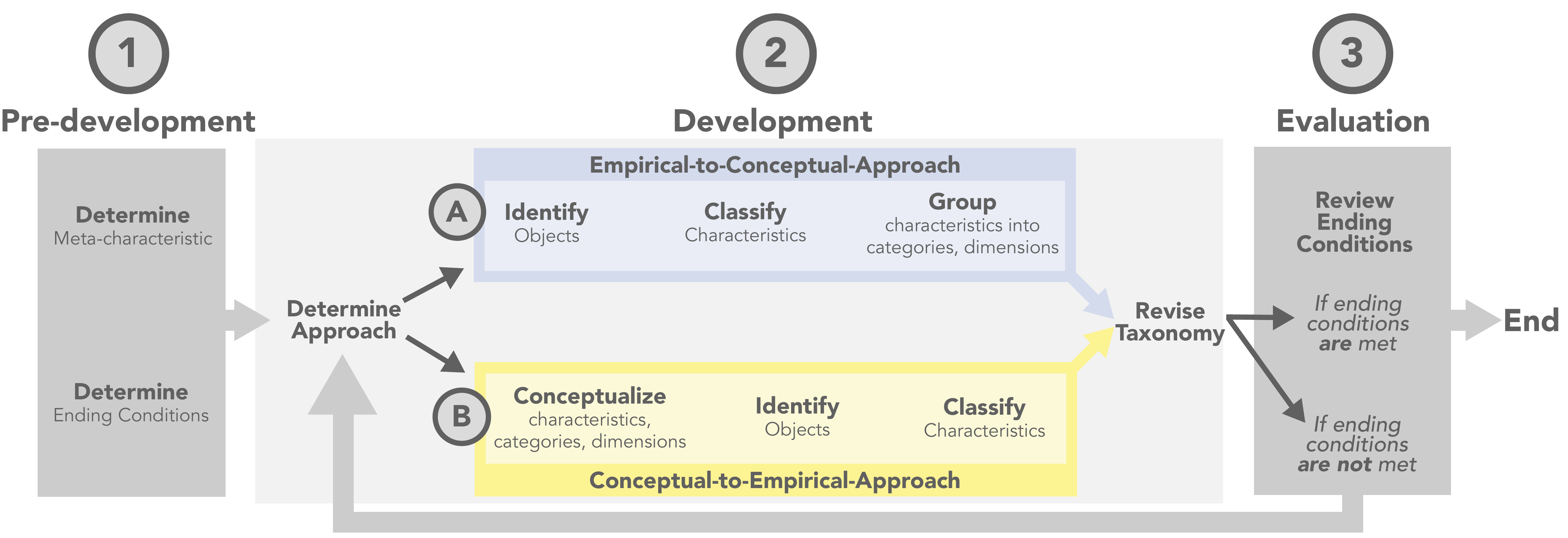}
    \caption{Overview of our taxonomy development methodology. We followed the methodology proposed by Nicerkson \textit{et al.}~\cite{Nickerson:taxonomy:2012}. \revCHIADD{The taxonomy development process consists of three stages. (1) The pre-development stage of the process involves defining a meta-characteristic and the ending criteria for development. (2) The development stage is repeated in the process until the ending conditions are met. This stage is done using one of two approaches, (A) empirical-to-conceptual or (B) conceptual-to-empirical. (3) We determine whether the ending criteria are met in the evaluation stage of the process. If the ending conditions are not met, we repeat the development stage until the ending conditions are satisfied.} 
    \revCHISUB{Each iteration of our taxonomy development concludes with a revision of the ending conditions. If those ending conditions are not met, another iteration is initiated. We primarily used an empirical-to-conceptual approach in the development stage. }
     \Description{A flow diagram showing the taxonomy methodology development method. We followed the methodology proposed by Nickerson \textit{et al.}~\cite{Nickerson:taxonomy:2012}. Each iteration of our taxonomy development concludes with a revision of the ending conditions. If those ending conditions are not met, another iteration is initiated. We primarily used an empirical-to-conceptual approach in the development stage.}}
    \label{fig:tax_method}
    \vspace{-5mm}
    
\end{figure*}

Taxonomies are a widely used system of knowledge organization that hierarchically groups concepts into logical associations based on shared qualities~\cite{Nickerson:taxonomy:2012,Prat:taxonomy:2015}. They provide a common language to speculate and build upon concepts that facilitate communication within a team of diverse experts~\cite{salmons2008expect}. Prior data visualization research has used taxonomies of tasks~(e.g.,~\cite{Brehmer:Typology:2013,Lee:Vis_Rec:2021,Valiati:tasks:2006}), data~(e.g.,\cite{Beck:tax_graphs:2017}), and visual techniques to motivate tool development. Taxonomies for AutoML and human-ML/AI collaboration have similarly been developed (see Section~\ref{related_work}), but their influence on tool development is tenuous, lacking a consistent mechanism for development. 
\revCHISUB{Across these different taxonomy development attempts, no consistent mechanism has emerged.}
\revCHIADD{As a result, the robustness of taxonomies in the literature can vary considerably in their quality and scope.  
Our taxonomy integrates and reconciles existing taxonomies, frameworks, ontologies, as well as artifacts of existing and theoretical systems, to provide a comprehensive set of AutoML artifacts.}
\revCHISUB{In creating our taxonomy, \textit{we integrate and reconcile existing taxonomies, frameworks, and ontologies} as well as outputs of existing and theoretical systems to arrive at a comprehensive set of artifacts that serves to inform our design process.} 
We have adopted a robust methodology from the information systems research that evaluates conciseness, robustness, comprehensiveness, extensiveness, and explainability~\cite{Nickerson:taxonomy:2012,Prat:taxonomy:2015}. As part of our taxonomy contribution, we describe our development approach, summarized in~\autoref{fig:tax_method}, to motivate the importance of robustness in taxonomy creation.

\subsection{Methodology Overview}
Nickerson \textit{et al.}~\cite{Nickerson:taxonomy:2012} and Prat \textit{et al.}~\cite{Prat:taxonomy:2015} define a multi-phased and integrated approach to defining and evaluating a taxonomy. Their approach is rooted in their definition of taxonomy as \textit{a set of objects classified according to taxonomic descriptors, which are a hierarchical set of dimensions, categories, and characteristics.}  Objects can refer to a variety of things, for example, living creatures, types of products sold in a store, or artifacts (as is the case here). 

They define three phases of taxonomy creation: pre-development, development, and evaluation. The \textbf{\textit{pre-development stage}} defines a meta-characteristic for the taxonomy objects and set of ending conditions for concluding taxonomy development. The subsequent \textbf{\textit{development stage}} takes either an empirical-to-conceptual or conceptual-to-empirical approach to define objects and their properties. Finally, in the \textbf{\textit{evaluation stage}} the taxonomy is assessed \revCHIADD{through an iterative process through a combination of objective and subjective criteria.} \revCHISUB{If the current iteration meets the end conditions, then taxonomy development concludes. Otherwise, the existing taxonomy is revised through a subsequent iteration of the development stage. We will go into more detail for each of these stages later in this section. The pre-development stage occurs only once at the onset of the taxonomy development, whereas the development and evaluative stages recur together until the ending conditions are met.}

Reflecting on their methodology, Nickerson \textit{et al.}~\cite{Nickerson:taxonomy:2012} emphasizes that a taxonomy is a `design search process' with an intractable solution. However, they argue, and we agree, that their methodology improves the resulting taxonomy's transparency, robustness, and extensibility. Here, we detail the choices we made through these taxonomy development stages. Artifacts of our research processes, which include notes, documents, and materials, generated across the 8 iterations of taxonomy development are \href{https://osf.io/3nmyj/?view_only=19962103d58b45d289b5c83421f48b36}{\textcolor{black}{\texttt{available online}}} \footnote{\url{https://osf.io/3nmyj/?view_only=19962103d58b45d289b5c83421f48b36}. This is an OSF view-only link for the review process, meaning it does not collect any data that could identify reviewers}. ~\revCHIADD{Due to limitations of space, additional details of our taxonomy and its  development are presented in the Supplemental Materials and annotated here with \texttt{[SM1]} (these are also clickable links). A full description of these supplemental materials appears at the end of this manuscript.}

\subsection{Pre-development Stage}

\subsubsection{Defining a Meta-Characteristic.} The taxonomy development process is initiated by delineating a concrete definition of a meta-characteristic that describes the objects under study (\autoref{fig:tax_method}.1). \textbf{In our research, we define an object in the taxonomy to be an AutoML artifact that is
generated and exchanged by a human or AutoML-driven task, which occurs across an end-to-end data science workflow encapsulating processes for data preparation, analysis, model deployment, and communication}.

\subsubsection{Defining Ending Conditions.} We defined an \textit{a priori} set of object and subjective ending criteria to evaluate our taxonomy upon each development stage (\autoref{fig:tax_method}.1). If these criteria are met in the evaluation stage, we conclude our taxonomy development. 
The taxonomy's structural stability across iterations is also part of the objective ending criteria. To meet this ending condition, our taxonomy should conform to the following criteria:
\begin{enumerate}
    \item No new dimensions, characteristics, or objects (artifacts) are added or modified from the previous iteration
    \item  No new dimensions, characteristics, or objects (artifact) were merged and split
    \item At least one object (artifact) is classified under every characteristic of each dimension
\end{enumerate}

The subjective ending conditions are defined by Nickerson~\textit{et als}~\cite{Nickerson:taxonomy:2012} as the minimum criteria for the utility of a taxonomy. These subjective conditions include conciseness, robustness, comprehensiveness, extensibility, and explanatory. 
As these are subjective criteria, they serve as a function to reflect on the taxonomy's internal validity. 

\subsection{Development Stage}\label{tax:dev-stage}
The development stage begins with either an  empirical-to-conceptual (\autoref{fig:tax_method}.2.A) or conceptual-to-empirical approach (\autoref{fig:tax_method}.2.B). In the former, objects are identified from an available data source, classified via quantitative (i.e., statistical clustering) or qualitative (i.e., thematic analysis) methodology, and grouped according to an emergent set of properties (characteristics, categories, dimensions). While in the latter approach, a set of properties are conceptualized and used to identify data sources and objects that are then subsequently classified. The approach taken can be different at the start of each development stage. We used primarily an empirical-to-conceptual identify objects for analysis.

\subsubsection{Literature Sources} We define both human and machine generated artifacts in automated data work from the research literature spanning Machine Learning, Human Computer Interaction, Computer Supported Collaborative Work, Information Visualization, and Visual Analytics. We sampled the research literature using two approaches. First, we gathered an initial set of 13 convenience sample papers, familiarized ourselves with the methodology, and created an initial taxonomy. The convenience sample was papers already known to the authors and from quick searches for ``artifacts AutoML'', ``taxonomy AutoML'', ``capturing AutoML'' and ``visualizing AutoML'' and subjectively selecting papers to discuss. Next, we identified a systematic set of published research and pre-prints on ``AutoML''. The search was current to June 14th, 2021, and retrieved 153 articles from venues such as KDD, AAAI, NeurIPs, CHI, and others. Most publications were retrieved from arXiv (100 of 153; 65\%) and dated within the past two years. A complete list of all sources used in our analysis and documentation on how they were used is found in online materials. We conducted an initial scan of all 153 papers. Based on this scan, we then developed inclusion and exclusion criteria. We excluded papers that were too narrow in scope because they focused on a highlight-specific technique. \revCHIADD{The list of literature sources in available in \suppTaxIter.} 


\subsubsection{Object Classification} We identified and extracted approximately 400 items from literature sources that could represent human or machine-generated artifacts. First, we coarsely classified these items into phases  of a data workflow (preparation, analysis, deployment, and communication)~\cite{Crisan:Batton:202}. Within these phases, we further classified items into artifact groups. Finally, we used this grouping to ideate a set of artifact properties.
We use open and axial coding techniques to derive the set of characteristics, categories, and dimensions that describe the artifact's properties. We used descriptions and definitions from the object's literature source text for this coding exercise.
We combined separate items as definitions for artifacts and their properties became clearer with each coding iteration (i.e., T-SNE and PCA were combined into mapping transformations artifacts because they both map data from a higher to lower dimensions). From the initial set of 400 items, we distilled into a set of 52 artifacts. \revCHIADD{A full list of artifacts and their classification is available in~\suppTaxSpread~and~\suppTaxSpreadFinal.}



\subsection{Evaluation Stage}\label{tax_method:eval}
After each development stage, we assessed whether we met our ending criteria (~\autoref{fig:tax_method}.3).
Per our definitions from ~\cite{Nickerson:taxonomy:2012} and~\cite{Prat:taxonomy:2015}, the taxonomy is \textit{concise}, \textit{robust} and \textit{comprehensive} if, at the conclusion of a development stage, objects can be comprehensively classified with a sufficient and not excessive, set of dimensions, categories, and characteristics.  It is \textit{extensible} if new dimensions, categories, and characteristics can be easily added throughout iterations. Finally, it is  \textit{explanatory} if it can be used to describe the nature of objects.

Our taxonomy development required eight iterations before it met the ending conditions. Both authors read the literature sources, extracted artifacts that met the definition of the meta characteristic, classified those items, and finally grouped them according to an evolving set of artifact properties. The authors met and discussed their individual classifications daily for a month. While we arrived at a consensus, we did not attempt to resolve all conflicts, ambiguities, or divergent interpretations exhaustively.

\section{AutoML Artifact Taxonomy}\label{taxonomy}
\begin{figure*}[ht!]
    \centering
    \includegraphics[width=\textwidth]{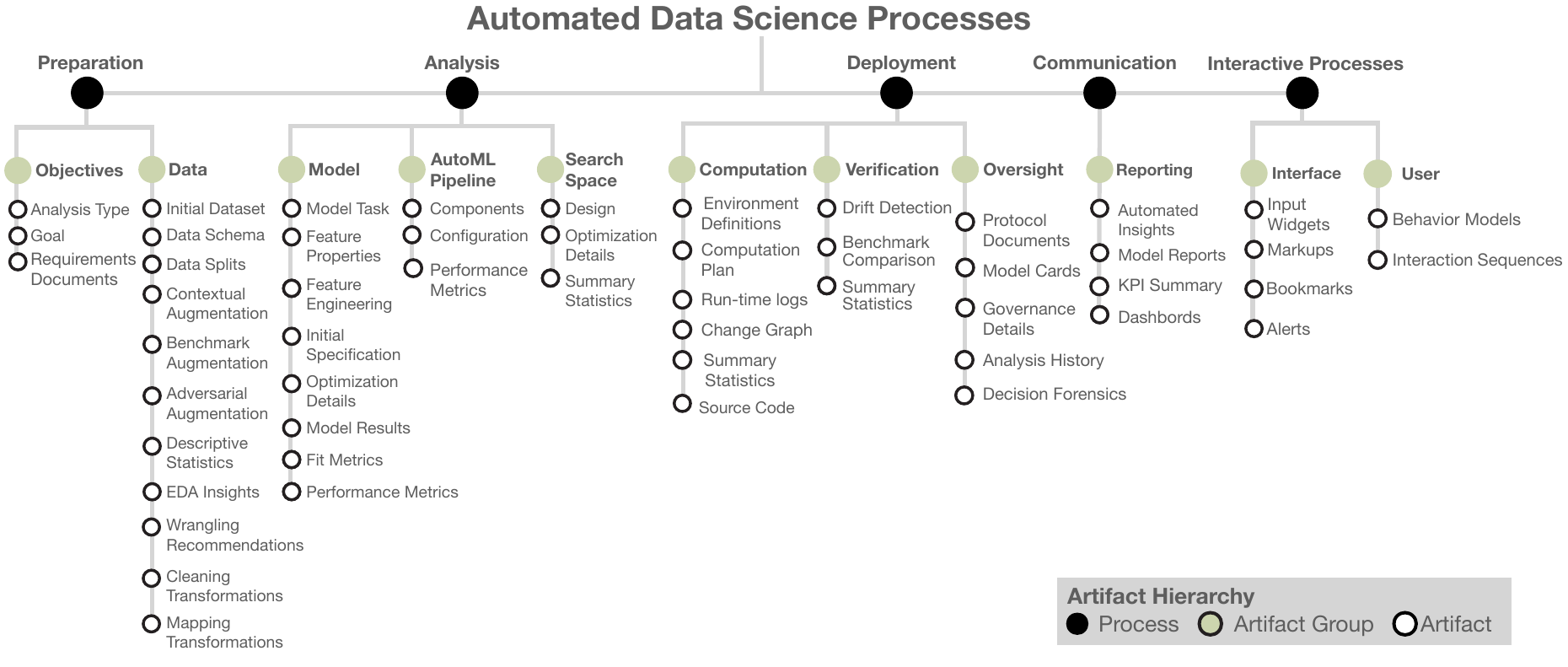}
    \caption{Artifacts elicited from AutoML toolkits, libraries, systems, and user studies. We summarized approximately 400 artifacts from these sources into 11 \artificatCat{Artifact Groups} and 52 artifacts. The properties of these artifacts are further delineated according to a taxonomy and a hierarchical set of \Dim{dimensions}, \Cat{categories}, and \Char{characteristics}.}
    \Description{A hierarchical diagram that summarizes Artifacts and examples elicited from AutoML toolkits, libraries, systems, and user studies. We summarized \~400 artifacts from these sources into 11 Artifact Groups and 52 artifacts. The artifact properties are further delineated according to a taxonomy and a hierarchical set of dimensions, categories, and characteristics. }
    \label{fig:tax_artifact}
    \vspace{-5mm}
\end{figure*}

Our taxonomy comprises 52 artifacts clustered within eleven groups by their properties. We defined the properties of these artifacts according to a set of 4 dimensions, 17 categories, and 41 characteristics. Importantly, no single AutoML system contains all of these artifacts~\cite{sant:automl:2021}. Instead, we rely on an amalgamation of design decisions made by individual AutoML toolkits, systems, and theoretical research papers. We argue that by looking broadly at existing systems, what they are, and what they aspire to be, our taxonomy can extend to systems not yet developed. A summary of artifacts, their groupings, and the data science processes they belong to (in addition to interactive processes) is in~\autoref{fig:tax_artifact}.

\subsection{AutoML Artifacts}\label{artifacts-final}

\subsubsection{Artifacts and Processes of Data Science Workflows}
As innovations in AutoML systems expand, so does the scope of task automation. As of this writing, many proposed systems do not exist for practical use~\cite{sant:automl:2021}. Leveraging a prior framework, we define an end-to-end data science workflow as comprising preparation, analysis, deployment, and communication processes. These stages also align with defined tasks and automation levels for AutoML systems proposed by Karmaker~\textit{et al.}~\cite{sant:automl:2021}. Likewise, AutoML systems composed of data science primitives~\cite{Heffetz:deepline:2020,pipelineprofiler2020} are similarly compartmentalized within these processes. While we imposed these processes on artifact classification (Section~\ref{tax:dev-stage}), we also found that most artifacts typically fit into one process. For example, the initial dataset is an artifact, typically supplied by a human, in the data preparation phase -- future AutoML systems may be able to find these datasets for data workers. The artifact would occupy that preparation phase, but its properties would reflect its machine progenitor. Conversely, a dashboard of the model's results is an artifact that exists in a communication process and likewise can be meticulously curated by a human or be automatically generated~\cite{Hu:VizML:2019}.  

AutoML artifacts are more than inputs and outputs to tasks within these data processes. Artifacts can also be metadata or other documentation created for or by data science processes. Prior work has examined metadata in machine learning or software systems and how they relate to provenance (Section~\ref{rel:prior-tax}). 
For example, organizational processes create human requirements documentation, a human-generated artifact that can directly dictate data analysis objectives and impact the choice of dataset or model. 


\subsubsection{Groups of Artifacts and Individual Artifacts} 
\begin{figure*}[ht!]
    \centering
    \includegraphics[width=0.9\textwidth]{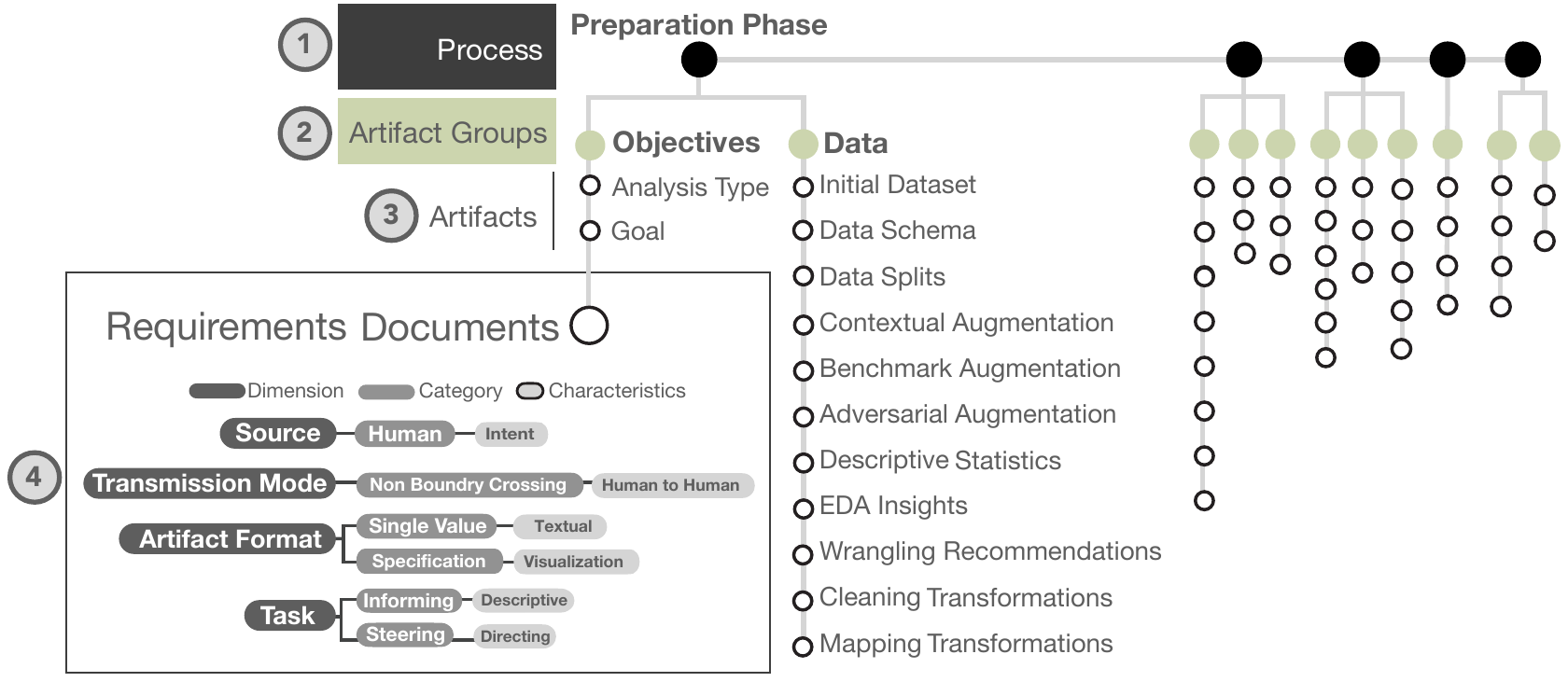}
    \caption{\revCHIADD{Breakdown of the hierarchy of information for the (1) \textit{preparation phase} process. Here we see the (2) artifact groups, and (3) artifacts for preparation phase. Each artifact represented by a white circle has its own dimensions, categories, and characteristics. This example shows the artifact properties for requirements documents (4). Each artifact has at least one example from the four dimensions; source, transmission mode, and tasks. Some artifacts have multiple categories and characteristics for each source. For example, requirements documents have two categories for both format and task.} }
    \label{fig:tax_prep}
    \vspace{-5mm}
\end{figure*}

We now describe artifact groups and examples of individual artifacts according to their data science processes. \revCHIADD{For an illustrated example of the artifact property hierarchy, see the breakdown for the ``requirements document'' artifact in \autoref{fig:tax_prep}.} While the processes are presented linearly here, in reality, they can occur in any order.

\texttt{\textbf{Preparation processes}} have two artifact groups: \CatText{objectives} and \CatText{data} (\autoref{fig:tax_prep}.2). Data work begins with some objective that can be expressed in the form of analysis goals, requirement specifications, or tasks~\cite{hong2020evaluating,gijsber:automl_benchmark:2019,Lam:GoalstoTasks:2018}. Goals can also be translated to tasks~\cite{sant:automl:2021,Brehmer:Typology:2013,Wong:NeuralAutoML:2018} and intents~\cite{gadhave:vis_intnet:2020,setlur:intent:2019} that further define specific analysis objectives. These objectives are necessary to define the dataset for analysis and any transformations or augmentations to the initial data and its schema representation~\cite{dellermann:future_collab:2021}. These transformations can result from data cleaning or wrangling operations~\cite{kasica:tablescraps:2021}, data splits~\cite{zhang:autocp:2020}, or mapping transformations. We also observed that additional datasets are recruited in the preparation stage to further benchmark model performance~\cite{Zoller:BenchmarkAS:2021,gijsber:automl_benchmark:2019} or evaluate its robustness. Augmentations to the data can include human-supplied semantic annotations~\cite{estevez:semantic_annots:2019}. We observed that the preparation stage is still largely dominated by the activities of a single human or multiple humans working together. These activities are presently the most time-intensive of data work~\cite{Crisan:AutoML:2021}, but also the most consequential~\cite{Nithya:datawork:2021}. As part of data preparation, we include exploratory data analysis that produces either automated or human-curated summaries, including descriptive statistics and visual summaries~\cite{wongsuphasawat:voayger:2016}.

\texttt{\textbf{Analysis processes}} are most extensively covered by prior literature and encapsulate what many consider to be AutoML's core functionality. We define four groups of artifacts of analysis: those pertaining to the individual \CatText{model}, an individual \CatText{AutoML pipeline} configuration, the \CatText{search space} of all possible pipeline configurations, and finally \CatText{computation}. The first set of artifacts concerns the \CatText{model}, which includes its task (i.e. classification, regression, clustering, or the various more nuanced tasks of neural networks) aspects of feature encoding~\cite{yao:humanout:2019,Yang:AutoML:2020,sant:automl:2021,Cambronero:Software_AutoML:2021}, generation\cite{wang:DS_automation:2021,Zimmer:Auto-Pytorch:2021}, and selection, as well as model optimization~\cite{Yang:OBOE:2019,Olson:tpot:2016,Thorton:AutoWeka:2013} (within which we include the architecture of a model like a deep neural network ~\cite{Haifeng:autokeras:2019,Zimmer:Auto-Pytorch:2021}), and performance assessment~\cite{wang:DS_automation:2021}.

However, the model is only one component~\cite{Yang:AutoML:2020,wang:vega:2020} or~primitive~\cite{Heffetz:deepline:2020,pipelineprofiler2020} of an AutoML pipeline. The pipeline itself is determined by a broader search space of possible alternative configurations~\cite{Zoller:BenchmarkAS:2021,Heffetz:deepline:2020,wang:atmseer:2018,wang:vega:2020,alletto:randomnet:2020,yao:humanout:2019,feurer:autosklearn:2020,pmlr-v64-salvador_adapting_2016}. Tools that visualize AutoML systems increasingly focus on the search space and pipeline configurations~\cite{wang:atmseer:2018,pipelineprofiler2020}. These two sources of artifacts compound the selection of the final model as they determine the scope of what form it may take. These three artifact groups, the model, pipeline, and search space, share similar artifacts, including preliminary configurations, performance assessments, optimization summaries, and a descriptive summary of the fit (or search) computation.

 More recent AutoML systems place computation more prominently in the analysis stage. While these can include source code~\cite{wang:DS_automation:2021,Cambronero:Software_AutoML:2021} (including analysis notebooks), they also include system configurations and environments~\cite{Cambronero:Software_AutoML:2021}. Recently, computational budgets~\cite{wang:atmseer:2018} are used to calibrate model performance against computation time.

We observed that AutoML systems automate as much of the analysis as is reasonable but include avenues for human intervention. The complexity of AutoML systems makes it increasingly difficult to trace how it arrived at the choice of a model unless the full spectrum of artifacts is considered. For example, a system that searches a space of possible AutoML pipeline configurations is dependent on both the initial configuration and the set of primitives available to it. Imposing a computational budget will also limit the extent of the search space explored. 

\texttt{\textbf{Deployment processes}} apply a final model to a production environment. We identified two groups of artifacts for deployment: those concerning \CatText{verification} and \CatText{oversight}. Verification artifacts result from monitoring the performance of a model (both before and after deployment)~\cite{wang:autods:2021}. They include the generation of summary statistics, explicit comparisons to existing benchmarks~\cite{Zoller:BenchmarkAS:2021,Zimmer:Auto-Pytorch:2021,zhang:autocp:2020}, and the detection of model drift or anomalies~\cite{Celik:Adaption:2021,elshawi2019automated,Spinner:explainer:2020}. These artifacts are important to capture changes in the model over time and frequently feed into the oversight artifacts. These oversight artifacts include documentation that describes the model's characteristics, for example, a model card~\cite{Mitchell:model_cards:2019}, decision forensic reports~\cite{wang:autods:2021}, provenance artifacts of use~\cite{Spinner:explainer:2020}, as well as documents governing the use the model~\cite{Crisan:AutoML:2021,wang:autods:2021}. Oversight artifacts provide a key point of knowledge sharing where humans monitor the model to ensure it is responsibly applied~\cite{wang:autods:2021}. Moreover, these artifacts, automatically generated by an analyst, provide important avenues for humans to intervene in automated work. For example, suppose a deployed model in production begins to exhibit poor performance on benchmark datasets. In that case, oversight artifacts can initiate a process where a human returns to the analysis and manually re-initiates the model fitting processes.

\texttt{\textbf{Communication processes}} artifacts in our taxonomy are primarily documents, both static (i.e., a report) or interactive (i.e., a dashboard) to \CatText{report} information. While communication encompasses humans communicating with each other, AutoML systems must also communicate with humans. Once again, there is an opportunity to learn from human-human communication to make human-machine communication more effective. Communication artifacts include an automated summary of insights or an explanation for the model's decision-making. Modeling explanations are automatically produced and are increasingly crucial for transparency ~\cite{wang:DS_automation:2021,Sperrle:HCE_STAR:2021,Brent:Explain:2019,Spinner:explainer:2020}.

\texttt{\textbf{Interactive processes}} are an outlier relative to other processes. We believed they should be treated separately as they represent distinctly human actions that can not be automated but seek to influence automated processes. Many artifacts in other phases can be generated  by some combination of human or machine actions. We separate interactive processes into the artifacts of the \CatText{graphical user interface} and the \CatText{user} themselves. Elements of the user interface include bookmarked or saved insights~\cite{wongsuphasawat:voayger:2016,Cutler:Trrack:2020}, annotations~\cite{Cutler:Trrack:2020,Renan:ml_prov:2019,estevez:semantic_annots:2019}. Humans can also trigger or modify automated processes ~\cite{Celik:Adaption:2021} across data science processes. Increasingly, these user actions are captured as behavioral graphs, interaction logs, or interaction sequences~\cite{battle:eva:2019,hong2020evaluating,Cashman:EMA:2019,stitz:kx_pearls:2020}, that can be visualized~\cite{stitz:kx_pearls:2020,Cutler:Trrack:2020}, to influence a machine learning component through semantic interactions~\cite{Gehrmann:SemanticInference:2020,Endert:2012}.

\subsection{Artifact Properties}\label{artifacts-properties-final}
\begin{figure}
    \centering
    \includegraphics[width=0.7\columnwidth]{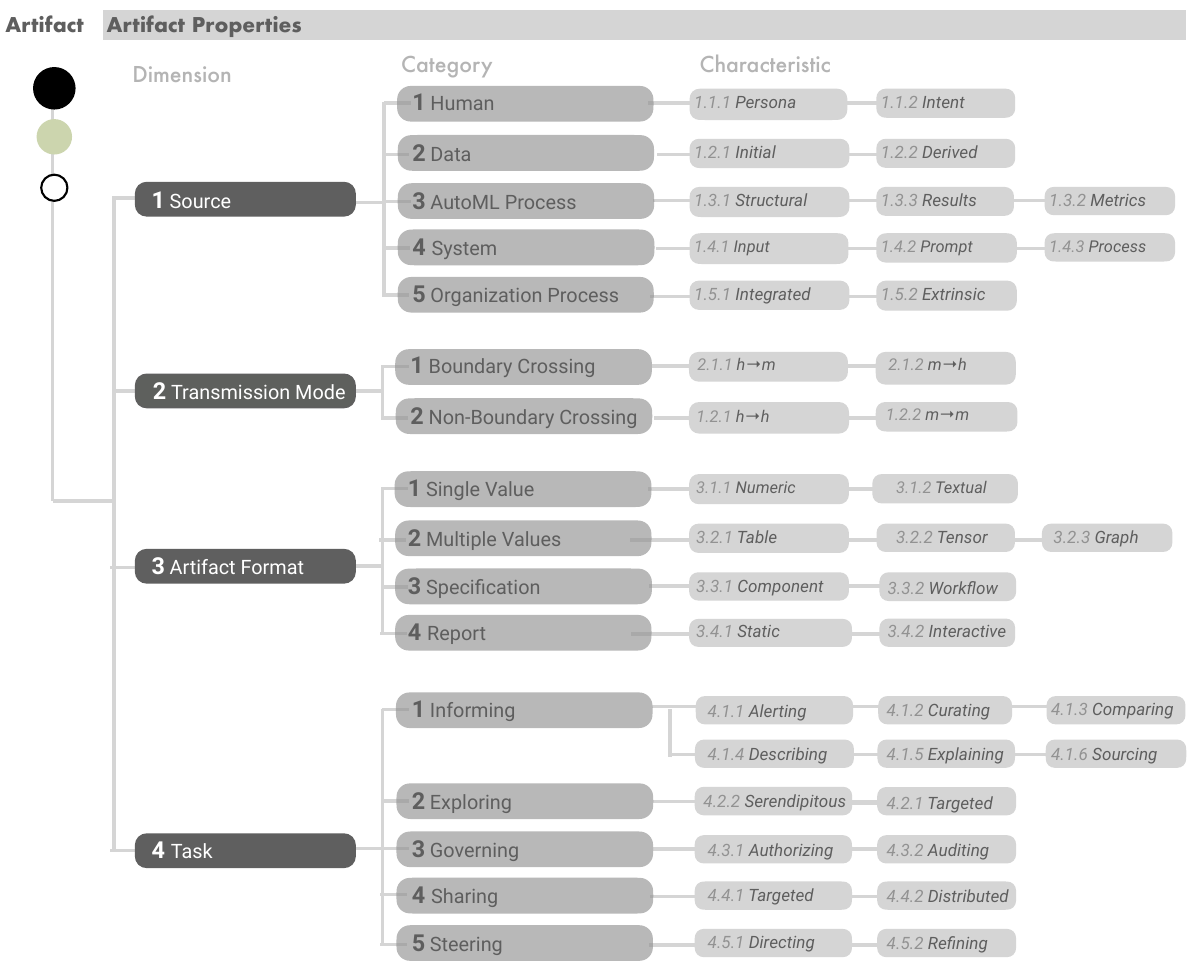}
    \caption{Artifact properties are a set of hierarchical taxonomic descriptors. The top level of this hierarchy is a dimension, followed by category, and finally characteristics.}
    \Description{A hierarchical diagram that summarizes Artifact properties are a set of hierarchical taxonomic descriptors. The top level of this hierarchy is a dimension, followed by category, and finally characteristics.}
    \label{fig:artifact_properties}
    \vspace{-5mm}
\end{figure}

The AutoML artifacts described in the previous section were determined by their properties. 
We used the initial set of 400 artifacts collected from the literature to derive a set of properties that allowed us to further group them into a smaller set.

The complete set of artifact properties is shown in~\autoref{fig:artifact_properties}, but to  avoid excessive repetition, a detailed breakdown of artifacts and their characteristics is in Appendix~\ref{taxonomy-extra}. While the initial goal of taxonomization was to \textit{describe} artifacts, we also found it useful for properties to be able to \textit{compare} them as well. For example, two AutoML pipelines may include a feature generation phase, which would produce a common artifact of a feature set. However, feature generation can be done automatically in one pipeline, whereas in the other, it is the job of a human. In both pipelines, the subsequent hyperparameter tuning may be done automatically. 
We endeavored for our taxonomy to describe a broad design space of AutoML systems; both implemented and theoretical. 

At the top level, our taxonomy has four dimensions that answer the following four questions: \textbf{\textit{``What generated the artifact?''}}~(Source), \textbf{\textit{``Does it cross the boundaries between human and AutoML processes''}}~(Transmission Mode), \textbf{\textit{``What shape does the artifact take?''}}~(Artifact Format), and finally, \textbf{\textit{``What is its intended purpose?''}}~(Task).

The \textbf{\texttt{Source}} of an artifact indicates by whom, or what, it was produced. We identified five sources: humans, an organization of humans, the data, AutoML processes, and the computational system. The first two sources distinguish between humans, acting individually and collaboratively, and a general set of organizational practices (i.e., business practices, legal or regulatory requirements) that can influence these people. Calculations, transformations, and other derivations from the initial dataset also produce new artifacts. Finally, AutoML processes and the computational infrastructure supporting that automation produce complementary but separate artifacts. For example, the former might produce a running summary of the model's loss, whereas the latter records and returns code failures or when computational budgets have been reached. 

The \texttt{\textbf{Transmission Mode}} properties describe whether the artifact has crossed boundaries between human and AutoML sources and in which direction. We have prioritized artifacts that are likely to transmit between humans and machines ($h \rightarrow m$) and vice-versa ($m \rightarrow h$); we determined directionality from reading the literature sources. Some artifacts that do not cross boundaries in the specific AUtoML system are critical to include as they add context to boundary cross artifacts.

The \texttt{\textbf{Artifact format}} property enables comparison between different AutoML pipelines. In our taxonomy development, we observed that artifact formats were closely tied to the design choices of AutoML systems. For example, AutoML systems that targeted an ML-expert end-user had artifacts limited to single values, texts, or tensors when displaying this information. Those that target domain experts presented the same data visually or interactively. We summarize four formats: single values, multiple values, specifications, and reports. Visualization systems and dashboards that we discuss in Section~\ref{related_work} are considered reports that have either static or interactive characteristics.

The \texttt{\textbf{Task}} describes the affordances of the artifact. We proposed four categories of tasks: informing, governing, sharing, and steering. Artifacts that inform, and describe the prior or current state of the data science pipeline. These can include reports, summary statistics, or a dashboard (among other possibilities). Governing artifacts are specific to regulating, auditing, and monitoring both automated and human-driven work. Sharing artifacts are intended to be distributed amongst humans, not just between analysis and the AutoML system. Finally, steering artifacts intervene anywhere in the data science pipeline to make a change. These artifacts result from human or automated processes acting on, for example, an alert to a data quality issue. 

\subsection{Further Extension}
The taxonomy itself can be further expanded over time, accommodating new artifacts that emerge as the capabilities of AutoML systems expand or to include highly bespoke qualities of specific system implementations. As we developed our taxonomy, we constantly reflected on its extensibility as part of our evaluation criteria. Specifically, as we merged the many different prior taxonomies specific to AutoML and machine learning~\cite{dellermann:future_collab:2021,tateman:repo_taxonomy:2018,von_Rueden:informedml:2021}, typologies of visual analysis~\cite{Brehmer:Typology:2013,Lam:GoalstoTasks:2018}, and other classification systems~\cite{Sacha:Vis2ML_ontology:2019,sant:automl:2021,Renan:ml_prov:2019,Sperrle:HCE_STAR:2021}, we scrutinized stability of our taxonomy to incorporate these changes. Moreover, our stopping criteria were predicated on the stability of the taxonomies structure.  We rely on future work to continuously reflect on its extensibility, as the present taxonomy incorporates currently available and relevant prior research.
\section{\SYSTEMNAME}\label{automl-trace}




We operationalize our artifact taxonomy through the creation of~\SYSTEMNAME, an interactive visual sketch~\cite{greenberg:usability_eval:2008}. \revCHIADD{Visual sketches are lower fidelity compared to more complex interactive prototypes but serve an important role in facilitating co-creation activities between researchers and their collaborators~\cite{Sanders:sketch:2014}. By comparison, Sanders~\textit{et. al.}~\cite{Sanders:sketch:2014} define prototypes to be  more mature in their conception and execution, which, in concurrence with Buxton and Greenberg~\cite{greenberg:usability_eval:2008}, can be counterproductive for co-creation and ideation. In this spirit, we develop~\~\SYSTEMNAME\, to investigate the utility of applying our taxonomy to the visual analysis of an existing AutoML system.}
\revCHISUB{The goals of AutoMLTrace are to facilitate a collaborative dialogue between researchers and developers of an AutoML system.} \revCHIADD{Although our goals are ultimately to develop~\SYSTEMNAME\ with the purpose of facilitating a dialogue with our collaborators (presented in Section~\ref{usage-scenario}),~\SYSTEMNAME, together with our taxonomy, can be repurposed to analyze AI/ML systems more generally. We especially aimed to emphasize the human element through the capture of artifacts and the delineation of their properties to illuminate human-ML/AI collaborative processes within AutoML systems. }
\revCHIADD{This section describes (1) how our taxonomy enables us to identify, classify, extract, and visualize both human and machine-derived artifacts (2) the overall design of our interactive sketch, ~\SYSTEMNAME, including the data and tasks it supports.}

\subsection{Operationalizing our Taxonomy}
Our AutoML artifact taxonomy captures human and machine-derived artifacts in an end-to-end pipeline of data work, from preparation to communication. Individual artifacts and their properties allow us to accommodate different degrees of automation, from human-driven to fully automated, and the hybrid modes in between~\cite{Parasuraman2000,Crisan:AutoML:2021,sant:automl:2021}. In hybrid automation modes, we capture the directionality of work --- from humans-to-machine processes ($h \rightarrow m$) and vice-versa ($m \rightarrow h$). With the addition of temporal information, we use our taxonomy to derive both the context and the time of artifacts' creation. By continuously capturing artifacts across an automated data work pipeline, we can show the evolution of data work and human-ML/AI collaborative processes over time.   

\subsubsection{Artifacts utilized in ~\SYSTEMNAME} 
\revCHIADD{We used artifacts captured from the enterprise team's AutoML pipeline within our interactive sketch. Using their artifacts directly not only provided an example of how our taxonomy can be operationalized with a live system, it promoted meaningful engagement with data important for the team's AutoML tool development.}
The first step to operationalize our taxonomy is to leverage it for identifying and characterizing artifacts from the AutoML system. Some artifacts can be captured programmatically as inputs to AutoML systems or outputs from different APIs. For example, a human can specify goals or targets through an interactive interface. Alternatively, AutoML processes can initialize and traverse a search space to find optimal sets of model parameters. Both the user input and the search space exploration can be captured from system logs. Other artifacts are manually captured. For example, documents that state a system's requirements or presentations communicating the results need to be captured from an existing document management system or other curation efforts. As these items are captured, either automatically or through curation efforts, the context of their creation (e.g., preparation, analysis, deployment, or communication stage) is provided through the taxonomy's structure and the artifacts' properties.

Our taxonomy allows us to identify the way in which these artifacts are generated and assign properties to them via manual annotation. For example, designers and ML/DS engineers can discuss the various inputs and outputs in the workflow, identify the type of artifact it may be, and describe them consistently with the taxonomy's controlled vocabulary. ~\SYSTEMNAME\ can support this process by defining a default template of artifacts and visually indicating what is captured or absent. However they are captured, the final result is a collection of artifacts traded between humans and automated processes in data work. \revCHIADD{Though we used a specific pipeline from the team's AutoML system in ~\SYSTEMNAME\, the captured artifacts are characterized by properties of our taxonomy developed for a range of AutoML systems. Considering the scaffolding provided by the taxonomy for artifacts, ~\SYSTEMNAME\ remains applicable to other pipelines. Future work would provide evaluation of the interactive sketch for engagement with other pipelines as well as provide further automation of the artifact annotation process.}

\revMR{\subsubsection{Tracing the chronology, dependencies, and variability of artifacts.}\label{chronology-dependencies} 
In addition to the creation context, we can collect a timestamp of artifact creation that enables us to examine the \textit{order} of their creation and dependencies. For example, feature generation artifacts serve as inputs to model fitting. We can also examine how \textit{artifacts change over time}. For example, say the initial set of features was generated automatically  by an AutoML algorithm, a human examining the artifact decides to update these features with their own manual selection. Now, two versions of the artifact exist. Through the artifact's properties, it is possible to identify that the first version of the artifact was created automatically, but the subsequent version resulted from human intervention.}

\vspace{-5mm}
\revMR{\subsubsection{Describing and comparing human-ML/AI collaborative analyses.}\label{describe-compare} Collaboration between human and ML/AI systems makes it hard to audit and compare analyses. We propose that by annotating analysis through our artifact taxonomy, we directly describe and compare the different analytic choices and deduce some level of automation, from full automation to none and varying degrees in between~\cite{Parasuraman2000,Lee2019AHP,sant:automl:2021}.}

\subsection{Data and Tasks}
We use both the individual artifacts and their collective metadata as an input dataset for~\SYSTEMNAME\ to visualize. Individual artifacts come in different formats that influence how they are captured and how they are visualized to the end users; we define these different formats in our taxonomy as part of the properties of an artifact (Figure~\ref{artifacts-properties-final}). The taxonomy along with additional information, such as timestamps and pipeline structure, define the metadata for a collection of artifacts. To facilitate an engaging, collaborative dialogue around these artifacts, we define a set of tasks that our interactive visual sketch should support:

\begin{itemize}
  \item \textbf{T1 \underline{Present} a Contextual Overview of Artifacts:} \revMR{The contextual overview ties the artifact creation with its specific data science phase (see Section~\ref{trace-define}). Whether an artifact was generated automatically or by a human was important -- this consideration would become a key component of the~\SYSTEMNAME design. The dependencies of artifacts on each other were also an important contextual component. }

 \item \textbf{T2 \underline{Locate} an Artifact:} Enable end-users to filter out artifacts they are not interested in and to focus on a specific artifact, or group of artifacts, that are of interest to them.
  \item\textbf{T3 \underline{Summarize} the Details of the Artifact:} Artifact details, like its properties and dependencies, should be progressively revealed to the end-user. Similarly, an artifact's taxonomic descriptors should reveal artifacts that share the same properties, not just those that a selected artifact depends on.
 \item \textbf{T4 \underline{Compare} an Artifact over its History: }  The end-user should be able to compare the states of an artifact over time and relative to its upstream and downstream dependencies. 
\end{itemize}

These tasks align with those for information seeking that were defined by Shneiderman~\cite{Shneiderman:TTT:1996} (Overview, Zoom, Filter, Details on Demand, Relate, Histories, and Extracts), but described using a terminology of more recent task typology defined by Brehmer and Munzner~\cite{Brehmer:Typology:2013}.

\subsection{\SYSTEMNAME~Interface}\label{prototype-ui}
\begin{figure*}
    \centering
    \includegraphics[width=\textwidth]{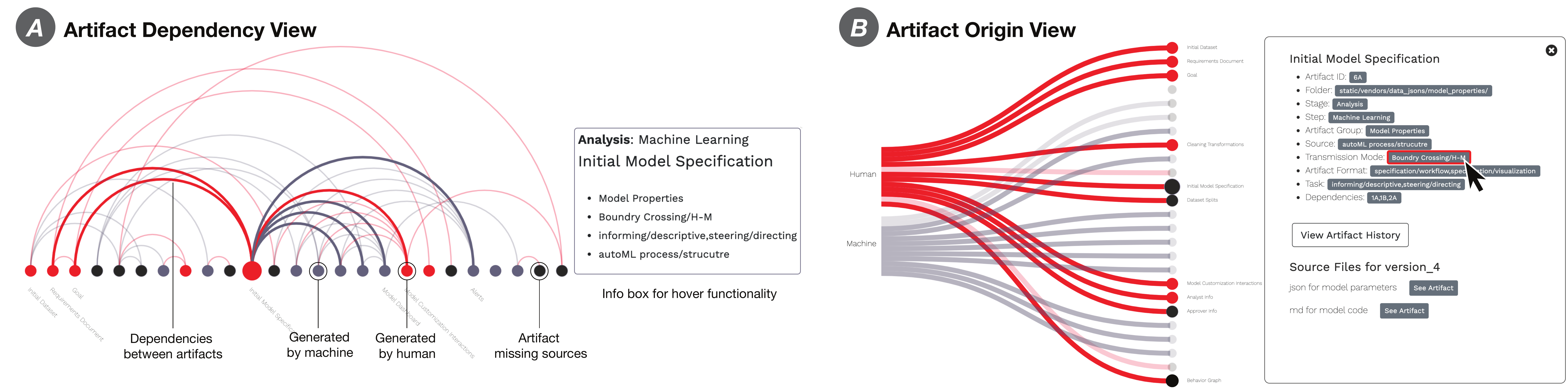}
    \caption{\revCHIADD{(A) Artifact Dependency View. This view shows the cascade of dependent artifacts in the context of the previously defined phases of data work. The color of the circles indicates whether the artifact was machine or human generated. The arcs illustrate explicit dependencies between one or more artifacts. When an artifact is selected, the artifact's dependencies are highlighted in the visualization, making it easier to track what is affected when a given artifact changes. In the example, the initial model specification is selected, showing the other artifact dependencies along with the artifact's characteristics from the taxonomy.  (B) Artifact Origin View shows what artifacts are human-generated versus machine-generated. In addition, this view allows the user to find commonalities in artifact properties defined by the taxonomy. Once again, the initial model specification is selected, and the user can hover over the selected artifact's parameters to see other artifacts sharing the same characteristic. In the example, the transmission mode indicates this artifact crosses the boundaries from human to machine. Hovering highlights all other artifacts that cross the boundary from human to machine.}
    \Description{Illustrations showing (A) Artifact Dependency View. This view shows what artifacts are dependent on one another. (B) Artifact Origin View shows what artifacts are human-generated versus machine-generated. This shows the Initial Model Specification artifact is selected, showing the info panel of the artifact's parameters.}}
    \label{fig:af_dep_view}
\end{figure*}

\begin{figure*}[h!]
    \centering
    \includegraphics[width=.8\textwidth]{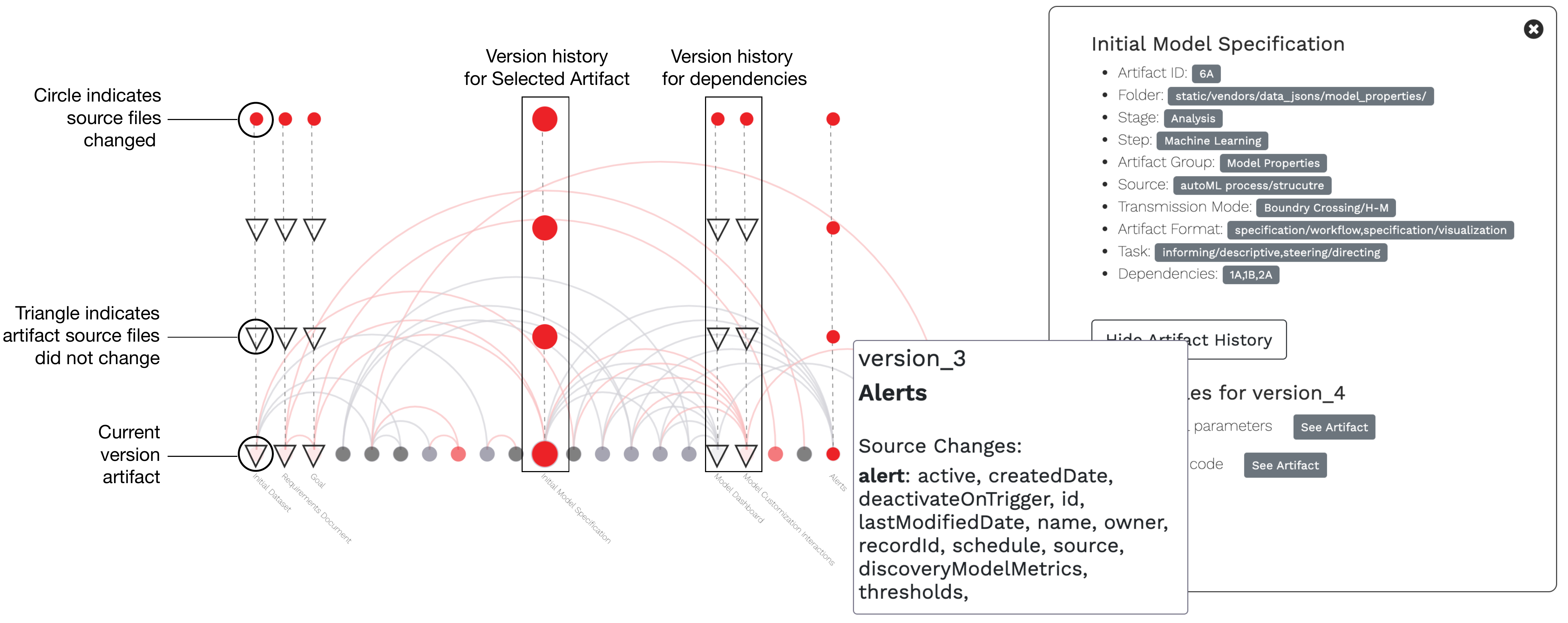}
    \caption{Artifact History View. This view shows the history of the selected artifact, differentiated by analysis versions. From the dependency view, the histories of the selected artifact dependencies are also shown. This example shows the history of the selected artifact ``Initial Model Specification''. The tool-tip shows details for the third version of the ``Alerts'' artifact, which is a dependency of ``Initial Model Specification''. }
   \Description{Artifact History View. This view shows the history of the selected artifact, differentiated by analysis versions. From the dependency view, the histories of the selected artifact dependencies are also shown. This example shows the history of the selected artifact ``Initial Model Specification''. The tool-tip shows details for the third version of the ``Alerts'' artifact, which is a dependency of ``Initial Model Specification''.}
    \label{fig:hist_view}
\end{figure*}

\SYSTEMNAME\ takes a collection of artifacts and their metadata as input for visualization. \revMR{It has three complementary views : Origin (\autoref{fig:af_dep_view}), Dependency (\ref{fig:af_dep_view}), and History views (\ref{fig:hist_view}). The encoding choices for the artifacts were the same for all views to maintain a consistent visual language.} The artifacts are represented as circles, color-coded by their origin (human or machine), and aligned by the Data Science phase (preparation, analysis, deployment, and communication). These views are inspired by the graph and network visual approaches from prior AutoML systems and studies (see Section~\ref{related_work}), although we did consider alternative designs (see Supplemental Materials -~\suppDesignExamples). \revMR{As this is an interactive sketch, we do not exhaustively compare it against other design alternatives.}

\vspace{1mm}
\noindent\textbf{Origin View: \textit{What artifacts are human versus machine- \\generated?}} The artifact origin view shows the artifacts collected from the AutoML system analysis in the context of whether they were generated by a human or automatically ~\ref{fig:af_dep_view}. We use an alluvial diagram to show the flow and trade-off between the origins of the artifact (\textbf{T1} (Present)). We emphasize human and machine-generated artifacts as a focal point of this view as a way to showcase the interleaving collaborative processes.  

Hovering triggers additional taxonomic details to be revealed on demand via an information card (\textbf{T3}). End-users can further hover on the taxonomic descriptors and contextual data such as dependencies and data science stage (\textbf{T2}). Once an artifact is selected, end-users can also view the raw source file outputs for the artifact.


\noindent\textbf{Dependency View: \textit{What artifacts are dependent on one another?}} The dependency view show the relationships between artifacts(Fig. \ref{fig:af_dep_view}). The design of this view is inspired by the illustration of Data Cascades~\cite{Nithya:datawork:2021}; indeed, this view is a direct response to surfacing those cascades through artifacts. Similar to the origin view, the end-user is presented with an overview (\textbf{T1}), and information is revealed via hover actions (\textbf{T3}). However, in this view, selecting an artifact highlights its dependencies~(\textbf{T2}). 

\vspace{1mm}
\noindent\textbf{Version History View: \textit{How did changes in one artifact influence changes in other artifacts?}}
This view is used to drill down into artifact histories and understand how changes in one artifact could influence changes in dependent artifacts (\autoref{fig:hist_view}). Users can view the artifact history by selecting a given artifact in either the Origin or Dependency view. This view enables end-users to \textbf{T4} (Compare) and the artifact itself over time as others. In \autoref{fig:hist_view}, there are four horizontal lines, which correspond to four revisions, or iterations, of the analysis. New artifacts or those modified by the update are represented as circles. Those that did not change are shown as a downward triangle. The dependencies for a selected artifact are also drawn. Like the previous two views, hovering reveals additional taxonomic descriptors of the artifact.

\section{Usage Scenario}\label{usage-scenario}
We present a usage scenario with a team of enterprise software developers where we use our taxonomy and~\SYSTEMNAME\ to explore and analyze their existing AutoML system. \revCHIADD{Our collaborators goal was to analyze their existing AutoML systems to understand when, what, and how end-users of their system intervened or overwrote the decisions of the AutoML system. At present, they had not mechanism for this interrogation.} We describe how~\SYSTEMNAME\ supported dialogue during discussions with the team to reflect on the systems' present capabilities and ideate around outstanding end-user needs. \revCHIADD{These discussions lasted one hour and occurred at a cadence of every two weeks for approximately a 3 month period. The entire team joined the meetings and discussions. In these conversations we presented collaborators with our analysis of their AutoML system as artifacts that we collected and annotated using our taxonomy. We refined both our collection of artifacts and the visual design of ~\SYSTEMNAME\ in response to collaborator feedback and discussion.}

\subsection{Collaboration Context}

\vspace{1mm}
\noindent\textbf{Overview of the existing system.} Their AutoML systems could automate aspects of data work from preparation to deployment (Section~\ref{trace-define}), including surfacing automatically flagged insights for exploring data, feature generation, and automated model selection.  A graphical user interface (GUI) guided end-users through the analysis and revisions of the results. The end-user could intervene to modify the analysis, for example, change the model type,  via input widgets and interactions through the GUI. Certain aspects of the system also required explicit human input \textit{before} initiating an automated process. For example, the systems would surface multicollinearity (in a non-technical manner) and required that the end-user confirm which features to remove from the analysis. The end-users could deploy a model to be used by others; automated processes would also monitor for concept drift and, if necessary, alert the end-user to trigger updates. 

\vspace{1mm}
\noindent\textbf{Team composition and collaboration goals.} Our collaboration had two primary goals. First, for the team to reflect on their existing system and better understand what AutoML systems are capable of. Our taxonomy created an avenue for this reflection by providing a structured vocabulary to describe their system and compare it to others. The second goal was to examine what the additional traceability would add to their system.
The project team consisted of software engineers, designers, user researchers, and a project manager. We also recruited one customer of their system for additional feedback. 
The team worked together to implement different components of AutoML work and implement the system.

\subsection{Artifact Identification, Classification, and Extraction}
We briefly describe how we analyzed our collaborator's existing system to develop a collection of artifacts visualized with~\SYSTEMNAME.

\vspace{1mm}
\noindent\textbf{Generating Artifacts.}
Within the GUI environment of the AutoML system, we created an end-to-end data analysis. We began with preparation and concluded with communication. 
During this process, we returned to earlier steps and made modifications. For example, we did not initially apply automatic data-cleaning recommendations but did so in a subsequent iteration. We also let the system pick features for the model in the first iteration and subsequently changed them. Carrying out this analysis had three goals. First, to produce a variety of possible artifacts, second, to document dependencies between artifacts, and finally, to observe how artifacts change in response to user interactions. The result was a set of artifacts derived from the same analysis that changed over time.

\vspace{1mm}
\noindent\textbf{Collecting Artifacts.}~\revMR{We used APIs developed by the team to collect a set of JSON files for our analysis. We used the API outputs over other approaches (i.e., usage logs) because these outputted the entire artifact, making it easier for us to classify the artifact according to our taxonomy. We additionally stored the order in which objects were created and could establish the dependencies of artifacts (Section~\ref{chronology-dependencies}). Like many existing AutoML systems, they did not explicate any human involvement. We had to manually record when an artifact was generated or modified by human intervention needed.} In the infrequent instances where we could not capture all aspects of the analysis, but we deemed an artifact was important, 
we took a screenshot of the artifact. For example, some of the automatically generated insights for data exploration had visualizations that could not be extracted from the APIs, so we took screenshots instead. 


\vspace{1mm}
\noindent\textbf{Classifying Artifacts} We annotated the files from the API calls or screenshots using our artifact taxonomy. 
However, in most instances, a single file contained multiple artifacts. For example, an API call for information on the initial dataset returned this information along with information on recommended wrangling transformations. \revMR{The authors first \textit{identified artifacts} from these APIs by manually inspecting them in a simple development environment - demarcating and marking up instances of artifacts. Next, the authors examined each of the artifacts individually and \textit{classified them according to our taxonomy}, modifying their properties as was pertinent to analysis (i.e., whether it was human or automatically generated). Finally, we examined and recorded the dependencies among artifacts. The authors repeated these two steps until they reached a consensus on the artifact type and its properties; we also engaged with our collaborators to verify that our artifacts were accurate.}
A final list of artifacts and their taxonomic annotations is available in the Supplemental Materials~(\suppDesignSpec); this list served as a backbone of our~\SYSTEMNAME~implementation. 

\subsection{Collaboration and Question Elicitation}\label{prototype-feedback}
\revMR{As a final step, we presented~\SYSTEMNAME~to our collaborators via chauffeured demonstrations~\cite{lloyd:chauffered_demo:2011} conducted over video conferencing platforms.}
We demonstrated the functionality and affordances of~\SYSTEMNAME~and our collaborators were given opportunities to provide feedback. \revMR{We iterated between discussing the analysis we conducted using their existing platform and the artifacts we harvested and visualized via~\SYSTEMNAME. This was an important step in our assessment, as it reinforced to our collaborators that traceability could be added to their existing system, as all artifacts of a real analysis were captured through their APIs. The team was excited to view their artifacts and their system's capabilities in this way.}

We wanted to collect the types of questions our prototype would elicit during these feedback sessions. The engagement was dynamic, with both the authors posing and responding to questions about the artifacts, their sources, dependencies, and changes over time. What our collaborators appreciated most was being able to see their system laid out according to our taxonomy. This new view of their system led them to examine aspects of their work from a perspective they had not previously considered. We summarize our discussion into three common themes: seeing and describing dependencies, comparing sequences analyses over time, and comparing how their system differed from others.

\vspace{2mm}
\noindent\textbf{Seeing and describing dependencies.} \revMR{Visualizing the dependencies of individual artifacts and the different types of artifacts} was something they had not been previously able to do. \revMR{They were especially interested and excited to see how the human and machine-generated processes interleaved through the analysis. This combination of the origin and dependency view allows them to infer} potential causal relationships between an artifact's current state and other actions. As we have previously indicated, many AutoML systems do not explicate the role of humans, but, with~\SYSTEMNAME\ the impact and effect of the human's role are undeniable. \revMR{The team saw the benefit of visualizing the analysis in this way as a way to reflect on the system's design. They also saw the benefit of surfacing such relationships to support governing an analytic pipeline. For example, if authorization is required to deploy a model, they saw~\SYSTEMNAME\ as a useful way to audit the existing analysis to either recommend or decline deployment.}


\vspace{2mm}
\noindent\textbf{Comparing sequences of analyses analyses.} 
\revMR{Our collaborators were also interested in using~\SYSTEMNAME\ to compare analyses conducted by multiple analysts over time. They specifically wanted to have multiple analysis sequences generated by different actors and to compare them. This scenario of asynchronous human collaboration, together with individual human-machine collaboration, is a promising sign of our taxonomy's utility for more complex problems.  While we can version artifacts, enabling detailed comparisons, our current design is not well optimized for multi-human collaboration - although, this again, points to fruitful directions for future work.}
 
One collaborator was particularly interested in understanding when humans took machine suggestions and applied them and when they ignored suggestions. A specific artifact sequence of interest began with the initial dataset, followed by wrangling transformation recommendations with a machine source. Then the recording of user actions would indicate whether the end-user applied any wrangling transformation recommendations. Finally, it concluded with any potential updates to the initial dataset. \revMR{This is yet another interesting usage scenario that not only enables us to understand a system's level of automation but, potentially defines signatures of automation per user. Moreover, it is also possible to assess whether some artifacts are modified more often than others. Collectively, these signatures could be leveraged to identify problematic features (for example, if the machine's results are constantly overwritten) or patterns of analysis behavior. }



\vspace{2mm}
\noindent\textbf{Comparing their system to others.} The taxonomy we developed is an amalgamation of various systems that span both human and machine processes. Our taxonomy provided a standard vocabulary for comparing these systems and reflecting on what artifacts might be missing relative to another system. For example, more recent advances in AutoML technology includes a computational budget to enable these automated processes to complete within a reasonable time frame and budgetary constraints. 
However, not all AutoML systems have such features.
Our taxonomy prompted a discussion of the design implications for our collaborator's system. They were first comforted to see that their existing system had elements that overlapped with others, but, could also see other interesting aspects that were absent in their current implementation.

\subsection{Summary}
In this second phase of research,  we probed the utility and ecological validity of taxonomy by collaborating with a team developing a complex AutoML system. The~\SYSTEMNAME\ sketch demonstrates that a taxonomy is a useful boundary object to engage with a team of software and ML/AI experts designing human-ML/AI collaborative systems. It also demonstrates that traceability has valuable applications to both human-machine and human-human collaborations. While our approach does not address all of the design challenges for evolving and adaptive systems~\cite{Yang:HAI_hard:2020}, it does take preliminary steps toward doing so.
\section{Discussion}\label{discussion}
Human collaboration with ML/AI systems will grow more ubiquitous as AutoML technology becomes increasingly integrated within data work. These systems lower the barrier for data work and help data scientists triage their work more effectively~\cite{wang:DS_automation:2021}. 
\revCHIADD{However, as existing systems still require human oversight and intervention, these semi-automated systems need to be observable and understandable. 
Recent work from the HCI community identifies challenges in scrutinizing ML/AI systems stemming from the complexities of human-ML/AI collaborative work and emphasizes the need for a common language for discourse in this space ~\cite{Yang:HAI_hard:2020}}.

\revCHIADD{Our work addresses limitations in human-ML/AI collaboration in several ways. First, we formalized a common language that accounts for human and machine aspects of these systems by creating an AutoML artifact taxonomy. Second, we operationalized this taxonomy in our interactive sketch AutoML Trace, characterizing these artifacts to facilitate a traceable workflow. Third, we characterized traceability for scrutinizing this highly complex and heterogeneous process. }
While prior research captures aspects of traceability through provenance tools (i.e.,~\cite{wang:atmseer:2018,pipelineprofiler2020,Liu:boba:2021}), they fail to differentiate between human and automated processes and frequently ignore human processes altogether.
Research in human-human collaboration and knowledge sharing has highlighted the importance of artifacts for capturing~\cite{Kreiner:tacit_artifact:2002,Lee:BO:artifacts:2007} and tracing~\cite{Tiwana:kx_system:2001,Fischer:kx_mgmt_og:2001,Mariano:kwx_mgmt_survey:2016} complex collaborative processes.
By considering traceability, we offer a different perspective on artifacts. We argue that traceability encourages a broader consideration of an artifact's lineage and the contextual factors of its generation and use. 
Moreover, through artifacts, our research acknowledges and elevates the sociotechnical relationships between humans and ML/AI systems. 

Beyond provenance, contemporary research is increasingly focused on the importance of transparency, interpretability, and explainability toward ML/AI systems~\cite{barrero:xai:2020,Harmanpreet:XAItechniques:2020,Brent:Explain:2019,Spinner:explainer:2020,Amershi:guidelines:2019}. However, this prior work focuses on the model itself and misses influential factors throughout the data cascade~\cite{Nithya:datawork:2021}. Our research expands the scope, capturing artifacts across an end-to-end pipeline of data science work through artifacts and taxonomies. We demonstrate that taxonomies can be robustly created and can serve as boundary objects for designing human-ML/AI collaborative systems. Our approach shows it is possible to have~\textit{``both transparency of process and transparency of product; the former refers to the transparency of the human processes of research and innovation, the latter to the transparency of [...] AI systems so developed.''}~\cite{Winfield:ethics_transparency:2018}.

Lastly, our research acknowledges and describes the difficulties of developing visual and interactive systems for human-ML/AI collaboration in data work. Design studies and other application-type research focus primarily on end-users, but complex systems still require the engagement of ML/AI experts. The collaboration between researchers and experts who are not the end-users remains complex and can require visualization tools as intermediaries to facilitate a dialogue~\cite{Yang:HAI_hard:2020}. \textit{\textbf{Absent reliable scaffolds for this dialogue, we took on the ambitious task of creating them. Developing an AutoML artifact taxonomy and~\SYSTEMNAME\ created boundary objects that we used to address these challenges. Our intent in describing our process is to provide possible avenues for other researchers facing similar challenges.}}


\subsection{Implications of Our Findings and Future Work}\label{future-work}

\subsubsection{On Design and Evaluation of Human-Centered AutoML Systems} Our artifact taxonomy can be used to reflect upon existing systems and ideate new ones. One of the limitations of existing guidelines for human-ML/AI interaction is that they target the initial ideation of the system and are less effective should a system already exist. In our case study, we observed an artifact taxonomy's potential to reflect design retrospectively and prospectively. This potential is essential to identify and modify ineffective approaches.
Our taxonomy serves to help researchers and practitioners ideate on new systems and speculate what an ML/AI system could do~\cite{Yang:HAI_hard:2020} while promoting reflection on the role of humans. 

\subsubsection{On Data Science Collaboration}\label{future-work:collab}
Different kinds of data workers are engaged across data work~\cite{Crisan:AutoML:2021,wang:autods:2021,Wang:hai_collab:2019}. 
Further work is needed to understand how different data science personas~\cite{Crisan:Batton:202}, from ML engineers to technical analysts, would use this taxonomy. Prior research shows that people trade-off aspects of data work amongst themselves~\cite{Crisan:Batton:202,zhang:ds_collab:2020}. Capturing and tracing artifacts can help a team of data workers understand what work was done and by whom (or what). Moreover, discussion around artifacts, visualized by tools like~\SYSTEMNAME\ can help teams of data workers make sense of and critique the analysis and its results~\cite{neff:critique_ds:2017}. Finally, while there exists some research exploring the relationship between data workers and levels of automation (i.e.,~Wang~\textit{et al.}~\cite{wang:DS_automation:2021}), the complex relationships of human-to-human with human-to-machine collaboration have not been explored. Our taxonomy may prove helpful for extending these prior studies to a more hybrid flow of data work. 

\subsubsection{On Data Visualization and Visual Analytics Tools}

\revCHIADD{Visual analysis tools leverage the advancement in machine learning to innovate on affordances visual analytic (VA) systems. Expanding our understanding of VA systems by capturing a more detailed catalog of artifacts would allow \textit{``users of the system to stay more engaged in the act of visual data exploration, as opposed to actively training the model and system,''}~\cite{endert2017state}. Inspired by Yang~\textit{et. al.}~\cite{Yang:HAI_hard:2020}, we were motivated to expand the view of what can be captured and surfaced from ML/AI pipelines beyond the model or analysis phase.  With the emphasis on a broader inclusion of the human element into what we capture and surface, our taxonomy is a step toward a more general view of human-generated artifacts captured across a workflow independent of a single system or pipeline. We hope others continue to utilize and expand on this in the space of VA systems.  Visualization and Human Computer Interaction researchers can build upon our research in two ways.
The first is expanding the scope of what can be visualized by VA systems for ML/AI. Our taxonomy
proposes a richer view of an AutoML pipeline that current work (Section~\ref{rel:prior-vis}) does not yet consider. Researchers can use our taxonomy to analyze and visualize other ML/AI systems, including but not limited AutoML systems, and even extend our taxonomy and contribute to expanding our catalog of ML/A artifacts. }
While our~\SYSTEMNAME\ interactive sketch proposes one possible visual approach, we believe there are rich opportunities to explore the space of visual designs.
The second is by expanding paradigms for human-ML/AI interaction. Data visualization tools are a medium for human and machine learning systems to work together~\cite{Spinner:explainer:2020,Sperrle:HCE_STAR:2021}. While interactions with these systems can be used to intervene with ML models~\cite{Endert:2012,Brown:2012,Gehrmann:SemanticInference:2020}, future work could extend this potential to other types of primitives and aspects of AutoML pipelines~\cite{Heffetz:deepline:2020,pipelineprofiler2020}.

\subsection{Limitations}
\revCHIADD{Data work involves a wide variety of different kinds of data workers.}
Further work is needed to understand how different data science personas~\cite{Crisan:Batton:202}, from ML engineers to technical analysts, would use this taxonomy. To re-purpose the adage about statistical models, \textit{``All taxonomies are wrong, some are useful.''} Like the taxonomies that came before ours, we strove to make our taxonomy useful to HCI, visualization, and machine learning researchers and practitioners. In service to this goal, we followed a rigorous process for taxonomy development proposed by Nickerson~\textit{et al.}~\cite{Nickerson:taxonomy:2012} with extensions from Prat~\textit{et al.}~\cite{Prat:taxonomy:2015}. We were diligent in documenting our taxonomy development and made artifacts of our research process available as supplementary materials so others might critique or extend our work. \revCHIADD{While this approach is more involved,  it also serves as an important alternative to \textit{ad hoc} approaches for taxonomy development that are difficult to interrogate and replicated.}
More generally, we argue that the research process brings greater attention to the importance of artifacts resulting from automation and human labor in data science work. 
Another limitation of our work is that our usage scenario excludes the ultimate end-user, the people conducting the analysis. The rationale for doing so was twofold. First, we needed some baseline to ground the development of a system like~\SYSTEMNAME. Absent this baseline, we needed to create one, hence, the primary contribution of our taxonomy.  The second rationale was that, for our present contributions, the developer team was the more appropriate group to conduct a preliminary assessment. In Section~\ref{usage-scenario}, we identify several fruitful ways to expand on our work and move toward end-user evaluations, including a broader investigation of the visual design space and an investigation of asynchronous, multi-human collaborations.

\revCHIADD{Finally, we reflect on our taxonomy development approach. Although our taxonomy was developed through a broad literature review, we assess it utility primarily through a collaboration with a single team. It is not uncommon for visualization research to focus on a core collaborator group, but, further work is required to assess its generality. We are optimistic that there is great potential
for using the taxonomy to compare disparate systems and pipelines from AutoML to visual analysis. We encourage the community to engage with the taxonomy operationalized in mediums they see fit for their data work. In addition, we look forward to seeing how this taxonomy can expand over time as  ML/AI technology advances. } 

\section{Conclusion}\label{the-end}
The growing ubiquity of AutoML technology enables a wider group of stakeholders to conduct data work but can also make it challenging to trace what was done and by whom (or what). Attempts to address these challenges are often stymied by the complexity of the systems themselves and a lack of available scaffolding for engaging with the ML and software developers that create these systems. We present a two-phased approach that explicates the collaborative relationships between humans and AutoML systems to carry out data work.  The first phase develops an artifact taxonomy that can be used to identify, classify, and describe artifacts from both socio- and technical sources. The second phase is a usage scenario with an enterprise development team that demonstrates the utility of our taxonomy as a boundary object. This usage scenario also reifies the value of tracing artifacts in the process of designing and developing collaborative human-ML/AI systems.
Our findings and contributions have implications for the design and evaluation of AutoML systems used to facilitate automation in data work. 


\vspace{1mm}
\noindent\textbf{Availability of Supplementary Materials.} Notes taken during the development of our taxonomy and the~\SYSTEMNAME~ prototype are \href{https://osf.io/3nmyj/?view_only=19962103d58b45d289b5c83421f48b36}{available online}\footnote{\url{https://osf.io/3nmyj/?view_only=19962103d58b45d289b5c83421f48b36}}. 

\begin{itemize}
    \item \suppTaxSpread~\textbf{AutoML Artifact Taxonomy Development :}~A spreadsheet of each artifact and their classification across each iteration of the taxonomy
   \item \suppTaxSpreadFinal~\textbf{AutoML Artifact Taxonomy Final :}~A spreadsheet of the final taxonomy, with artifacts, properties, and their classification. A summary is also shown in 
   \item \suppTaxIter~\textbf{Taxonomy Iterations Details :}~A summary of our decision process for adjusting the taxonomic dimensions, categories, and objects. It has a detailed annotation of each iteration, what changed, and how we assess our end conditions for taxonomy development.
   \item \suppDesignSpec~\textbf{Design Specification :}~Design specifications for AutoML Trace prototype that provide additional context for the content in Section~\ref{automl-trace}.
   \item \suppDesignExamples~\textbf{Design Alternatives :}~Design alternatives that were considered and discussed with our collaborators.
\end{itemize}


\bibliographystyle{ACM-Reference-Format}
\bibliography{main.bib}


\begin{thebibliography}{120}


\ifx \showCODEN    \undefined \def \showCODEN     #1{\unskip}     \fi
\ifx \showDOI      \undefined \def \showDOI       #1{#1}\fi
\ifx \showISBNx    \undefined \def \showISBNx     #1{\unskip}     \fi
\ifx \showISBNxiii \undefined \def \showISBNxiii  #1{\unskip}     \fi
\ifx \showISSN     \undefined \def \showISSN      #1{\unskip}     \fi
\ifx \showLCCN     \undefined \def \showLCCN      #1{\unskip}     \fi
\ifx \shownote     \undefined \def \shownote      #1{#1}          \fi
\ifx \showarticletitle \undefined \def \showarticletitle #1{#1}   \fi
\ifx \showURL      \undefined \def \showURL       {\relax}        \fi
\providecommand\bibfield[2]{#2}
\providecommand\bibinfo[2]{#2}
\providecommand\natexlab[1]{#1}
\providecommand\showeprint[2][]{arXiv:#2}

\bibitem[\protect\citeauthoryear{Agarwal, Ge, Shakeri, and Al-Rfou}{Agarwal
  et~al\mbox{.}}{2021}]%
        {agarwal:knowledge:2021}
\bibfield{author}{\bibinfo{person}{Oshin Agarwal}, \bibinfo{person}{Heming Ge},
  \bibinfo{person}{Siamak Shakeri}, {and} \bibinfo{person}{Rami Al-Rfou}.}
  \bibinfo{year}{2021}\natexlab{}.
\newblock \bibinfo{title}{Knowledge Graph Based Synthetic Corpus Generation for
  Knowledge-Enhanced Language Model Pre-training}.
\newblock
\newblock
\urldef\tempurl%
\url{https://arxiv.org/abs/2010.12688}
\showURL{%
\tempurl}


\bibitem[\protect\citeauthoryear{Alletto, Huang, Francois-Lavet, Nakata, and
  Rabusseau}{Alletto et~al\mbox{.}}{2020}]%
        {alletto:randomnet:2020}
\bibfield{author}{\bibinfo{person}{Stefano Alletto}, \bibinfo{person}{Shenyang
  Huang}, \bibinfo{person}{Vincent Francois-Lavet}, \bibinfo{person}{Yohei
  Nakata}, {and} \bibinfo{person}{Guillaume Rabusseau}.}
  \bibinfo{year}{2020}\natexlab{}.
\newblock \bibinfo{title}{RandomNet: Towards Fully Automatic Neural
  Architecture Design for Multimodal Learning}.
\newblock
\newblock
\showeprint{2003.01181}
\urldef\tempurl%
\url{https://arxiv.org/abs/2003.01181}
\showURL{%
\tempurl}


\bibitem[\protect\citeauthoryear{Amershi, Weld, Vorvoreanu, Fourney, Nushi,
  Collisson, Suh, Iqbal, Bennett, Inkpen, Teevan, Kikin-Gil, and
  Horvitz}{Amershi et~al\mbox{.}}{2019}]%
        {Amershi:guidelines:2019}
\bibfield{author}{\bibinfo{person}{Saleema Amershi}, \bibinfo{person}{Dan
  Weld}, \bibinfo{person}{Mihaela Vorvoreanu}, \bibinfo{person}{Adam Fourney},
  \bibinfo{person}{Besmira Nushi}, \bibinfo{person}{Penny Collisson},
  \bibinfo{person}{Jina Suh}, \bibinfo{person}{Shamsi Iqbal},
  \bibinfo{person}{Paul~N. Bennett}, \bibinfo{person}{Kori Inkpen},
  \bibinfo{person}{Jaime Teevan}, \bibinfo{person}{Ruth Kikin-Gil}, {and}
  \bibinfo{person}{Eric Horvitz}.} \bibinfo{year}{2019}\natexlab{}.
\newblock \showarticletitle{Guidelines for Human-AI Interaction}. In
  \bibinfo{booktitle}{\emph{Proc. CHI'19}}. \bibinfo{pages}{1–13}.
\newblock
\urldef\tempurl%
\url{https://doi.org/10.1145/3290605.3300233}
\showDOI{\tempurl}


\bibitem[\protect\citeauthoryear{Bansal, Nushi, Kamar, Weld, Lasecki, and
  Horvitz}{Bansal et~al\mbox{.}}{2019}]%
        {Horvitz:tradoff:2019}
\bibfield{author}{\bibinfo{person}{Gagan Bansal}, \bibinfo{person}{Besmira
  Nushi}, \bibinfo{person}{Ece Kamar}, \bibinfo{person}{Daniel~S. Weld},
  \bibinfo{person}{Walter~S. Lasecki}, {and} \bibinfo{person}{Eric Horvitz}.}
  \bibinfo{year}{2019}\natexlab{}.
\newblock \showarticletitle{Updates in Human-AI Teams: Understanding and
  Addressing the Performance/Compatibility Tradeoff}.
\newblock \bibinfo{journal}{\emph{Proc AAAI'19}} \bibinfo{volume}{33},
  \bibinfo{number}{01} (\bibinfo{date}{Jul.} \bibinfo{year}{2019}),
  \bibinfo{pages}{2429--2437}.
\newblock
\urldef\tempurl%
\url{https://doi.org/10.1609/aaai.v33i01.33012429}
\showDOI{\tempurl}


\bibitem[\protect\citeauthoryear{{Barredo Arrieta}, Díaz-Rodríguez, {Del
  Ser}, Bennetot, Tabik, Barbado, Garcia, Gil-Lopez, Molina, Benjamins,
  Chatila, and Herrera}{{Barredo Arrieta} et~al\mbox{.}}{2020}]%
        {barrero:xai:2020}
\bibfield{author}{\bibinfo{person}{Alejandro {Barredo Arrieta}},
  \bibinfo{person}{Natalia Díaz-Rodríguez}, \bibinfo{person}{Javier {Del
  Ser}}, \bibinfo{person}{Adrien Bennetot}, \bibinfo{person}{Siham Tabik},
  \bibinfo{person}{Alberto Barbado}, \bibinfo{person}{Salvador Garcia},
  \bibinfo{person}{Sergio Gil-Lopez}, \bibinfo{person}{Daniel Molina},
  \bibinfo{person}{Richard Benjamins}, \bibinfo{person}{Raja Chatila}, {and}
  \bibinfo{person}{Francisco Herrera}.} \bibinfo{year}{2020}\natexlab{}.
\newblock \showarticletitle{Explainable Artificial Intelligence (XAI):
  Concepts, taxonomies, opportunities and challenges toward responsible AI}.
\newblock \bibinfo{journal}{\emph{Information Fusion}}  \bibinfo{volume}{58}
  (\bibinfo{year}{2020}), \bibinfo{pages}{82--115}.
\newblock
\showISSN{1566-2535}
\urldef\tempurl%
\url{https://doi.org//10.1016/j.inffus.2019.12.012}
\showDOI{\tempurl}


\bibitem[\protect\citeauthoryear{Battle and Heer}{Battle and Heer}{2019}]%
        {battle:eva:2019}
\bibfield{author}{\bibinfo{person}{Leilani Battle} {and}
  \bibinfo{person}{Jeffrey Heer}.} \bibinfo{year}{2019}\natexlab{}.
\newblock \showarticletitle{Characterizing Exploratory Visual Analysis: A
  Literature Review and Evaluation of Analytic Provenance in Tableau}.
\newblock \bibinfo{journal}{\emph{Computer Graphics Forum}}
  \bibinfo{volume}{38}, \bibinfo{number}{3} (\bibinfo{year}{2019}),
  \bibinfo{pages}{145--159}.
\newblock
\urldef\tempurl%
\url{https://doi.org/10.1111/cgf.13678}
\showDOI{\tempurl}


\bibitem[\protect\citeauthoryear{Beck, Burch, Diehl, and Weiskopf}{Beck
  et~al\mbox{.}}{2017}]%
        {Beck:tax_graphs:2017}
\bibfield{author}{\bibinfo{person}{Fabian Beck}, \bibinfo{person}{Michael
  Burch}, \bibinfo{person}{Stephan Diehl}, {and} \bibinfo{person}{Daniel
  Weiskopf}.} \bibinfo{year}{2017}\natexlab{}.
\newblock \showarticletitle{A Taxonomy and Survey of Dynamic Graph
  Visualization}.
\newblock \bibinfo{journal}{\emph{Computer Graphics Forum}}
  \bibinfo{volume}{36}, \bibinfo{number}{1} (\bibinfo{year}{2017}),
  \bibinfo{pages}{133--159}.
\newblock
\urldef\tempurl%
\url{https://doi.org/10.1111/cgf.12791}
\showDOI{\tempurl}


\bibitem[\protect\citeauthoryear{Boukhelifa, Bezerianos, Chang, Collins,
  Drucker, Endert, Hullman, North, and Sedlmair}{Boukhelifa
  et~al\mbox{.}}{2020}]%
        {boukhelifa2020challenges}
\bibfield{author}{\bibinfo{person}{Nadia Boukhelifa},
  \bibinfo{person}{Anastasia Bezerianos}, \bibinfo{person}{Remco Chang},
  \bibinfo{person}{Christopher Collins}, \bibinfo{person}{Steven Drucker},
  \bibinfo{person}{Alexander Endert}, \bibinfo{person}{Jessica Hullman},
  \bibinfo{person}{Chris North}, {and} \bibinfo{person}{Michael Sedlmair}.}
  \bibinfo{year}{2020}\natexlab{}.
\newblock \showarticletitle{Challenges in evaluating interactive visual machine
  learning systems}.
\newblock \bibinfo{journal}{\emph{IEEE Computer Graphics and Applications}}
  \bibinfo{volume}{40}, \bibinfo{number}{6} (\bibinfo{year}{2020}),
  \bibinfo{pages}{88--96}.
\newblock


\bibitem[\protect\citeauthoryear{Brehmer and Munzner}{Brehmer and
  Munzner}{2013}]%
        {Brehmer:Typology:2013}
\bibfield{author}{\bibinfo{person}{Matthew Brehmer} {and}
  \bibinfo{person}{Tamara Munzner}.} \bibinfo{year}{2013}\natexlab{}.
\newblock \showarticletitle{A Multi-Level Typology of Abstract Visualization
  Tasks}.
\newblock \bibinfo{journal}{\emph{IEEE Transactions on Visualization and
  Computer Graphics}} \bibinfo{volume}{19}, \bibinfo{number}{12}
  (\bibinfo{year}{2013}), \bibinfo{pages}{2376--2385}.
\newblock
\urldef\tempurl%
\url{https://doi.org/10.1109/TVCG.2013.124}
\showDOI{\tempurl}


\bibitem[\protect\citeauthoryear{Brown, Lie, Brodely, and Chang}{Brown
  et~al\mbox{.}}{2012}]%
        {Brown:2012}
\bibfield{author}{\bibinfo{person}{Eli~T. Brown}, \bibinfo{person}{Jingjing
  Lie}, \bibinfo{person}{Carla~E. Brodely}, {and} \bibinfo{person}{Remco
  Chang}.} \bibinfo{year}{2012}\natexlab{}.
\newblock \showarticletitle{Dis-function: Learning Distance Functions
  Interactively}. In \bibinfo{booktitle}{\emph{2012 IEEE Conference on Visual
  Analytics Science and Technology (VAST)}}. \bibinfo{pages}{83--92}.
\newblock
\urldef\tempurl%
\url{https://doi.org/10.1109/VAST.2012.6400486}
\showDOI{\tempurl}


\bibitem[\protect\citeauthoryear{Cambronero}{Cambronero}{2021}]%
        {Cambronero:Software_AutoML:2021}
\bibfield{author}{\bibinfo{person}{Jos\'{e}~Pablo Cambronero}.}
  \bibinfo{year}{2021}\natexlab{}.
\newblock \emph{\bibinfo{title}{Mining Software Artifacts for use in Automated
  Machine Learning}}.
\newblock \bibinfo{thesistype}{Ph.D. Dissertation}.
  \bibinfo{school}{MIT-CSAIL}.
\newblock
\urldef\tempurl%
\url{https://www.josecambronero.com/pdf/JCambronero-PhD-EECS-June2021.pdf}
\showURL{%
\tempurl}


\bibitem[\protect\citeauthoryear{Cashman, Humayoun, Heimerl, Park, Das,
  Thompson, Saket, Mosca, Stasko, Endert, Gleicher, and Chang}{Cashman
  et~al\mbox{.}}{2019}]%
        {Cashman:EMA:2019}
\bibfield{author}{\bibinfo{person}{Dylan Cashman}, \bibinfo{person}{Shah~Rukh
  Humayoun}, \bibinfo{person}{Florian Heimerl}, \bibinfo{person}{Kendall Park},
  \bibinfo{person}{Subhajit Das}, \bibinfo{person}{John Thompson},
  \bibinfo{person}{Bahador Saket}, \bibinfo{person}{Abigail Mosca},
  \bibinfo{person}{John Stasko}, \bibinfo{person}{Alex Endert},
  \bibinfo{person}{Michael Gleicher}, {and} \bibinfo{person}{Remco Chang}.}
  \bibinfo{year}{2019}\natexlab{}.
\newblock \showarticletitle{A User-based Visual Analytics Workflow for
  Exploratory Model Analysis}.
\newblock \bibinfo{journal}{\emph{Computer Graphics Forum}}
  \bibinfo{volume}{38}, \bibinfo{number}{3} (\bibinfo{year}{2019}),
  \bibinfo{pages}{185--199}.
\newblock
\urldef\tempurl%
\url{https://doi.org/10.1111/cgf.13681}
\showDOI{\tempurl}


\bibitem[\protect\citeauthoryear{Celik and Vanschoren}{Celik and
  Vanschoren}{2021}]%
        {Celik:Adaption:2021}
\bibfield{author}{\bibinfo{person}{Bilge Celik} {and} \bibinfo{person}{Joaquin
  Vanschoren}.} \bibinfo{year}{2021}\natexlab{}.
\newblock \showarticletitle{Adaptation Strategies for Automated Machine
  Learning on Evolving Data}.
\newblock \bibinfo{journal}{\emph{IEEE Transactions on Pattern Analysis and
  Machine Intelligence}} (\bibinfo{year}{2021}), \bibinfo{pages}{1–1}.
\newblock
\urldef\tempurl%
\url{https://doi.org/10.1109/TPAMI.2021.3062900}
\showDOI{\tempurl}


\bibitem[\protect\citeauthoryear{Crisan and Fiore-Gartland}{Crisan and
  Fiore-Gartland}{2021}]%
        {Crisan:AutoML:2021}
\bibfield{author}{\bibinfo{person}{Anamaria Crisan} {and}
  \bibinfo{person}{Brittany Fiore-Gartland}.} \bibinfo{year}{2021}\natexlab{}.
\newblock \showarticletitle{Fits and Starts: Enterprise Use of AutoML and the
  Role of Humans in the Loop}. In \bibinfo{booktitle}{\emph{Proc CHI'21}}.
  Article \bibinfo{articleno}{601}, \bibinfo{numpages}{15}~pages.
\newblock


\bibitem[\protect\citeauthoryear{Crisan, Fiore-Gartland, and Tory}{Crisan
  et~al\mbox{.}}{2021}]%
        {Crisan:Batton:202}
\bibfield{author}{\bibinfo{person}{Anamaria Crisan}, \bibinfo{person}{Brittany
  Fiore-Gartland}, {and} \bibinfo{person}{Melanie Tory}.}
  \bibinfo{year}{2021}\natexlab{}.
\newblock \showarticletitle{Passing the Data Baton : A Retrospective Analysis
  on Data Science Work and Workers}.
\newblock \bibinfo{journal}{\emph{IEEE Transactions on Visualization and
  Computer Graphics}} \bibinfo{volume}{27}, \bibinfo{number}{2}
  (\bibinfo{year}{2021}), \bibinfo{pages}{1860--1870}.
\newblock
\urldef\tempurl%
\url{https://doi.org/10.1109/TVCG.2020.3030340}
\showDOI{\tempurl}


\bibitem[\protect\citeauthoryear{Cutler, Gadhave, and Lex}{Cutler
  et~al\mbox{.}}{2020}]%
        {Cutler:Trrack:2020}
\bibfield{author}{\bibinfo{person}{Zach Cutler}, \bibinfo{person}{Kiran
  Gadhave}, {and} \bibinfo{person}{Alexander Lex}.}
  \bibinfo{year}{2020}\natexlab{}.
\newblock \showarticletitle{Trrack: A Library for Provenance-Tracking in
  Web-Based Visualizations}. In \bibinfo{booktitle}{\emph{2020 IEEE
  Visualization Conference (VIS)}}. \bibinfo{pages}{116--120}.
\newblock
\urldef\tempurl%
\url{https://doi.org/10.1109/VIS47514.2020.00030}
\showDOI{\tempurl}


\bibitem[\protect\citeauthoryear{Davies and Frank}{Davies and Frank}{2013}]%
        {Davis:rawdata:2013}
\bibfield{author}{\bibinfo{person}{Tim Davies} {and} \bibinfo{person}{Mark
  Frank}.} \bibinfo{year}{2013}\natexlab{}.
\newblock \showarticletitle{'There's No Such Thing as Raw Data': Exploring the
  Socio-Technical Life of a Government Dataset}. In
  \bibinfo{booktitle}{\emph{Proc WebSci '13}}. \bibinfo{pages}{75–78}.
\newblock
\urldef\tempurl%
\url{https://doi.org/10.1145/2464464.2464472}
\showDOI{\tempurl}


\bibitem[\protect\citeauthoryear{De~Bie, De~Raedt, Hern\'{a}ndez-Orallo, Hoos,
  Smyth, and Williams}{De~Bie et~al\mbox{.}}{2022}]%
        {Bie:AutoDS:2022}
\bibfield{author}{\bibinfo{person}{Tijl De~Bie}, \bibinfo{person}{Luc
  De~Raedt}, \bibinfo{person}{Jos\'{e} Hern\'{a}ndez-Orallo},
  \bibinfo{person}{Holger~H. Hoos}, \bibinfo{person}{Padhraic Smyth}, {and}
  \bibinfo{person}{Christopher K.~I. Williams}.}
  \bibinfo{year}{2022}\natexlab{}.
\newblock \showarticletitle{Automating Data Science}.
\newblock \bibinfo{journal}{\emph{Commun. ACM}} \bibinfo{volume}{65},
  \bibinfo{number}{3} (\bibinfo{date}{feb} \bibinfo{year}{2022}),
  \bibinfo{pages}{76–87}.
\newblock
\urldef\tempurl%
\url{https://doi.org/10.1145/3495256}
\showDOI{\tempurl}


\bibitem[\protect\citeauthoryear{Dellermann, Calma, Lipusch, Weber, Weigel, and
  Ebel}{Dellermann et~al\mbox{.}}{2021}]%
        {dellermann:future_collab:2021}
\bibfield{author}{\bibinfo{person}{Dominik Dellermann}, \bibinfo{person}{Adrian
  Calma}, \bibinfo{person}{Nikolaus Lipusch}, \bibinfo{person}{Thorsten Weber},
  \bibinfo{person}{Sascha Weigel}, {and} \bibinfo{person}{Philipp Ebel}.}
  \bibinfo{year}{2021}\natexlab{}.
\newblock \bibinfo{title}{The future of human-AI collaboration: a taxonomy of
  design knowledge for hybrid intelligence systems}.
\newblock
\newblock
\showeprint{2105.03354}
\urldef\tempurl%
\url{https://arxiv.org/abs/2105.03354}
\showURL{%
\tempurl}


\bibitem[\protect\citeauthoryear{Domova and Vrotsou}{Domova and
  Vrotsou}{2022}]%
        {domova2022model}
\bibfield{author}{\bibinfo{person}{Veronika Domova} {and}
  \bibinfo{person}{Katerina Vrotsou}.} \bibinfo{year}{2022}\natexlab{}.
\newblock \showarticletitle{A Model for Types and Levels of Automation in
  Visual Analytics: a Survey, a Taxonomy, and Examples}.
\newblock \bibinfo{journal}{\emph{IEEE Transactions on Visualization and
  Computer Graphics}} (\bibinfo{year}{2022}).
\newblock


\bibitem[\protect\citeauthoryear{Drozdal, Weisz, Wang, Dass, Yao, Zhao, Muller,
  Ju, and Su}{Drozdal et~al\mbox{.}}{2020}]%
        {Drozdal:trust:2020}
\bibfield{author}{\bibinfo{person}{Jaimie Drozdal}, \bibinfo{person}{Justin
  Weisz}, \bibinfo{person}{Dakuo Wang}, \bibinfo{person}{Gaurav Dass},
  \bibinfo{person}{Bingsheng Yao}, \bibinfo{person}{Changruo Zhao},
  \bibinfo{person}{Michael Muller}, \bibinfo{person}{Lin Ju}, {and}
  \bibinfo{person}{Hui Su}.} \bibinfo{year}{2020}\natexlab{}.
\newblock \showarticletitle{Trust in AutoML: Exploring Information Needs for
  Establishing Trust in Automated Machine Learning Systems}. In
  \bibinfo{booktitle}{\emph{Proc. IUI'20}}. \bibinfo{pages}{297–307}.
\newblock
\urldef\tempurl%
\url{https://doi.org/10.1145/3377325.3377501}
\showDOI{\tempurl}


\bibitem[\protect\citeauthoryear{Elshawi, Maher, and Sakr}{Elshawi
  et~al\mbox{.}}{2019}]%
        {elshawi2019automated}
\bibfield{author}{\bibinfo{person}{Radwa Elshawi}, \bibinfo{person}{Mohamed
  Maher}, {and} \bibinfo{person}{Sherif Sakr}.}
  \bibinfo{year}{2019}\natexlab{}.
\newblock \bibinfo{title}{Automated Machine Learning: State-of-The-Art and Open
  Challenges}.
\newblock
\newblock
\showeprint{1906.02287}
\urldef\tempurl%
\url{https://arxiv.org/abs/1906.02287}
\showURL{%
\tempurl}


\bibitem[\protect\citeauthoryear{Endert, Fiaux, and North}{Endert
  et~al\mbox{.}}{2012}]%
        {Endert:2012}
\bibfield{author}{\bibinfo{person}{Alex Endert}, \bibinfo{person}{Patrick
  Fiaux}, {and} \bibinfo{person}{Chris North}.}
  \bibinfo{year}{2012}\natexlab{}.
\newblock \showarticletitle{Semantic Interaction for Visual Text Analytics}. In
  \bibinfo{booktitle}{\emph{Proc CHI'12}}. \bibinfo{pages}{473–482}.
\newblock
\urldef\tempurl%
\url{https://doi.org/10.1145/2207676.2207741}
\showDOI{\tempurl}


\bibitem[\protect\citeauthoryear{Endert, Ribarsky, Turkay, Wong, Nabney,
  Blanco, and Rossi}{Endert et~al\mbox{.}}{2017}]%
        {endert2017state}
\bibfield{author}{\bibinfo{person}{Alex Endert}, \bibinfo{person}{William
  Ribarsky}, \bibinfo{person}{Cagatay Turkay}, \bibinfo{person}{BL~William
  Wong}, \bibinfo{person}{Ian Nabney}, \bibinfo{person}{I~D{\'\i}az Blanco},
  {and} \bibinfo{person}{Fabrice Rossi}.} \bibinfo{year}{2017}\natexlab{}.
\newblock \showarticletitle{The state of the art in integrating machine
  learning into visual analytics}. In \bibinfo{booktitle}{\emph{Computer
  Graphics Forum}}, Vol.~\bibinfo{volume}{36}. Wiley Online Library,
  \bibinfo{pages}{458--486}.
\newblock


\bibitem[\protect\citeauthoryear{Estevez-Velarde, Guti{\'e}rrez, Montoyo, and
  Almeida-Cruz}{Estevez-Velarde et~al\mbox{.}}{2019}]%
        {estevez:semantic_annots:2019}
\bibfield{author}{\bibinfo{person}{Suilan Estevez-Velarde},
  \bibinfo{person}{Yoan Guti{\'e}rrez}, \bibinfo{person}{Andr{\'e}s Montoyo},
  {and} \bibinfo{person}{Yudivi{\'a}n Almeida-Cruz}.}
  \bibinfo{year}{2019}\natexlab{}.
\newblock \showarticletitle{{A}uto{ML} Strategy Based on Grammatical Evolution:
  A Case Study about Knowledge Discovery from Text}. In
  \bibinfo{booktitle}{\emph{Proc ACL'19}}. \bibinfo{pages}{4356--4365}.
\newblock
\urldef\tempurl%
\url{https://doi.org/10.18653/v1/P19-1428}
\showDOI{\tempurl}


\bibitem[\protect\citeauthoryear{Feurer, Eggensperger, Falkner, Lindauer, and
  Hutter}{Feurer et~al\mbox{.}}{2020}]%
        {feurer:autosklearn:2020}
\bibfield{author}{\bibinfo{person}{Matthias Feurer}, \bibinfo{person}{Katharina
  Eggensperger}, \bibinfo{person}{Stefan Falkner}, \bibinfo{person}{Marius
  Lindauer}, {and} \bibinfo{person}{Frank Hutter}.}
  \bibinfo{year}{2020}\natexlab{}.
\newblock \bibinfo{title}{Auto-Sklearn 2.0: The Next Generation}.
\newblock
\newblock
\showeprint{2007.04074}
\urldef\tempurl%
\url{https://arxiv.org/abs/2007.04074}
\showURL{%
\tempurl}


\bibitem[\protect\citeauthoryear{Fischer and Otswald}{Fischer and
  Otswald}{2001}]%
        {Fischer:kx_mgmt_og:2001}
\bibfield{author}{\bibinfo{person}{Gerhard~. Fischer} {and}
  \bibinfo{person}{Jonathan. Otswald}.} \bibinfo{year}{2001}\natexlab{}.
\newblock \showarticletitle{Knowledge management: problems, promises,
  realities, and challenges}.
\newblock \bibinfo{journal}{\emph{IEEE Intelligent Systems}}
  \bibinfo{volume}{16}, \bibinfo{number}{1} (\bibinfo{year}{2001}),
  \bibinfo{pages}{60--72}.
\newblock
\urldef\tempurl%
\url{https://doi.org/10.1109/5254.912386}
\showDOI{\tempurl}


\bibitem[\protect\citeauthoryear{Gadhave, Görtler, Cutler, Nobre, Deussen,
  Meyer, Phillips, and Lex}{Gadhave et~al\mbox{.}}{2020}]%
        {gadhave:vis_intnet:2020}
\bibfield{author}{\bibinfo{person}{Kiran Gadhave}, \bibinfo{person}{Jochen
  Görtler}, \bibinfo{person}{Zach Cutler}, \bibinfo{person}{Carolina Nobre},
  \bibinfo{person}{Oliver Deussen}, \bibinfo{person}{Miriah Meyer},
  \bibinfo{person}{Jeff Phillips}, {and} \bibinfo{person}{Alexander Lex}.}
  \bibinfo{year}{2020}\natexlab{}.
\newblock \bibinfo{title}{Capturing User Intent when Brushing in Scatterplots}.
\newblock
\newblock
\urldef\tempurl%
\url{https://doi.org/10.31219/osf.io/mq2rk}
\showDOI{\tempurl}


\bibitem[\protect\citeauthoryear{Gehrmann, Strobelt, Krüger, Pfister, and
  Rush}{Gehrmann et~al\mbox{.}}{2020}]%
        {Gehrmann:SemanticInference:2020}
\bibfield{author}{\bibinfo{person}{Sebastian Gehrmann},
  \bibinfo{person}{Hendrik Strobelt}, \bibinfo{person}{Robert Krüger},
  \bibinfo{person}{Hanspeter Pfister}, {and} \bibinfo{person}{Alexander~M.
  Rush}.} \bibinfo{year}{2020}\natexlab{}.
\newblock \showarticletitle{Visual Interaction with Deep Learning Models
  through Collaborative Semantic Inference}.
\newblock \bibinfo{journal}{\emph{IEEE Transactions on Visualization and
  Computer Graphics}} \bibinfo{volume}{26}, \bibinfo{number}{1}
  (\bibinfo{year}{2020}), \bibinfo{pages}{884--894}.
\newblock
\urldef\tempurl%
\url{https://doi.org/10.1109/TVCG.2019.2934595}
\showDOI{\tempurl}


\bibitem[\protect\citeauthoryear{Gijsbers, LeDell, Thomas, Poirier, Bischl, and
  Vanschoren}{Gijsbers et~al\mbox{.}}{2019}]%
        {gijsber:automl_benchmark:2019}
\bibfield{author}{\bibinfo{person}{Pieter Gijsbers}, \bibinfo{person}{Erin
  LeDell}, \bibinfo{person}{Janek Thomas}, \bibinfo{person}{Sébastien
  Poirier}, \bibinfo{person}{Bernd Bischl}, {and} \bibinfo{person}{Joaquin
  Vanschoren}.} \bibinfo{year}{2019}\natexlab{}.
\newblock \bibinfo{title}{An Open Source AutoML Benchmark}.
\newblock
\newblock
\showeprint{1907.00909}
\urldef\tempurl%
\url{https://arxiv.org/abs/1907.00909}
\showURL{%
\tempurl}


\bibitem[\protect\citeauthoryear{Gitelman}{Gitelman}{2013}]%
        {Gitelman:rawdata:2013}
\bibfield{author}{\bibinfo{person}{Lisa Gitelman}.}
  \bibinfo{year}{2013}\natexlab{}.
\newblock \bibinfo{booktitle}{\emph{``Raw Data'' Is an Oxymoron}}.
\newblock \bibinfo{publisher}{MIT Press}, \bibinfo{address}{Cambridge, USA}.
\newblock


\bibitem[\protect\citeauthoryear{Gleicher}{Gleicher}{2018}]%
        {Gliecher:Comparison_2:2018}
\bibfield{author}{\bibinfo{person}{Michael Gleicher}.}
  \bibinfo{year}{2018}\natexlab{}.
\newblock \showarticletitle{Considerations for Visualizing Comparison}.
\newblock \bibinfo{journal}{\emph{IEEE Transactions on Visualization and
  Computer Graphics}} \bibinfo{volume}{24}, \bibinfo{number}{1}
  (\bibinfo{year}{2018}), \bibinfo{pages}{413--423}.
\newblock
\urldef\tempurl%
\url{https://doi.org/10.1109/TVCG.2017.2744199}
\showDOI{\tempurl}


\bibitem[\protect\citeauthoryear{Gleicher, Albers, Walker, Jusufi, Hansen, and
  Roberts}{Gleicher et~al\mbox{.}}{2011}]%
        {Gliecher:comparison_taxonomy:2011}
\bibfield{author}{\bibinfo{person}{Michael Gleicher}, \bibinfo{person}{Danielle
  Albers}, \bibinfo{person}{Rick Walker}, \bibinfo{person}{Ilir Jusufi},
  \bibinfo{person}{Charles~D. Hansen}, {and} \bibinfo{person}{Jonathan~C.
  Roberts}.} \bibinfo{year}{2011}\natexlab{}.
\newblock \showarticletitle{Visual Comparison for Information Visualization}.
\newblock \bibinfo{journal}{\emph{Information Visualization}}
  \bibinfo{volume}{10}, \bibinfo{number}{4} (\bibinfo{date}{Oct.}
  \bibinfo{year}{2011}), \bibinfo{pages}{289–309}.
\newblock
\urldef\tempurl%
\url{https://doi.org/10.1177/1473871611416549}
\showDOI{\tempurl}


\bibitem[\protect\citeauthoryear{Golovin, Solnik, Moitra, Kochanski, Karro, and
  Sculley}{Golovin et~al\mbox{.}}{2017}]%
        {golovin:googleVizer:2017}
\bibfield{author}{\bibinfo{person}{Daniel Golovin}, \bibinfo{person}{Benjamin
  Solnik}, \bibinfo{person}{Subhodeep Moitra}, \bibinfo{person}{Greg
  Kochanski}, \bibinfo{person}{John Karro}, {and} \bibinfo{person}{D.
  Sculley}.} \bibinfo{year}{2017}\natexlab{}.
\newblock \showarticletitle{Google Vizier: A Service for Black-Box
  Optimization}. In \bibinfo{booktitle}{\emph{Proc KDD'17}}.
  \bibinfo{pages}{1487–1495}.
\newblock
\urldef\tempurl%
\url{https://doi.org/10.1145/3097983.3098043}
\showDOI{\tempurl}


\bibitem[\protect\citeauthoryear{Greenberg and Buxton}{Greenberg and
  Buxton}{2008}]%
        {greenberg:usability_eval:2008}
\bibfield{author}{\bibinfo{person}{Saul Greenberg} {and}
  \bibinfo{person}{William Buxton}.} \bibinfo{year}{2008}\natexlab{}.
\newblock \showarticletitle{Usability evaluation considered harmful (some of
  the time)}.
\newblock \bibinfo{journal}{\emph{Proc. CHI'08}}, \bibinfo{pages}{111--120}.
\newblock
\urldef\tempurl%
\url{https://doi.org/10.1145/1357054.1357074}
\showDOI{\tempurl}


\bibitem[\protect\citeauthoryear{He, Zhao, and Chu}{He et~al\mbox{.}}{2021}]%
        {He:AutoML_Stota:2021}
\bibfield{author}{\bibinfo{person}{Xin He}, \bibinfo{person}{Kaiyong Zhao},
  {and} \bibinfo{person}{Xiaowen Chu}.} \bibinfo{year}{2021}\natexlab{}.
\newblock \showarticletitle{AutoML: A survey of the state-of-the-art}.
\newblock \bibinfo{journal}{\emph{Knowledge-Based Systems}}
  \bibinfo{volume}{212} (\bibinfo{date}{Jan} \bibinfo{year}{2021}),
  \bibinfo{pages}{106622}.
\newblock
\urldef\tempurl%
\url{https://doi.org/10.1016/j.knosys.2020.106622}
\showDOI{\tempurl}


\bibitem[\protect\citeauthoryear{Heer}{Heer}{2019}]%
        {Heer:agency:2019}
\bibfield{author}{\bibinfo{person}{Jeffrey Heer}.}
  \bibinfo{year}{2019}\natexlab{}.
\newblock \showarticletitle{Agency Plus Automation: Designing Artificial
  Intelligence into Interactive Systems}.
\newblock \bibinfo{journal}{\emph{Proceedings of the National Academy of
  Sciences}} \bibinfo{volume}{116}, \bibinfo{number}{6} (\bibinfo{year}{2019}),
  \bibinfo{pages}{1844–1850}.
\newblock
\urldef\tempurl%
\url{https://doi.org/10.1073/pnas.1807184115}
\showDOI{\tempurl}


\bibitem[\protect\citeauthoryear{Heffetz, Vainshtein, Katz, and Rokach}{Heffetz
  et~al\mbox{.}}{2020}]%
        {Heffetz:deepline:2020}
\bibfield{author}{\bibinfo{person}{Yuval Heffetz}, \bibinfo{person}{Roman
  Vainshtein}, \bibinfo{person}{Gilad Katz}, {and} \bibinfo{person}{Lior
  Rokach}.} \bibinfo{year}{2020}\natexlab{}.
\newblock \showarticletitle{DeepLine: AutoML Tool for Pipelines Generation
  Using Deep Reinforcement Learning and Hierarchical Actions Filtering}. In
  \bibinfo{booktitle}{\emph{Proc KDD '20}}. \bibinfo{pages}{2103–2113}.
\newblock
\urldef\tempurl%
\url{https://doi.org/10.1145/3394486.3403261}
\showDOI{\tempurl}


\bibitem[\protect\citeauthoryear{Hong, Castelo, D'Orazio, Benthune, Santos,
  Langevin, Jonker, Bertini, and Freire}{Hong et~al\mbox{.}}{2020a}]%
        {hong2020evaluating}
\bibfield{author}{\bibinfo{person}{Sungsoo~Ray Hong}, \bibinfo{person}{Sonia
  Castelo}, \bibinfo{person}{Vito D'Orazio}, \bibinfo{person}{Christopher
  Benthune}, \bibinfo{person}{Aecio Santos}, \bibinfo{person}{Scott Langevin},
  \bibinfo{person}{David Jonker}, \bibinfo{person}{Enrico Bertini}, {and}
  \bibinfo{person}{Juliana Freire}.} \bibinfo{year}{2020}\natexlab{a}.
\newblock \bibinfo{title}{Towards Evaluating Exploratory Model Building Process
  with AutoML Systems}.
\newblock
\newblock
\showeprint{2009.00449}
\urldef\tempurl%
\url{https://arxiv.org/abs/2009.00449}
\showURL{%
\tempurl}


\bibitem[\protect\citeauthoryear{Hong, Hullman, and Bertini}{Hong
  et~al\mbox{.}}{2020b}]%
        {Hong:interpret:2020}
\bibfield{author}{\bibinfo{person}{Sungsoo~Ray Hong}, \bibinfo{person}{Jessica
  Hullman}, {and} \bibinfo{person}{Enrico Bertini}.}
  \bibinfo{year}{2020}\natexlab{b}.
\newblock \showarticletitle{Human Factors in Model Interpretability: Industry
  Practices, Challenges, and Needs}.
\newblock \bibinfo{journal}{\emph{Proc. CSCW'20.}} \bibinfo{volume}{4},
  \bibinfo{number}{CSCW1}, Article \bibinfo{articleno}{068}
  (\bibinfo{date}{May} \bibinfo{year}{2020}), \bibinfo{numpages}{26}~pages.
\newblock
\urldef\tempurl%
\url{https://doi.org/10.1145/3392878}
\showDOI{\tempurl}


\bibitem[\protect\citeauthoryear{Hu, Bakker, Li, Kraska, and Hidalgo}{Hu
  et~al\mbox{.}}{2019}]%
        {Hu:VizML:2019}
\bibfield{author}{\bibinfo{person}{Kevin Hu}, \bibinfo{person}{Michiel~A.
  Bakker}, \bibinfo{person}{Stephen Li}, \bibinfo{person}{Tim Kraska}, {and}
  \bibinfo{person}{C\'{e}sar Hidalgo}.} \bibinfo{year}{2019}\natexlab{}.
\newblock \showarticletitle{VizML: A Machine Learning Approach to Visualization
  Recommendation}. In \bibinfo{booktitle}{\emph{Proc. CHI'19}}.
  \bibinfo{publisher}{Association for Computing Machinery},
  \bibinfo{address}{New York, NY, USA}, \bibinfo{pages}{1–12}.
\newblock
\showISBNx{9781450359702}
\urldef\tempurl%
\url{https://doi.org/10.1145/3290605.3300358}
\showURL{%
\tempurl}


\bibitem[\protect\citeauthoryear{Jin, Song, and Hu}{Jin et~al\mbox{.}}{2019}]%
        {Haifeng:autokeras:2019}
\bibfield{author}{\bibinfo{person}{Haifeng Jin}, \bibinfo{person}{Qingquan
  Song}, {and} \bibinfo{person}{Xia Hu}.} \bibinfo{year}{2019}\natexlab{}.
\newblock \showarticletitle{Auto-Keras: An Efficient Neural Architecture Search
  System}. In \bibinfo{booktitle}{\emph{Proc KDD '19}} (Anchorage, AK, USA).
  \bibinfo{publisher}{Association for Computing Machinery},
  \bibinfo{address}{New York, NY, USA}, \bibinfo{pages}{1946–1956}.
\newblock
\showISBNx{9781450362016}
\urldef\tempurl%
\url{https://doi.org/10.1145/3292500.3330648}
\showDOI{\tempurl}


\bibitem[\protect\citeauthoryear{John, Bass, Kazman, and Chen}{John
  et~al\mbox{.}}{2004}]%
        {john2004identifying}
\bibfield{author}{\bibinfo{person}{Bonnie~E John}, \bibinfo{person}{Len Bass},
  \bibinfo{person}{Rick Kazman}, {and} \bibinfo{person}{Eugene Chen}.}
  \bibinfo{year}{2004}\natexlab{}.
\newblock \showarticletitle{Identifying gaps between HCI, software engineering,
  and design, and boundary objects to bridge them}. In
  \bibinfo{booktitle}{\emph{CHI'04 extended abstracts on Human factors in
  computing systems}}. \bibinfo{pages}{1723--1724}.
\newblock


\bibitem[\protect\citeauthoryear{Karmaker, Hassan, Smith, Xu, Zhai, and
  Veeramachaneni}{Karmaker et~al\mbox{.}}{2021}]%
        {sant:automl:2021}
\bibfield{author}{\bibinfo{person}{Shubhra~Kanti Karmaker},
  \bibinfo{person}{Md.~Mahadi Hassan}, \bibinfo{person}{Micah~J. Smith},
  \bibinfo{person}{Lei Xu}, \bibinfo{person}{ChengXiang Zhai}, {and}
  \bibinfo{person}{Kalyan Veeramachaneni}.} \bibinfo{year}{2021}\natexlab{}.
\newblock \bibinfo{title}{AutoML to Date and Beyond: Challenges and
  Opportunities}.
\newblock
\newblock
\showeprint{2010.10777}
\urldef\tempurl%
\url{https://arxiv.org/abs/2010.10777}
\showURL{%
\tempurl}


\bibitem[\protect\citeauthoryear{Kasica, Berret, and Munzner}{Kasica
  et~al\mbox{.}}{2021}]%
        {kasica:tablescraps:2021}
\bibfield{author}{\bibinfo{person}{Stephen Kasica}, \bibinfo{person}{Charles
  Berret}, {and} \bibinfo{person}{Tamara Munzner}.}
  \bibinfo{year}{2021}\natexlab{}.
\newblock \showarticletitle{Table Scraps: An Actionable Framework for
  Multi-Table Data Wrangling From An Artifact Study of Computational
  Journalism}.
\newblock \bibinfo{journal}{\emph{IEEE Transactions on Visualization and
  Computer Graphics}} \bibinfo{volume}{27}, \bibinfo{number}{2}
  (\bibinfo{year}{2021}), \bibinfo{pages}{957--966}.
\newblock
\urldef\tempurl%
\url{https://doi.org/10.1109/TVCG.2020.3030462}
\showDOI{\tempurl}


\bibitem[\protect\citeauthoryear{Kaur, Nori, Jenkins, Caruana, Wallach, and
  Wortman~Vaughan}{Kaur et~al\mbox{.}}{2020}]%
        {Harmanpreet:XAItechniques:2020}
\bibfield{author}{\bibinfo{person}{Harmanpreet Kaur}, \bibinfo{person}{Harsha
  Nori}, \bibinfo{person}{Samuel Jenkins}, \bibinfo{person}{Rich Caruana},
  \bibinfo{person}{Hanna Wallach}, {and} \bibinfo{person}{Jennifer
  Wortman~Vaughan}.} \bibinfo{year}{2020}\natexlab{}.
\newblock \bibinfo{booktitle}{\emph{Interpreting Interpretability:
  Understanding Data Scientists' Use of Interpretability Tools for Machine
  Learning}}.
\newblock \bibinfo{pages}{1–14}.
\newblock
\urldef\tempurl%
\url{https://doi.org/h10.1145/3313831.3376219}
\showDOI{\tempurl}


\bibitem[\protect\citeauthoryear{Kreiner}{Kreiner}{2002}]%
        {Kreiner:tacit_artifact:2002}
\bibfield{author}{\bibinfo{person}{Kristian Kreiner}.}
  \bibinfo{year}{2002}\natexlab{}.
\newblock \showarticletitle{Tacit knowledge management: the role of artifacts}.
\newblock \bibinfo{journal}{\emph{Journal of Knowledge Management}}
  \bibinfo{volume}{6}, \bibinfo{number}{2} (\bibinfo{date}{01 Jan}
  \bibinfo{year}{2002}), \bibinfo{pages}{112--123}.
\newblock
\urldef\tempurl%
\url{https://doi.org/10.1108/13673270210424648}
\showDOI{\tempurl}


\bibitem[\protect\citeauthoryear{Lam, Tory, and Munzner}{Lam
  et~al\mbox{.}}{2018}]%
        {Lam:GoalstoTasks:2018}
\bibfield{author}{\bibinfo{person}{Heidi Lam}, \bibinfo{person}{Melanie Tory},
  {and} \bibinfo{person}{Tamara Munzner}.} \bibinfo{year}{2018}\natexlab{}.
\newblock \showarticletitle{Bridging from Goals to Tasks with Design Study
  Analysis Reports}.
\newblock \bibinfo{journal}{\emph{IEEE Transactions on Visualization and
  Computer Graphics}} \bibinfo{volume}{24}, \bibinfo{number}{1}
  (\bibinfo{year}{2018}), \bibinfo{pages}{435--445}.
\newblock
\urldef\tempurl%
\url{https://doi.org/10.1109/TVCG.2017.2744319}
\showDOI{\tempurl}


\bibitem[\protect\citeauthoryear{Lee}{Lee}{2007}]%
        {Lee:BO:artifacts:2007}
\bibfield{author}{\bibinfo{person}{Charlotte~P. Lee}.}
  \bibinfo{year}{2007}\natexlab{}.
\newblock \showarticletitle{Boundary Negotiating Artifacts: Unbinding the
  Routine of Boundary Objects and Embracing Chaos in Collaborative Work}.
\newblock \bibinfo{journal}{\emph{Proc. CSCW'07}} \bibinfo{volume}{16},
  \bibinfo{number}{3} (\bibinfo{date}{01 Jun} \bibinfo{year}{2007}),
  \bibinfo{pages}{307--339}.
\newblock
\urldef\tempurl%
\url{https://doi.org/10.1007/s10606-007-9044-5}
\showDOI{\tempurl}


\bibitem[\protect\citeauthoryear{Lee, Macke, Xin, Lee, Huang, and
  Parameswaran}{Lee et~al\mbox{.}}{2019}]%
        {Lee2019AHP}
\bibfield{author}{\bibinfo{person}{D. Lee}, \bibinfo{person}{Stephen Macke},
  \bibinfo{person}{Doris Xin}, \bibinfo{person}{Angela Lee},
  \bibinfo{person}{Silu Huang}, {and} \bibinfo{person}{Aditya~G.
  Parameswaran}.} \bibinfo{year}{2019}\natexlab{}.
\newblock \showarticletitle{A Human-in-the-loop Perspective on AutoML:
  Milestones and the Road Ahead}.
\newblock \bibinfo{journal}{\emph{IEEE Data Eng. Bull.}} \bibinfo{volume}{42},
  \bibinfo{number}{2} (\bibinfo{year}{2019}), \bibinfo{pages}{59--70}.
\newblock
\urldef\tempurl%
\url{http://sites.computer.org/debull/A19june/p59.pdf}
\showURL{%
\tempurl}


\bibitem[\protect\citeauthoryear{Lee, Setlur, Tory, Karahalios, and
  Parameswaran}{Lee et~al\mbox{.}}{2021a}]%
        {lee2021deconstructing}
\bibfield{author}{\bibinfo{person}{Doris Jung-Lin Lee}, \bibinfo{person}{Vidya
  Setlur}, \bibinfo{person}{Melanie Tory}, \bibinfo{person}{Karrie~G
  Karahalios}, {and} \bibinfo{person}{Aditya Parameswaran}.}
  \bibinfo{year}{2021}\natexlab{a}.
\newblock \showarticletitle{Deconstructing categorization in visualization
  recommendation: A taxonomy and comparative study}.
\newblock \bibinfo{journal}{\emph{IEEE Transactions on Visualization and
  Computer Graphics}} (\bibinfo{year}{2021}).
\newblock


\bibitem[\protect\citeauthoryear{Lee, Setlur, Tory, Karahalios, and
  Parameswaran}{Lee et~al\mbox{.}}{2021b}]%
        {Lee:Vis_Rec:2021}
\bibfield{author}{\bibinfo{person}{Doris Jung-Lin Lee}, \bibinfo{person}{Vidya
  Setlur}, \bibinfo{person}{Melanie Tory}, \bibinfo{person}{Karrie~G.
  Karahalios}, {and} \bibinfo{person}{Aditya Parameswaran}.}
  \bibinfo{year}{2021}\natexlab{b}.
\newblock \showarticletitle{Deconstructing Categorization in Visualization
  Recommendation: A Taxonomy and Comparative Study}.
\newblock \bibinfo{journal}{\emph{IEEE Transactions on Visualization and
  Computer Graphics}} (\bibinfo{year}{2021}), \bibinfo{pages}{1--1}.
\newblock
\urldef\tempurl%
\url{https://doi.org/10.1109/TVCG.2021.3085751}
\showDOI{\tempurl}


\bibitem[\protect\citeauthoryear{Lee, Tang, Agarwal, Boonmark, Chen, Kang,
  Mukhopadhyay, Song, Yong, Hearst, and Parameswaran}{Lee
  et~al\mbox{.}}{2021c}]%
        {lee:lux:2021}
\bibfield{author}{\bibinfo{person}{Doris Jung-Lin Lee}, \bibinfo{person}{Dixin
  Tang}, \bibinfo{person}{Kunal Agarwal}, \bibinfo{person}{Thyne Boonmark},
  \bibinfo{person}{Caitlyn Chen}, \bibinfo{person}{Jake Kang},
  \bibinfo{person}{Ujjaini Mukhopadhyay}, \bibinfo{person}{Jerry Song},
  \bibinfo{person}{Micah Yong}, \bibinfo{person}{Marti~A. Hearst}, {and}
  \bibinfo{person}{Aditya~G. Parameswaran}.} \bibinfo{year}{2021}\natexlab{c}.
\newblock \bibinfo{title}{Lux: Always-on Visualization Recommendations for
  Exploratory Data Science}.
\newblock
\newblock
\urldef\tempurl%
\url{https://arxiv.org/abs/2105.00121}
\showURL{%
\tempurl}


\bibitem[\protect\citeauthoryear{Liu, Wang, Liu, and Zhu}{Liu
  et~al\mbox{.}}{2017}]%
        {liu:mlvis_analysis:2017}
\bibfield{author}{\bibinfo{person}{Shixia Liu}, \bibinfo{person}{Xiting Wang},
  \bibinfo{person}{Mengchen Liu}, {and} \bibinfo{person}{Jun Zhu}.}
  \bibinfo{year}{2017}\natexlab{}.
\newblock \showarticletitle{Towards better analysis of machine learning models:
  A visual analytics perspective}.
\newblock \bibinfo{journal}{\emph{Visual Informatics}} \bibinfo{volume}{1},
  \bibinfo{number}{1} (\bibinfo{year}{2017}), \bibinfo{pages}{48--56}.
\newblock
\showISSN{2468-502X}
\urldef\tempurl%
\url{https://doi.org/10.1016/j.visinf.2017.01.006}
\showDOI{\tempurl}


\bibitem[\protect\citeauthoryear{Liu, Althoff, and Heer}{Liu
  et~al\mbox{.}}{2020}]%
        {Liu:Paths:2020}
\bibfield{author}{\bibinfo{person}{Yang Liu}, \bibinfo{person}{Tim Althoff},
  {and} \bibinfo{person}{Jeffrey Heer}.} \bibinfo{year}{2020}\natexlab{}.
\newblock \showarticletitle{Paths Explored, Paths Omitted, Paths Obscured:
  Decision Points; Selective Reporting in End-to-End Data Analysis}. In
  \bibinfo{booktitle}{\emph{Proc. CHI'20}}. \bibinfo{pages}{1–14}.
\newblock
\urldef\tempurl%
\url{https://doi.org/10.1145/3313831.3376533}
\showDOI{\tempurl}


\bibitem[\protect\citeauthoryear{Liu, Kale, Althoff, and Heer}{Liu
  et~al\mbox{.}}{2021}]%
        {Liu:boba:2021}
\bibfield{author}{\bibinfo{person}{Yang Liu}, \bibinfo{person}{Alex Kale},
  \bibinfo{person}{Tim Althoff}, {and} \bibinfo{person}{Jeffrey Heer}.}
  \bibinfo{year}{2021}\natexlab{}.
\newblock \showarticletitle{Boba: Authoring and Visualizing Multiverse
  Analyses}.
\newblock \bibinfo{journal}{\emph{IEEE Transactions on Visualization and
  Computer Graphics}} \bibinfo{volume}{27}, \bibinfo{number}{2}
  (\bibinfo{date}{Feb} \bibinfo{year}{2021}), \bibinfo{pages}{1753–1763}.
\newblock
\showISSN{2160-9306}
\urldef\tempurl%
\url{https://doi.org/10.1109/tvcg.2020.3028985}
\showDOI{\tempurl}


\bibitem[\protect\citeauthoryear{Lloyd and Dykes}{Lloyd and Dykes}{2011}]%
        {lloyd:chauffered_demo:2011}
\bibfield{author}{\bibinfo{person}{David Lloyd} {and} \bibinfo{person}{Jason
  Dykes}.} \bibinfo{year}{2011}\natexlab{}.
\newblock \showarticletitle{Human-Centered Approaches in Geovisualization
  Design: Investigating Multiple Methods Through a Long-Term Case Study}.
\newblock \bibinfo{journal}{\emph{IEEE Transactions on Visualization and
  Computer Graphics}} \bibinfo{volume}{17}, \bibinfo{number}{12}
  (\bibinfo{year}{2011}), \bibinfo{pages}{2498--2507}.
\newblock
\urldef\tempurl%
\url{https://doi.org/10.1109/TVCG.2011.209}
\showDOI{\tempurl}


\bibitem[\protect\citeauthoryear{Loyola-González}{Loyola-González}{2019}]%
        {ocatavio:black-v-white:2019}
\bibfield{author}{\bibinfo{person}{Octavio Loyola-González}.}
  \bibinfo{year}{2019}\natexlab{}.
\newblock \showarticletitle{Black-Box vs. White-Box: Understanding Their
  Advantages and Weaknesses From a Practical Point of View}.
\newblock \bibinfo{journal}{\emph{IEEE Access}}  \bibinfo{volume}{7}
  (\bibinfo{year}{2019}), \bibinfo{pages}{154096--154113}.
\newblock
\urldef\tempurl%
\url{https://doi.org/10.1109/ACCESS.2019.2949286}
\showDOI{\tempurl}


\bibitem[\protect\citeauthoryear{Mariano and Awazu}{Mariano and Awazu}{2016}]%
        {Mariano:kwx_mgmt_survey:2016}
\bibfield{author}{\bibinfo{person}{Stefania Mariano} {and}
  \bibinfo{person}{Yukika Awazu}.} \bibinfo{year}{2016}\natexlab{}.
\newblock \showarticletitle{Artifacts in knowledge management research: a
  systematic literature review and future research directions}.
\newblock \bibinfo{journal}{\emph{Journal of Knowledge Management}}
  \bibinfo{volume}{20}, \bibinfo{number}{6} (\bibinfo{date}{01 Jan}
  \bibinfo{year}{2016}), \bibinfo{pages}{1333--1352}.
\newblock
\showISSN{1367-3270}
\urldef\tempurl%
\url{https://doi.org/10.1108/JKM-05-2016-0199}
\showDOI{\tempurl}


\bibitem[\protect\citeauthoryear{Mishra and Rzeszotarski}{Mishra and
  Rzeszotarski}{2021}]%
        {Swati:design_nonexperts:2021}
\bibfield{author}{\bibinfo{person}{Swati Mishra} {and}
  \bibinfo{person}{Jeffrey~M Rzeszotarski}.} \bibinfo{year}{2021}\natexlab{}.
\newblock \showarticletitle{Designing Interactive Transfer Learning Tools for
  ML Non-Experts}. In \bibinfo{booktitle}{\emph{Proc. CHI'21}}. Article
  \bibinfo{articleno}{364}, \bibinfo{numpages}{15}~pages.
\newblock
\urldef\tempurl%
\url{https://doi.org/10.1145/3411764.3445096}
\showDOI{\tempurl}


\bibitem[\protect\citeauthoryear{Mitchell, Wu, Zaldivar, Barnes, Vasserman,
  Hutchinson, Spitzer, Raji, and Gebru}{Mitchell et~al\mbox{.}}{2019}]%
        {Mitchell:model_cards:2019}
\bibfield{author}{\bibinfo{person}{Margaret Mitchell}, \bibinfo{person}{Simone
  Wu}, \bibinfo{person}{Andrew Zaldivar}, \bibinfo{person}{Parker Barnes},
  \bibinfo{person}{Lucy Vasserman}, \bibinfo{person}{Ben Hutchinson},
  \bibinfo{person}{Elena Spitzer}, \bibinfo{person}{Inioluwa~Deborah Raji},
  {and} \bibinfo{person}{Timnit Gebru}.} \bibinfo{year}{2019}\natexlab{}.
\newblock \showarticletitle{Model Cards for Model Reporting}. In
  \bibinfo{booktitle}{\emph{Proc. FAccT'19}}.
\newblock
\urldef\tempurl%
\url{https://doi.org/10.1145/3287560.3287596}
\showDOI{\tempurl}


\bibitem[\protect\citeauthoryear{Mittelstadt, Russell, and Wachter}{Mittelstadt
  et~al\mbox{.}}{2019}]%
        {Brent:Explain:2019}
\bibfield{author}{\bibinfo{person}{Brent Mittelstadt}, \bibinfo{person}{Chris
  Russell}, {and} \bibinfo{person}{Sandra Wachter}.}
  \bibinfo{year}{2019}\natexlab{}.
\newblock \showarticletitle{Explaining Explanations in AI}. In
  \bibinfo{booktitle}{\emph{Proc. FAccT'19}}. \bibinfo{pages}{279–288}.
\newblock
\urldef\tempurl%
\url{https://doi.org/10.1145/3287560.3287574}
\showDOI{\tempurl}


\bibitem[\protect\citeauthoryear{Mora-Cantallops, Sánchez-Alonso,
  García-Barriocanal, and Sicilia}{Mora-Cantallops et~al\mbox{.}}{2021}]%
        {mora:trace_and_trust:2021}
\bibfield{author}{\bibinfo{person}{Marçal Mora-Cantallops},
  \bibinfo{person}{Salvador Sánchez-Alonso}, \bibinfo{person}{Elena
  García-Barriocanal}, {and} \bibinfo{person}{Miguel-Angel Sicilia}.}
  \bibinfo{year}{2021}\natexlab{}.
\newblock \showarticletitle{Traceability for Trustworthy AI: A Review of Models
  and Tools}.
\newblock \bibinfo{journal}{\emph{Big Data and Cognitive Computing}}
  \bibinfo{volume}{5}, \bibinfo{number}{2} (\bibinfo{date}{May}
  \bibinfo{year}{2021}), \bibinfo{pages}{20}.
\newblock
\urldef\tempurl%
\url{https://doi.org/10.3390/bdcc5020020}
\showDOI{\tempurl}


\bibitem[\protect\citeauthoryear{Narkar, Zhang, Liao, Wang, and Weisz}{Narkar
  et~al\mbox{.}}{2021}]%
        {Narkar:Model_lineupper:2021}
\bibfield{author}{\bibinfo{person}{Shweta Narkar}, \bibinfo{person}{Yunfeng
  Zhang}, \bibinfo{person}{Q.~Vera Liao}, \bibinfo{person}{Dakuo Wang}, {and}
  \bibinfo{person}{Justin~D. Weisz}.} \bibinfo{year}{2021}\natexlab{}.
\newblock \showarticletitle{Model LineUpper: Supporting Interactive Model
  Comparison at Multiple Levels for AutoML}. In \bibinfo{booktitle}{\emph{Proc.
  IUI'21}}. \bibinfo{pages}{170–174}.
\newblock
\urldef\tempurl%
\url{https://doi.org/10.1145/3397481.3450658}
\showDOI{\tempurl}


\bibitem[\protect\citeauthoryear{Neff, Tanweer, Fiore-Gartland, and
  Osburn}{Neff et~al\mbox{.}}{2017}]%
        {neff:critique_ds:2017}
\bibfield{author}{\bibinfo{person}{Gina Neff}, \bibinfo{person}{Anissa
  Tanweer}, \bibinfo{person}{Brittany Fiore-Gartland}, {and}
  \bibinfo{person}{Laura Osburn}.} \bibinfo{year}{2017}\natexlab{}.
\newblock \showarticletitle{Critique and contribute: A practice-based framework
  for improving critical data studies and data science}.
\newblock \bibinfo{journal}{\emph{Big data}} \bibinfo{volume}{5},
  \bibinfo{number}{2} (\bibinfo{year}{2017}), \bibinfo{pages}{85--97}.
\newblock


\bibitem[\protect\citeauthoryear{Neto, Alves, and Campos}{Neto
  et~al\mbox{.}}{2020}]%
        {neto:nasirt:2020}
\bibfield{author}{\bibinfo{person}{Habib~Asseiss Neto}, \bibinfo{person}{Ronnie
  C.~O. Alves}, {and} \bibinfo{person}{Sergio V.~A. Campos}.}
  \bibinfo{year}{2020}\natexlab{}.
\newblock \bibinfo{title}{NASirt: AutoML based learning with instance-level
  complexity information}.
\newblock
\newblock
\showeprint{2008.11846}
\urldef\tempurl%
\url{https://arxiv.org/abs/2008.11846}
\showURL{%
\tempurl}


\bibitem[\protect\citeauthoryear{Nickerson, Varshney, and Muntermann}{Nickerson
  et~al\mbox{.}}{2013}]%
        {Nickerson:taxonomy:2012}
\bibfield{author}{\bibinfo{person}{Robert Nickerson}, \bibinfo{person}{Upkar
  Varshney}, {and} \bibinfo{person}{Jan Muntermann}.}
  \bibinfo{year}{2013}\natexlab{}.
\newblock \showarticletitle{A Method for Taxonomy Development and its
  Application in Information Systems}.
\newblock \bibinfo{journal}{\emph{European Journal of Information Systems}}
  \bibinfo{volume}{22} (\bibinfo{date}{05} \bibinfo{year}{2013}).
\newblock
\urldef\tempurl%
\url{https://doi.org/10.1057/ejis.2012.26}
\showDOI{\tempurl}


\bibitem[\protect\citeauthoryear{Nikitin, Vychuzhanin, Sarafanov, Polonskaia,
  Revin, Barabanova, Maximov, Kalyuzhnaya, and Boukhanovsky}{Nikitin
  et~al\mbox{.}}{2021}]%
        {nikitin:automated:2021}
\bibfield{author}{\bibinfo{person}{Nikolay~O. Nikitin}, \bibinfo{person}{Pavel
  Vychuzhanin}, \bibinfo{person}{Mikhail Sarafanov}, \bibinfo{person}{Iana~S.
  Polonskaia}, \bibinfo{person}{Ilia Revin}, \bibinfo{person}{Irina~V.
  Barabanova}, \bibinfo{person}{Gleb Maximov}, \bibinfo{person}{Anna~V.
  Kalyuzhnaya}, {and} \bibinfo{person}{Alexander Boukhanovsky}.}
  \bibinfo{year}{2021}\natexlab{}.
\newblock \bibinfo{title}{Automated Evolutionary Approach for the Design of
  Composite Machine Learning Pipelines}.
\newblock
\newblock
\urldef\tempurl%
\url{https://arxiv.org/abs/2106.15397}
\showURL{%
\tempurl}


\bibitem[\protect\citeauthoryear{Olson, Bartley, Urbanowicz, and Moore}{Olson
  et~al\mbox{.}}{2016}]%
        {Olson:tpot:2016}
\bibfield{author}{\bibinfo{person}{Randal~S. Olson}, \bibinfo{person}{Nathan
  Bartley}, \bibinfo{person}{Ryan~J. Urbanowicz}, {and}
  \bibinfo{person}{Jason~H. Moore}.} \bibinfo{year}{2016}\natexlab{}.
\newblock \showarticletitle{Evaluation of a Tree-Based Pipeline Optimization
  Tool for Automating Data Science}. In \bibinfo{booktitle}{\emph{Proc
  GECCO'16}}. \bibinfo{pages}{485–492}.
\newblock
\urldef\tempurl%
\url{https://doi.org/10.1145/2908812.2908918}
\showDOI{\tempurl}


\bibitem[\protect\citeauthoryear{Ono, Castelo, Lopez, Bertini, Freire, and
  Silva}{Ono et~al\mbox{.}}{2021}]%
        {pipelineprofiler2020}
\bibfield{author}{\bibinfo{person}{Jorge~Piazentin Ono}, \bibinfo{person}{Sonia
  Castelo}, \bibinfo{person}{Roque Lopez}, \bibinfo{person}{Enrico Bertini},
  \bibinfo{person}{Juliana Freire}, {and} \bibinfo{person}{Claudio Silva}.}
  \bibinfo{year}{2021}\natexlab{}.
\newblock \showarticletitle{PipelineProfiler: A Visual Analytics Tool for the
  Exploration of AutoML Pipelines}.
\newblock \bibinfo{journal}{\emph{IEEE Transactions on Visualization and
  Computer Graphics}} \bibinfo{volume}{27}, \bibinfo{number}{2}
  (\bibinfo{year}{2021}), \bibinfo{pages}{390--400}.
\newblock
\urldef\tempurl%
\url{https://doi.org/10.1109/TVCG.2020.3030361}
\showDOI{\tempurl}


\bibitem[\protect\citeauthoryear{Pan and Yang}{Pan and Yang}{2010}]%
        {Pan:transfer_learning_survey:2010}
\bibfield{author}{\bibinfo{person}{Sinno~Jialin Pan} {and}
  \bibinfo{person}{Qiang Yang}.} \bibinfo{year}{2010}\natexlab{}.
\newblock \showarticletitle{A Survey on Transfer Learning}.
\newblock \bibinfo{journal}{\emph{IEEE Transactions on Knowledge and Data
  Engineering}} \bibinfo{volume}{22}, \bibinfo{number}{10}
  (\bibinfo{year}{2010}), \bibinfo{pages}{1345--1359}.
\newblock
\urldef\tempurl%
\url{https://doi.org/10.1109/TKDE.2009.191}
\showDOI{\tempurl}


\bibitem[\protect\citeauthoryear{Parasuraman, Sheridan, and
  Wickens}{Parasuraman et~al\mbox{.}}{2000}]%
        {Parasuraman2000}
\bibfield{author}{\bibinfo{person}{Raja Parasuraman},
  \bibinfo{person}{Thomas~B. Sheridan}, {and} \bibinfo{person}{Christopher~D.
  Wickens}.} \bibinfo{year}{2000}\natexlab{}.
\newblock \showarticletitle{A Model for Types and Levels of Human Interaction
  with Automation}.
\newblock \bibinfo{journal}{\emph{IEEE Transactions on Systems, Man, and
  Cybernetics - Part A: Systems and Humans}} \bibinfo{volume}{30},
  \bibinfo{number}{3} (\bibinfo{year}{2000}), \bibinfo{pages}{286--297}.
\newblock
\urldef\tempurl%
\url{https://doi.org/10.1109/3468.844354}
\showDOI{\tempurl}


\bibitem[\protect\citeauthoryear{Prat, Comyn-Wattiau, and Akoka}{Prat
  et~al\mbox{.}}{2015}]%
        {Prat:taxonomy:2015}
\bibfield{author}{\bibinfo{person}{Nicolas Prat}, \bibinfo{person}{Isabelle
  Comyn-Wattiau}, {and} \bibinfo{person}{Jacky Akoka}.}
  \bibinfo{year}{2015}\natexlab{}.
\newblock \showarticletitle{A Taxonomy of Evaluation Methods for Information
  Systems Artifacts}.
\newblock \bibinfo{journal}{\emph{Journal of Management Information Systems}}
  \bibinfo{volume}{32}, \bibinfo{number}{3} (\bibinfo{year}{2015}),
  \bibinfo{pages}{229--267}.
\newblock
\urldef\tempurl%
\url{https://doi.org/10.1080/07421222.2015.1099390}
\showDOI{\tempurl}


\bibitem[\protect\citeauthoryear{Publio, Esteves, Lawrynowicz, Panov,
  Soldatova, Soru, Vanschoren, and Zafar}{Publio et~al\mbox{.}}{2018}]%
        {Publio:MLSchemaET:2018}
\bibfield{author}{\bibinfo{person}{G. Publio}, \bibinfo{person}{Diego Esteves},
  \bibinfo{person}{Agnieszka Lawrynowicz}, \bibinfo{person}{P. Panov},
  \bibinfo{person}{L. Soldatova}, \bibinfo{person}{Tommaso Soru},
  \bibinfo{person}{J. Vanschoren}, {and} \bibinfo{person}{Hamid Zafar}.}
  \bibinfo{year}{2018}\natexlab{}.
\newblock \bibinfo{title}{ML-Schema: Exposing the Semantics of Machine Learning
  with Schemas and Ontologies}.
\newblock
\newblock
\urldef\tempurl%
\url{https://openreview.net/forum?id=B1e8MrXVxQ}
\showURL{%
\tempurl}


\bibitem[\protect\citeauthoryear{Ragan, Endert, Sanyal, and Chen}{Ragan
  et~al\mbox{.}}{2015}]%
        {ragan2015characterizing}
\bibfield{author}{\bibinfo{person}{Eric~D Ragan}, \bibinfo{person}{Alex
  Endert}, \bibinfo{person}{Jibonananda Sanyal}, {and} \bibinfo{person}{Jian
  Chen}.} \bibinfo{year}{2015}\natexlab{}.
\newblock \showarticletitle{Characterizing provenance in visualization and data
  analysis: an organizational framework of provenance types and purposes}.
\newblock \bibinfo{journal}{\emph{IEEE transactions on visualization and
  computer graphics}} \bibinfo{volume}{22}, \bibinfo{number}{1}
  (\bibinfo{year}{2015}), \bibinfo{pages}{31--40}.
\newblock


\bibitem[\protect\citeauthoryear{Rogers, Patton, Harmon, Lex, and Meyer}{Rogers
  et~al\mbox{.}}{2021}]%
        {rogers:trrrace:2021}
\bibfield{author}{\bibinfo{person}{Jen Rogers}, \bibinfo{person}{Austin~H.
  Patton}, \bibinfo{person}{Luke Harmon}, \bibinfo{person}{Alexander Lex},
  {and} \bibinfo{person}{Miriah Meyer}.} \bibinfo{year}{2021}\natexlab{}.
\newblock \showarticletitle{Insights From Experiments With Rigor in an EvoBio
  Design Study}.
\newblock \bibinfo{journal}{\emph{IEEE Transactions on Visualization and
  Computer Graphics}} \bibinfo{volume}{27}, \bibinfo{number}{2}
  (\bibinfo{year}{2021}), \bibinfo{pages}{1106--1116}.
\newblock
\urldef\tempurl%
\url{https://doi.org/10.1109/TVCG.2020.3030405}
\showDOI{\tempurl}


\bibitem[\protect\citeauthoryear{Sacha, Kraus, Keim, and Chen}{Sacha
  et~al\mbox{.}}{2019}]%
        {Sacha:Vis2ML_ontology:2019}
\bibfield{author}{\bibinfo{person}{Dominik Sacha}, \bibinfo{person}{Matthias
  Kraus}, \bibinfo{person}{Daniel~A. Keim}, {and} \bibinfo{person}{Min Chen}.}
  \bibinfo{year}{2019}\natexlab{}.
\newblock \showarticletitle{VIS4ML: An Ontology for Visual Analytics Assisted
  Machine Learning}.
\newblock \bibinfo{journal}{\emph{IEEE Transactions on Visualization and
  Computer Graphics}} \bibinfo{volume}{25}, \bibinfo{number}{1}
  (\bibinfo{year}{2019}), \bibinfo{pages}{385--395}.
\newblock
\urldef\tempurl%
\url{https://doi.org/10.1109/TVCG.2018.2864838}
\showDOI{\tempurl}


\bibitem[\protect\citeauthoryear{Sacha, Sedlmair, Zhang, Lee, Peltonen,
  Weiskopf, North, and Keim}{Sacha et~al\mbox{.}}{2017}]%
        {sacha:wywc:2017}
\bibfield{author}{\bibinfo{person}{Dominik Sacha}, \bibinfo{person}{Michael
  Sedlmair}, \bibinfo{person}{Leishi Zhang}, \bibinfo{person}{John~A. Lee},
  \bibinfo{person}{Jaakko Peltonen}, \bibinfo{person}{Daniel Weiskopf},
  \bibinfo{person}{Stephen~C. North}, {and} \bibinfo{person}{Daniel~A. Keim}.}
  \bibinfo{year}{2017}\natexlab{}.
\newblock \showarticletitle{What you see is what you can change: Human-centered
  machine learning by interactive visualization}.
\newblock \bibinfo{journal}{\emph{Neurocomputing}}  \bibinfo{volume}{268}
  (\bibinfo{year}{2017}), \bibinfo{pages}{164--175}.
\newblock
\urldef\tempurl%
\url{https://doi.org/10.1016/j.neucom.2017.01.105}
\showDOI{\tempurl}


\bibitem[\protect\citeauthoryear{Salmons}{Salmons}{2008}]%
        {salmons2008expect}
\bibfield{author}{\bibinfo{person}{Janet Salmons}.}
  \bibinfo{year}{2008}\natexlab{}.
\newblock \showarticletitle{Expect originality! Using taxonomies to structure
  assignments that support original work}.
\newblock In \bibinfo{booktitle}{\emph{Student plagiarism in an online world:
  Problems and solutions}}. \bibinfo{publisher}{IGI Global},
  \bibinfo{pages}{208--227}.
\newblock


\bibitem[\protect\citeauthoryear{Salvador, Budka, and Gabrys}{Salvador
  et~al\mbox{.}}{2016}]%
        {pmlr-v64-salvador_adapting_2016}
\bibfield{author}{\bibinfo{person}{Manuel~Martin Salvador},
  \bibinfo{person}{Marcin Budka}, {and} \bibinfo{person}{Bogdan Gabrys}.}
  \bibinfo{year}{2016}\natexlab{}.
\newblock \showarticletitle{Adapting Multicomponent Predictive Systems using
  Hybrid Adaptation Strategies with Auto-WEKA in Process Industry}. In
  \bibinfo{booktitle}{\emph{Proceedings of the Workshop on Automatic Machine
  Learning}} \emph{(\bibinfo{series}{Proceedings of Machine Learning Research},
  Vol.~\bibinfo{volume}{64})}, \bibfield{editor}{\bibinfo{person}{Frank
  Hutter}, \bibinfo{person}{Lars Kotthoff}, {and} \bibinfo{person}{Joaquin
  Vanschoren}} (Eds.). \bibinfo{publisher}{PMLR}, \bibinfo{address}{New York,
  New York, USA}, \bibinfo{pages}{48--57}.
\newblock
\urldef\tempurl%
\url{http://proceedings.mlr.press/v64/salvador_adapting_2016.html}
\showURL{%
\tempurl}


\bibitem[\protect\citeauthoryear{Sambasivan, Kapania, Highfill, Akrong,
  Paritosh, and Aroyo}{Sambasivan et~al\mbox{.}}{2021}]%
        {Nithya:datawork:2021}
\bibfield{author}{\bibinfo{person}{Nithya Sambasivan}, \bibinfo{person}{Shivani
  Kapania}, \bibinfo{person}{Hannah Highfill}, \bibinfo{person}{Diana Akrong},
  \bibinfo{person}{Praveen Paritosh}, {and} \bibinfo{person}{Lora~M Aroyo}.}
  \bibinfo{year}{2021}\natexlab{}.
\newblock \showarticletitle{``Everyone Wants to Do the Model Work, Not the Data
  Work'': Data Cascades in High-Stakes AI}. In \bibinfo{booktitle}{\emph{Proc.
  CHI'21}}. Article \bibinfo{articleno}{39}, \bibinfo{numpages}{15}~pages.
\newblock
\urldef\tempurl%
\url{https://doi.org/10.1145/3411764.3445518}
\showDOI{\tempurl}


\bibitem[\protect\citeauthoryear{Sanders and Stappers}{Sanders and
  Stappers}{2014}]%
        {Sanders:sketch:2014}
\bibfield{author}{\bibinfo{person}{Elizabeth B.-N. Sanders} {and}
  \bibinfo{person}{Pieter~Jan Stappers}.} \bibinfo{year}{2014}\natexlab{}.
\newblock \showarticletitle{Probes, toolkits and prototypes: three approaches
  to making in codesigning}.
\newblock \bibinfo{journal}{\emph{CoDesign}} \bibinfo{volume}{10},
  \bibinfo{number}{1} (\bibinfo{year}{2014}), \bibinfo{pages}{5--14}.
\newblock
\urldef\tempurl%
\url{https://doi.org/10.1080/15710882.2014.888183}
\showDOI{\tempurl}


\bibitem[\protect\citeauthoryear{Santos, Castelo, Felix, Ono, Yu, Hong, Silva,
  Bertini, and Freire}{Santos et~al\mbox{.}}{2019}]%
        {Santos:Visus:2019}
\bibfield{author}{\bibinfo{person}{A\'{e}cio Santos}, \bibinfo{person}{Sonia
  Castelo}, \bibinfo{person}{Cristian Felix}, \bibinfo{person}{Jorge~Piazentin
  Ono}, \bibinfo{person}{Bowen Yu}, \bibinfo{person}{Sungsoo~Ray Hong},
  \bibinfo{person}{Cl\'{a}udio~T. Silva}, \bibinfo{person}{Enrico Bertini},
  {and} \bibinfo{person}{Juliana Freire}.} \bibinfo{year}{2019}\natexlab{}.
\newblock \showarticletitle{Visus: An Interactive System for Automatic Machine
  Learning Model Building and Curation}. In \bibinfo{booktitle}{\emph{Proc.
  HILDA'19}}. Article \bibinfo{articleno}{6}, \bibinfo{numpages}{7}~pages.
\newblock
\urldef\tempurl%
\url{https://doi.org/10.1145/3328519.3329134}
\showDOI{\tempurl}


\bibitem[\protect\citeauthoryear{Schelter, Boese, Kirschnick, Klein, and
  Seufert}{Schelter et~al\mbox{.}}{2017}]%
        {schelter:MLProv:2017}
\bibfield{author}{\bibinfo{person}{Sebastian Schelter},
  \bibinfo{person}{Joos-Hendrik Boese}, \bibinfo{person}{Johannes Kirschnick},
  \bibinfo{person}{Thoralf Klein}, {and} \bibinfo{person}{Stephan Seufert}.}
  \bibinfo{year}{2017}\natexlab{}.
\newblock \showarticletitle{Automatically tracking metadata and provenance of
  machine learning experiments}. In \bibinfo{booktitle}{\emph{Machine Learning
  Systems Workshop at NIPS}}. \bibinfo{pages}{27--29}.
\newblock


\bibitem[\protect\citeauthoryear{Setlur, Tory, and Djalali}{Setlur
  et~al\mbox{.}}{2019}]%
        {setlur:intent:2019}
\bibfield{author}{\bibinfo{person}{Vidya Setlur}, \bibinfo{person}{Melanie
  Tory}, {and} \bibinfo{person}{Alex Djalali}.}
  \bibinfo{year}{2019}\natexlab{}.
\newblock \showarticletitle{Inferencing Underspecified Natural Language
  Utterances in Visual Analysis}. In \bibinfo{booktitle}{\emph{Proc IUI '19}}.
  \bibinfo{pages}{40–51}.
\newblock
\urldef\tempurl%
\url{https://doi.org/10.1145/3301275.3302270}
\showDOI{\tempurl}


\bibitem[\protect\citeauthoryear{Shneiderman}{Shneiderman}{1996}]%
        {Shneiderman:TTT:1996}
\bibfield{author}{\bibinfo{person}{B. Shneiderman}.}
  \bibinfo{year}{1996}\natexlab{}.
\newblock \showarticletitle{The eyes have it: a task by data type taxonomy for
  information visualizations}. In \bibinfo{booktitle}{\emph{Proc. VLHCC'96}}.
  \bibinfo{pages}{336--343}.
\newblock
\urldef\tempurl%
\url{https://doi.org/10.1109/VL.1996.545307}
\showDOI{\tempurl}


\bibitem[\protect\citeauthoryear{Souza, Valduriez, Mattoso, Cerqueira, Netto,
  Azevedo, Lourenço, F.~de S.~Soares, Melo, Brandão, Salles~Civitarese,
  Vital~Brazil, and Ferreira~Moreno}{Souza et~al\mbox{.}}{2019}]%
        {Renan:ml_prov:2019}
\bibfield{author}{\bibinfo{person}{Renan Souza}, \bibinfo{person}{Patrick
  Valduriez}, \bibinfo{person}{Marta Mattoso}, \bibinfo{person}{Renato
  Cerqueira}, \bibinfo{person}{Marco Netto}, \bibinfo{person}{Leonardo
  Azevedo}, \bibinfo{person}{Vítor Lourenço}, \bibinfo{person}{Elton F.~de
  S.~Soares}, \bibinfo{person}{Raphael Melo}, \bibinfo{person}{Rafael
  Brandão}, \bibinfo{person}{Daniel Salles~Civitarese},
  \bibinfo{person}{Emilio Vital~Brazil}, {and} \bibinfo{person}{Marcio
  Ferreira~Moreno}.} \bibinfo{year}{2019}\natexlab{}.
\newblock \showarticletitle{Provenance Data in the Machine Learning Lifecycle
  in Computational Science and Engineering}. In \bibinfo{booktitle}{\emph{Proc.
  WORKS'19}}. \bibinfo{pages}{1--10}.
\newblock
\urldef\tempurl%
\url{https://doi.org/10.1109/WORKS49585.2019.00006}
\showDOI{\tempurl}


\bibitem[\protect\citeauthoryear{Sperrle, El-Assady, Guo, Borgo, Chau, Endert,
  and Keim}{Sperrle et~al\mbox{.}}{2021}]%
        {Sperrle:HCE_STAR:2021}
\bibfield{author}{\bibinfo{person}{Fabian Sperrle},
  \bibinfo{person}{Mennatallah El-Assady}, \bibinfo{person}{Grace Guo},
  \bibinfo{person}{Rita Borgo}, \bibinfo{person}{Duen~Horng Chau},
  \bibinfo{person}{Alex Endert}, {and} \bibinfo{person}{Daniel Keim}.}
  \bibinfo{year}{2021}\natexlab{}.
\newblock \showarticletitle{A Survey of Human-Centered Evaluations in
  Human-Centered Machine Learning}.
\newblock \bibinfo{journal}{\emph{Computer Graphics Forum}}
  \bibinfo{volume}{40}, \bibinfo{number}{3} (\bibinfo{year}{2021}),
  \bibinfo{pages}{543--567}.
\newblock
\urldef\tempurl%
\url{https://doi.org/10.1111/cgf.14329}
\showDOI{\tempurl}


\bibitem[\protect\citeauthoryear{Sperrle, Jeitler, Bernard, Keim, and
  El-Assady}{Sperrle et~al\mbox{.}}{2020}]%
        {Sperrle:EuroHILML:2020}
\bibfield{author}{\bibinfo{person}{Fabian Sperrle}, \bibinfo{person}{Astrik
  Jeitler}, \bibinfo{person}{Jürgen Bernard}, \bibinfo{person}{Daniel~A.
  Keim}, {and} \bibinfo{person}{Mennatallah El-Assady}.}
  \bibinfo{year}{2020}\natexlab{}.
\newblock \showarticletitle{{Learning and Teaching in Co-Adaptive Guidance for
  Mixed-Initiative Visual Analytics}}. In \bibinfo{booktitle}{\emph{EuroVis
  Workshop on Visual Analytics (EuroVA)}}. \bibinfo{publisher}{The Eurographics
  Association}.
\newblock
\urldef\tempurl%
\url{https://doi.org/10.2312/eurova.20201088}
\showDOI{\tempurl}


\bibitem[\protect\citeauthoryear{Spinner, Schlegel, Schäfer, and
  El-Assady}{Spinner et~al\mbox{.}}{2020}]%
        {Spinner:explainer:2020}
\bibfield{author}{\bibinfo{person}{Thilo Spinner}, \bibinfo{person}{Udo
  Schlegel}, \bibinfo{person}{Hanna Schäfer}, {and}
  \bibinfo{person}{Mennatallah El-Assady}.} \bibinfo{year}{2020}\natexlab{}.
\newblock \showarticletitle{explAIner: A Visual Analytics Framework for
  Interactive and Explainable Machine Learning}.
\newblock \bibinfo{journal}{\emph{IEEE Transactions on Visualization and
  Computer Graphics}} \bibinfo{volume}{26}, \bibinfo{number}{1}
  (\bibinfo{year}{2020}), \bibinfo{pages}{1064--1074}.
\newblock
\urldef\tempurl%
\url{https://doi.org/10.1109/TVCG.2019.2934629}
\showDOI{\tempurl}


\bibitem[\protect\citeauthoryear{Stitz, Gratzl, Piringer, Zichner, and
  Streit}{Stitz et~al\mbox{.}}{2019}]%
        {stitz:kx_pearls:2020}
\bibfield{author}{\bibinfo{person}{Holger Stitz}, \bibinfo{person}{Samuel
  Gratzl}, \bibinfo{person}{Harald Piringer}, \bibinfo{person}{Thomas Zichner},
  {and} \bibinfo{person}{Marc Streit}.} \bibinfo{year}{2019}\natexlab{}.
\newblock \showarticletitle{KnowledgePearls: Provenance-Based Visualization
  Retrieval}.
\newblock \bibinfo{journal}{\emph{IEEE Transactions on Visualization and
  Computer Graphics}} \bibinfo{volume}{25}, \bibinfo{number}{1}
  (\bibinfo{year}{2019}), \bibinfo{pages}{120--130}.
\newblock
\urldef\tempurl%
\url{https://doi.org/10.1109/TVCG.2018.2865024}
\showDOI{\tempurl}


\bibitem[\protect\citeauthoryear{Sundaram, Hayes, Dekhtyar, and
  Holbrook}{Sundaram et~al\mbox{.}}{2010}]%
        {Sundaram:traceaing:2010}
\bibfield{author}{\bibinfo{person}{Senthil~Karthikeyan Sundaram},
  \bibinfo{person}{Jane~Huffman Hayes}, \bibinfo{person}{Alex Dekhtyar}, {and}
  \bibinfo{person}{E.~Ashlee Holbrook}.} \bibinfo{year}{2010}\natexlab{}.
\newblock \showarticletitle{Assessing traceability of software engineering
  artifacts}.
\newblock \bibinfo{journal}{\emph{Requirements Engineering}}
  \bibinfo{volume}{15}, \bibinfo{number}{3} (\bibinfo{date}{01 Sep}
  \bibinfo{year}{2010}), \bibinfo{pages}{313--335}.
\newblock
\showISSN{1432-010X}
\urldef\tempurl%
\url{https://doi.org/10.1007/s00766-009-0096-6}
\showDOI{\tempurl}


\bibitem[\protect\citeauthoryear{Taman, VanderPlas, and Dane}{Taman
  et~al\mbox{.}}{2018}]%
        {tateman:repo_taxonomy:2018}
\bibfield{author}{\bibinfo{person}{Rachel Taman}, \bibinfo{person}{Jake
  VanderPlas}, {and} \bibinfo{person}{Sohier Dane}.}
  \bibinfo{year}{2018}\natexlab{}.
\newblock \bibinfo{title}{A Practical Taxonomy of Reproducibility for Machine
  Learning Research}.
\newblock
\newblock
\urldef\tempurl%
\url{https://openreview.net/forum?id=B1eYYK5QgX}
\showURL{%
\tempurl}


\bibitem[\protect\citeauthoryear{Thornton, Hutter, Hoos, and
  Leyton-Brown}{Thornton et~al\mbox{.}}{2013}]%
        {Thorton:AutoWeka:2013}
\bibfield{author}{\bibinfo{person}{Chris Thornton}, \bibinfo{person}{Frank
  Hutter}, \bibinfo{person}{Holger~H. Hoos}, {and} \bibinfo{person}{Kevin
  Leyton-Brown}.} \bibinfo{year}{2013}\natexlab{}.
\newblock \showarticletitle{Auto-WEKA: Combined Selection and Hyperparameter
  Optimization of Classification Algorithms}. In
  \bibinfo{booktitle}{\emph{Proc. KDD'13}}. \bibinfo{pages}{847–855}.
\newblock
\urldef\tempurl%
\url{https://doi.org/10.1145/2487575.2487629}
\showDOI{\tempurl}


\bibitem[\protect\citeauthoryear{Tiwana and Ramesh}{Tiwana and Ramesh}{2001}]%
        {Tiwana:kx_system:2001}
\bibfield{author}{\bibinfo{person}{Amrit Tiwana} {and}
  \bibinfo{person}{Balasubramaniam Ramesh}.} \bibinfo{year}{2001}\natexlab{}.
\newblock \showarticletitle{A design knowledge management system to support
  collaborative information product evolution}.
\newblock \bibinfo{journal}{\emph{Decision Support Systems}}
  \bibinfo{volume}{31}, \bibinfo{number}{2} (\bibinfo{year}{2001}),
  \bibinfo{pages}{241--262}.
\newblock
\showISSN{0167-9236}
\urldef\tempurl%
\url{https://doi.org/10.1016/S0167-9236(00)00134-2}
\showDOI{\tempurl}


\bibitem[\protect\citeauthoryear{Valiati, Pimenta, and Freitas}{Valiati
  et~al\mbox{.}}{2006}]%
        {Valiati:tasks:2006}
\bibfield{author}{\bibinfo{person}{Eliane R.~A. Valiati},
  \bibinfo{person}{Marcelo~S. Pimenta}, {and} \bibinfo{person}{Carla M. D.~S.
  Freitas}.} \bibinfo{year}{2006}\natexlab{}.
\newblock \showarticletitle{A Taxonomy of Tasks for Guiding the Evaluation of
  Multidimensional Visualizations}. In \bibinfo{booktitle}{\emph{Proceedings of
  the 2006 AVI Workshop on BEyond Time and Errors: Novel Evaluation Methods for
  Information Visualization}} (Venice, Italy) \emph{(\bibinfo{series}{BELIV
  '06})}. \bibinfo{publisher}{Association for Computing Machinery},
  \bibinfo{address}{New York, NY, USA}, \bibinfo{pages}{1–6}.
\newblock
\showISBNx{1595935622}
\urldef\tempurl%
\url{https://doi.org/10.1145/1168149.1168169}
\showDOI{\tempurl}


\bibitem[\protect\citeauthoryear{von Rueden, Mayer, Beckh, Georgiev,
  Giesselbach, Heese, Kirsch, Walczak, Pfrommer, Pick, and et~al.}{von Rueden
  et~al\mbox{.}}{2021}]%
        {von_Rueden:informedml:2021}
\bibfield{author}{\bibinfo{person}{Laura von Rueden},
  \bibinfo{person}{Sebastian Mayer}, \bibinfo{person}{Katharina Beckh},
  \bibinfo{person}{Bogdan Georgiev}, \bibinfo{person}{Sven Giesselbach},
  \bibinfo{person}{Raoul Heese}, \bibinfo{person}{Birgit Kirsch},
  \bibinfo{person}{Michal Walczak}, \bibinfo{person}{Julius Pfrommer},
  \bibinfo{person}{Annika Pick}, {and} \bibinfo{person}{et al.}}
  \bibinfo{year}{2021}\natexlab{}.
\newblock \showarticletitle{Informed Machine Learning - A Taxonomy and Survey
  of Integrating Prior Knowledge into Learning Systems}.
\newblock \bibinfo{journal}{\emph{IEEE Transactions on Knowledge and Data
  Engineering}} (\bibinfo{year}{2021}), \bibinfo{pages}{1–1}.
\newblock
\urldef\tempurl%
\url{https://doi.org/10.1109/tkde.2021.3079836}
\showDOI{\tempurl}


\bibitem[\protect\citeauthoryear{Wang, Mittal, Brooks, and Oney}{Wang
  et~al\mbox{.}}{2019b}]%
        {Wang_April:Notebook:2019b}
\bibfield{author}{\bibinfo{person}{April~Yi Wang}, \bibinfo{person}{Anant
  Mittal}, \bibinfo{person}{Christopher Brooks}, {and} \bibinfo{person}{Steve
  Oney}.} \bibinfo{year}{2019}\natexlab{b}.
\newblock \showarticletitle{How Data Scientists Use Computational Notebooks for
  Real-Time Collaboration}.
\newblock \bibinfo{journal}{\emph{Proc CSCW'19}}  \bibinfo{volume}{3}
  (\bibinfo{date}{Nov} \bibinfo{year}{2019}), \bibinfo{pages}{1–30}.
\newblock
\urldef\tempurl%
\url{https://doi.org/10.1145/3359141}
\showDOI{\tempurl}


\bibitem[\protect\citeauthoryear{Wang, Xu, Zhang, Chen, Fang, Xu, Kang, Hong,
  Jiang, Cai, Li, Zhou, Li, Liu, Chen, Han, Shu, Song, Wang, Zhang, Xu, Li,
  Liu, and Zhang}{Wang et~al\mbox{.}}{2020}]%
        {wang:vega:2020}
\bibfield{author}{\bibinfo{person}{Bochao Wang}, \bibinfo{person}{Hang Xu},
  \bibinfo{person}{Jiajin Zhang}, \bibinfo{person}{Chen Chen},
  \bibinfo{person}{Xiaozhi Fang}, \bibinfo{person}{Yixing Xu},
  \bibinfo{person}{Ning Kang}, \bibinfo{person}{Lanqing Hong},
  \bibinfo{person}{Chenhan Jiang}, \bibinfo{person}{Xinyue Cai},
  \bibinfo{person}{Jiawei Li}, \bibinfo{person}{Fengwei Zhou},
  \bibinfo{person}{Yong Li}, \bibinfo{person}{Zhicheng Liu},
  \bibinfo{person}{Xinghao Chen}, \bibinfo{person}{Kai Han},
  \bibinfo{person}{Han Shu}, \bibinfo{person}{Dehua Song},
  \bibinfo{person}{Yunhe Wang}, \bibinfo{person}{Wei Zhang},
  \bibinfo{person}{Chunjing Xu}, \bibinfo{person}{Zhenguo Li},
  \bibinfo{person}{Wenzhi Liu}, {and} \bibinfo{person}{Tong Zhang}.}
  \bibinfo{year}{2020}\natexlab{}.
\newblock \bibinfo{title}{VEGA: Towards an End-to-End Configurable AutoML
  Pipeline}.
\newblock
\newblock
\showeprint{2011.01507}
\urldef\tempurl%
\url{https://arxiv.org/abs/2011.01507}
\showURL{%
\tempurl}


\bibitem[\protect\citeauthoryear{Wang, Andres, Weisz, Oduor, and Dugan}{Wang
  et~al\mbox{.}}{2021a}]%
        {wang:autods:2021}
\bibfield{author}{\bibinfo{person}{Dakuo Wang}, \bibinfo{person}{Josh Andres},
  \bibinfo{person}{Justin~D. Weisz}, \bibinfo{person}{Erick Oduor}, {and}
  \bibinfo{person}{Casey Dugan}.} \bibinfo{year}{2021}\natexlab{a}.
\newblock \showarticletitle{AutoDS: Towards Human-Centered Automation of Data
  Science}. In \bibinfo{booktitle}{\emph{Proc. CHI'21}}. Article
  \bibinfo{articleno}{79}, \bibinfo{numpages}{12}~pages.
\newblock
\urldef\tempurl%
\url{https://doi.org/10.1145/3411764.3445526}
\showDOI{\tempurl}


\bibitem[\protect\citeauthoryear{Wang, Liao, Zhang, Khurana, Samulowitz, Park,
  Muller, and Amini}{Wang et~al\mbox{.}}{2021b}]%
        {wang:DS_automation:2021}
\bibfield{author}{\bibinfo{person}{Dakuo Wang}, \bibinfo{person}{Q.~Vera Liao},
  \bibinfo{person}{Yunfeng Zhang}, \bibinfo{person}{Udayan Khurana},
  \bibinfo{person}{Horst Samulowitz}, \bibinfo{person}{Soya Park},
  \bibinfo{person}{Michael Muller}, {and} \bibinfo{person}{Lisa Amini}.}
  \bibinfo{year}{2021}\natexlab{b}.
\newblock \bibinfo{title}{How Much Automation Does a Data Scientist Want?}
\newblock
\newblock
\showeprint{2101.03970}
\urldef\tempurl%
\url{https://arxiv.org/abs/2101.03970}
\showURL{%
\tempurl}


\bibitem[\protect\citeauthoryear{Wang, Weisz, Muller, Ram, Geyer, Dugan,
  Tausczik, Samulowitz, and Gray}{Wang et~al\mbox{.}}{2019c}]%
        {Wang:hai_collab:2019}
\bibfield{author}{\bibinfo{person}{Dakuo Wang}, \bibinfo{person}{Justin~D.
  Weisz}, \bibinfo{person}{Michael Muller}, \bibinfo{person}{Parikshit Ram},
  \bibinfo{person}{Werner Geyer}, \bibinfo{person}{Casey Dugan},
  \bibinfo{person}{Yla Tausczik}, \bibinfo{person}{Horst Samulowitz}, {and}
  \bibinfo{person}{Alexander Gray}.} \bibinfo{year}{2019}\natexlab{c}.
\newblock \showarticletitle{Human-AI Collaboration in Data Science: Exploring
  Data Scientists’ Perceptions of Automated AI}.
\newblock \bibinfo{journal}{\emph{Proc. CSCW'19}},
  \bibinfo{numpages}{24}~pages.
\newblock
\urldef\tempurl%
\url{https://doi.org/10.1145/3359313}
\showDOI{\tempurl}


\bibitem[\protect\citeauthoryear{Wang, Ming, Jin, Shen, Liu, Smith,
  Veeramachaneni, and Qu}{Wang et~al\mbox{.}}{2019a}]%
        {wang:atmseer:2018}
\bibfield{author}{\bibinfo{person}{Qianwen Wang}, \bibinfo{person}{Yao Ming},
  \bibinfo{person}{Zhihua Jin}, \bibinfo{person}{Qiaomu Shen},
  \bibinfo{person}{Dongyu Liu}, \bibinfo{person}{Micah~J. Smith},
  \bibinfo{person}{Kalyan Veeramachaneni}, {and} \bibinfo{person}{Huamin Qu}.}
  \bibinfo{year}{2019}\natexlab{a}.
\newblock \showarticletitle{ATMSeer: Increasing Transparency and
  Controllability in Automated Machine Learning}. In
  \bibinfo{booktitle}{\emph{Proc CHI'19}}. \bibinfo{pages}{1–12}.
\newblock
\urldef\tempurl%
\url{https://doi.org/10.1145/3290605.3300911}
\showDOI{\tempurl}


\bibitem[\protect\citeauthoryear{Weidele, Weisz, Oduor, Muller, Andres, Gray,
  and Wang}{Weidele et~al\mbox{.}}{2020}]%
        {Weidele:autovizai:2020}
\bibfield{author}{\bibinfo{person}{Daniel Karl~I. Weidele},
  \bibinfo{person}{Justin~D. Weisz}, \bibinfo{person}{Erick Oduor},
  \bibinfo{person}{Michael Muller}, \bibinfo{person}{Josh Andres},
  \bibinfo{person}{Alexander Gray}, {and} \bibinfo{person}{Dakuo Wang}.}
  \bibinfo{year}{2020}\natexlab{}.
\newblock \showarticletitle{AutoAIViz: Opening the Blackbox of Automated
  Artificial Intelligence with Conditional Parallel Coordinates}. In
  \bibinfo{booktitle}{\emph{Proc. IUI'20}}. \bibinfo{pages}{308–312}.
\newblock
\urldef\tempurl%
\url{https://doi.org/10.1145/3377325.3377538}
\showDOI{\tempurl}


\bibitem[\protect\citeauthoryear{Winfield and Jirotka}{Winfield and
  Jirotka}{2018}]%
        {Winfield:ethics_transparency:2018}
\bibfield{author}{\bibinfo{person}{Alan F.~T. Winfield} {and}
  \bibinfo{person}{Marina Jirotka}.} \bibinfo{year}{2018}\natexlab{}.
\newblock \showarticletitle{Ethical governance is essential to building trust
  in robotics and artificial intelligence systems}.
\newblock \bibinfo{journal}{\emph{Philosophical Transactions of the Royal
  Society A: Mathematical, Physical and Engineering Sciences}}
  \bibinfo{volume}{376}, \bibinfo{number}{2133} (\bibinfo{year}{2018}),
  \bibinfo{pages}{20180085}.
\newblock
\urldef\tempurl%
\url{https://doi.org/10.1098/rsta.2018.0085}
\showDOI{\tempurl}


\bibitem[\protect\citeauthoryear{Wong, Houlsby, Lu, and Gesmundo}{Wong
  et~al\mbox{.}}{2018}]%
        {Wong:NeuralAutoML:2018}
\bibfield{author}{\bibinfo{person}{Catherine Wong}, \bibinfo{person}{Neil
  Houlsby}, \bibinfo{person}{Yifeng Lu}, {and} \bibinfo{person}{Andrea
  Gesmundo}.} \bibinfo{year}{2018}\natexlab{}.
\newblock \showarticletitle{Transfer Learning with Neural AutoML}. In
  \bibinfo{booktitle}{\emph{Proc NeurIPS'18}}. \bibinfo{publisher}{Curran
  Associates Inc.}, \bibinfo{address}{Red Hook, NY, USA},
  \bibinfo{pages}{8366–8375}.
\newblock


\bibitem[\protect\citeauthoryear{Wongsuphasawat, Moritz, Anand, Mackinlay,
  Howe, and Heer}{Wongsuphasawat et~al\mbox{.}}{2016}]%
        {wongsuphasawat:voayger:2016}
\bibfield{author}{\bibinfo{person}{K. Wongsuphasawat}, \bibinfo{person}{D.
  Moritz}, \bibinfo{person}{A. Anand}, \bibinfo{person}{J. Mackinlay},
  \bibinfo{person}{D. Howe}, {and} \bibinfo{person}{J. Heer}.}
  \bibinfo{year}{2016}\natexlab{}.
\newblock \showarticletitle{Voyager: Exploratory Analysis via Faceted Browsing
  of Visualization Recommendations}.
\newblock \bibinfo{journal}{\emph{IEEE Transactions on Visualization and
  Computer Graphics}} \bibinfo{volume}{22}, \bibinfo{number}{1}
  (\bibinfo{date}{Jan} \bibinfo{year}{2016}), \bibinfo{pages}{649--658}.
\newblock
\showISSN{1077-2626}
\urldef\tempurl%
\url{https://doi.org/10.1109/TVCG.2015.2467191}
\showDOI{\tempurl}


\bibitem[\protect\citeauthoryear{Xin, Ma, Liu, Macke, Song, and
  Parameswaran}{Xin et~al\mbox{.}}{2018}]%
        {Xin:Analysis_DAG:2018}
\bibfield{author}{\bibinfo{person}{Doris Xin}, \bibinfo{person}{Litian Ma},
  \bibinfo{person}{Jialin Liu}, \bibinfo{person}{Stephen Macke},
  \bibinfo{person}{Shuchen Song}, {and} \bibinfo{person}{Aditya Parameswaran}.}
  \bibinfo{year}{2018}\natexlab{}.
\newblock \showarticletitle{Accelerating Human-in-the-Loop Machine Learning:
  Challenges and Opportunities}. In \bibinfo{booktitle}{\emph{Proc. DEEM'18}}.
\newblock
\urldef\tempurl%
\url{https://doi.org/10.1145/3209889.3209897}
\showDOI{\tempurl}


\bibitem[\protect\citeauthoryear{Xin, Wu, Lee, Salehi, and Parameswaran}{Xin
  et~al\mbox{.}}{2021}]%
        {Xin:wither_automl:2021}
\bibfield{author}{\bibinfo{person}{Doris Xin}, \bibinfo{person}{Eva~Yiwei Wu},
  \bibinfo{person}{Doris Jung-Lin Lee}, \bibinfo{person}{Niloufar Salehi},
  {and} \bibinfo{person}{Aditya Parameswaran}.}
  \bibinfo{year}{2021}\natexlab{}.
\newblock \bibinfo{booktitle}{\emph{Whither AutoML? Understanding the Role of
  Automation in Machine Learning Workflows}}.
\newblock
\urldef\tempurl%
\url{https://doi.org/10.1145/3411764.3445306}
\showDOI{\tempurl}


\bibitem[\protect\citeauthoryear{Yang, Akimoto, Kim, and Udell}{Yang
  et~al\mbox{.}}{2019}]%
        {Yang:OBOE:2019}
\bibfield{author}{\bibinfo{person}{Chengrun Yang}, \bibinfo{person}{Yuji
  Akimoto}, \bibinfo{person}{Dae~Won Kim}, {and} \bibinfo{person}{Madeleine
  Udell}.} \bibinfo{year}{2019}\natexlab{}.
\newblock \showarticletitle{OBOE: Collaborative Filtering for AutoML Model
  Selection}. In \bibinfo{booktitle}{\emph{Proc KDD '19}} (Anchorage, AK, USA).
  \bibinfo{pages}{1173–1183}.
\newblock
\urldef\tempurl%
\url{https://doi.org/10.1145/3292500.3330909}
\showDOI{\tempurl}


\bibitem[\protect\citeauthoryear{Yang, Fan, Wu, and Udell}{Yang
  et~al\mbox{.}}{2020a}]%
        {Yang:AutoML:2020}
\bibfield{author}{\bibinfo{person}{Chengrun Yang}, \bibinfo{person}{Jicong
  Fan}, \bibinfo{person}{Ziyang Wu}, {and} \bibinfo{person}{Madeleine Udell}.}
  \bibinfo{year}{2020}\natexlab{a}.
\newblock \showarticletitle{AutoML Pipeline Selection: Efficiently Navigating
  the Combinatorial Space}. In \bibinfo{booktitle}{\emph{Proc KDD '20}}.
  \bibinfo{pages}{1446–1456}.
\newblock
\urldef\tempurl%
\url{https://doi.org/10.1145/3394486.3403197}
\showDOI{\tempurl}


\bibitem[\protect\citeauthoryear{Yang, Steinfeld, Ros\'{e}, and Zimmerman}{Yang
  et~al\mbox{.}}{2020b}]%
        {Yang:HAI_hard:2020}
\bibfield{author}{\bibinfo{person}{Qian Yang}, \bibinfo{person}{Aaron
  Steinfeld}, \bibinfo{person}{Carolyn Ros\'{e}}, {and} \bibinfo{person}{John
  Zimmerman}.} \bibinfo{year}{2020}\natexlab{b}.
\newblock \bibinfo{booktitle}{\emph{Re-Examining Whether, Why, and How Human-AI
  Interaction Is Uniquely Difficult to Design}}.
\newblock \bibinfo{pages}{1–13}.
\newblock
\urldef\tempurl%
\url{https://doi.org/10.1145/3313831.3376301}
\showDOI{\tempurl}


\bibitem[\protect\citeauthoryear{Yao, Wang, Chen, Dai, Li, Tu, Yang, and
  Yu}{Yao et~al\mbox{.}}{2019}]%
        {yao:humanout:2019}
\bibfield{author}{\bibinfo{person}{Quanming Yao}, \bibinfo{person}{Mengshuo
  Wang}, \bibinfo{person}{Yuqiang Chen}, \bibinfo{person}{Wenyuan Dai},
  \bibinfo{person}{Yu-Feng Li}, \bibinfo{person}{Wei-Wei Tu},
  \bibinfo{person}{Qiang Yang}, {and} \bibinfo{person}{Yang Yu}.}
  \bibinfo{year}{2019}\natexlab{}.
\newblock \bibinfo{title}{Taking Human out of Learning Applications: A Survey
  on Automated Machine Learning}.
\newblock
\newblock
\showeprint{1810.13306}
\urldef\tempurl%
\url{https://arxiv.org/abs/1810.13306}
\showURL{%
\tempurl}


\bibitem[\protect\citeauthoryear{Zhang, Muller, and Wang}{Zhang
  et~al\mbox{.}}{2020a}]%
        {zhang:ds_collab:2020}
\bibfield{author}{\bibinfo{person}{Amy~X. Zhang}, \bibinfo{person}{Michael
  Muller}, {and} \bibinfo{person}{Dakuo Wang}.}
  \bibinfo{year}{2020}\natexlab{a}.
\newblock \showarticletitle{How Do Data Science Workers Collaborate? Roles,
  Workflows, and Tools}.
\newblock \bibinfo{journal}{\emph{Proc CSCW'2020}}, Article
  \bibinfo{articleno}{022} (\bibinfo{date}{May} \bibinfo{year}{2020}),
  \bibinfo{numpages}{23}~pages.
\newblock
\urldef\tempurl%
\url{https://doi.org/10.1145/3392826}
\showDOI{\tempurl}


\bibitem[\protect\citeauthoryear{Zhang, Anderljung, Kahn, Dreksler, Horowitz,
  and Dafoe}{Zhang et~al\mbox{.}}{2021}]%
        {zhangB:ethics:2021}
\bibfield{author}{\bibinfo{person}{Baobao Zhang}, \bibinfo{person}{Markus
  Anderljung}, \bibinfo{person}{Lauren Kahn}, \bibinfo{person}{Noemi Dreksler},
  \bibinfo{person}{Michael~C. Horowitz}, {and} \bibinfo{person}{Allan Dafoe}.}
  \bibinfo{year}{2021}\natexlab{}.
\newblock \bibinfo{title}{Ethics and Governance of Artificial Intelligence:
  Evidence from a Survey of Machine Learning Researchers}.
\newblock
\newblock
\showeprint[arxiv]{2105.02117}~[cs.CY]


\bibitem[\protect\citeauthoryear{Zhang, Zame, and van~der Schaar}{Zhang
  et~al\mbox{.}}{2020b}]%
        {zhang:autocp:2020}
\bibfield{author}{\bibinfo{person}{Yao Zhang}, \bibinfo{person}{William Zame},
  {and} \bibinfo{person}{Mihaela van~der Schaar}.}
  \bibinfo{year}{2020}\natexlab{b}.
\newblock \bibinfo{title}{AutoCP: Automated Pipelines for Accurate Prediction
  Intervals}.
\newblock
\newblock
\showeprint{2006.14099}
\urldef\tempurl%
\url{https://arxiv.org/abs/2006.14099}
\showURL{%
\tempurl}


\bibitem[\protect\citeauthoryear{Zhuang, Qi, Duan, Xi, Zhu, Zhu, Xiong, and
  He}{Zhuang et~al\mbox{.}}{2021}]%
        {Zhuang:transfer_learning_survey:2021}
\bibfield{author}{\bibinfo{person}{Fuzhen Zhuang}, \bibinfo{person}{Zhiyuan
  Qi}, \bibinfo{person}{Keyu Duan}, \bibinfo{person}{Dongbo Xi},
  \bibinfo{person}{Yongchun Zhu}, \bibinfo{person}{Hengshu Zhu},
  \bibinfo{person}{Hui Xiong}, {and} \bibinfo{person}{Qing He}.}
  \bibinfo{year}{2021}\natexlab{}.
\newblock \showarticletitle{A Comprehensive Survey on Transfer Learning}.
\newblock \bibinfo{journal}{\emph{Proc. IEEE}} \bibinfo{volume}{109},
  \bibinfo{number}{1} (\bibinfo{year}{2021}), \bibinfo{pages}{43--76}.
\newblock
\urldef\tempurl%
\url{https://doi.org/10.1109/JPROC.2020.3004555}
\showDOI{\tempurl}


\bibitem[\protect\citeauthoryear{Zimmer, Lindauer, and Hutter}{Zimmer
  et~al\mbox{.}}{2021}]%
        {Zimmer:Auto-Pytorch:2021}
\bibfield{author}{\bibinfo{person}{Lucas Zimmer}, \bibinfo{person}{Marius
  Lindauer}, {and} \bibinfo{person}{Frank Hutter}.}
  \bibinfo{year}{2021}\natexlab{}.
\newblock \showarticletitle{Auto-Pytorch: Multi-Fidelity MetaLearning for
  Efficient and Robust AutoDL}.
\newblock \bibinfo{journal}{\emph{IEEE Transactions on Pattern Analysis \&
  Machine Intelligence}} \bibinfo{number}{01} (\bibinfo{year}{2021}),
  \bibinfo{pages}{1--1}.
\newblock
\urldef\tempurl%
\url{https://doi.org/10.1109/TPAMI.2021.3067763}
\showDOI{\tempurl}


\bibitem[\protect\citeauthoryear{Z{\"o}ller and Huber}{Z{\"o}ller and
  Huber}{2021}]%
        {Zoller:BenchmarkAS:2021}
\bibfield{author}{\bibinfo{person}{Marc-Andr{\'e} Z{\"o}ller} {and}
  \bibinfo{person}{M. Huber}.} \bibinfo{year}{2021}\natexlab{}.
\newblock \showarticletitle{Benchmark and Survey of Automated Machine Learning
  Frameworks}.
\newblock \bibinfo{journal}{\emph{J. Artif. Intell. Res.}}
  \bibinfo{volume}{70} (\bibinfo{year}{2021}), \bibinfo{pages}{409--472}.
\newblock


\bibitem[\protect\citeauthoryear{Zoph and Le}{Zoph and Le}{2017}]%
        {Zoph:NAS:2017}
\bibfield{author}{\bibinfo{person}{Barret Zoph} {and} \bibinfo{person}{Quoc~V.
  Le}.} \bibinfo{year}{2017}\natexlab{}.
\newblock \showarticletitle{Neural Architecture Search with Reinforcement
  Learning}.
\newblock
\urldef\tempurl%
\url{https://arxiv.org/abs/1611.01578}
\showURL{%
\tempurl}


\end{thebibliography}

\appendix
\section{Appendix: AutoML Artifact Taxonomy Additional Details}\label{taxonomy-extra}

We describe the artifact properties according to our taxonomy. We use color highlighting through this subsection to to emphasize the \Dim{dimensions}, \Cat{categories}, and \Char{characteristics} of our taxonomy (see Section~\ref{taxonomy-method}). The exposition of our taxonomy proceeds in a hierarchical order, beginning with a dimension down to its respective characteristics. 
\newline

\begin{figure}[h!]
    \centering
    \includegraphics[width=0.8\columnwidth]{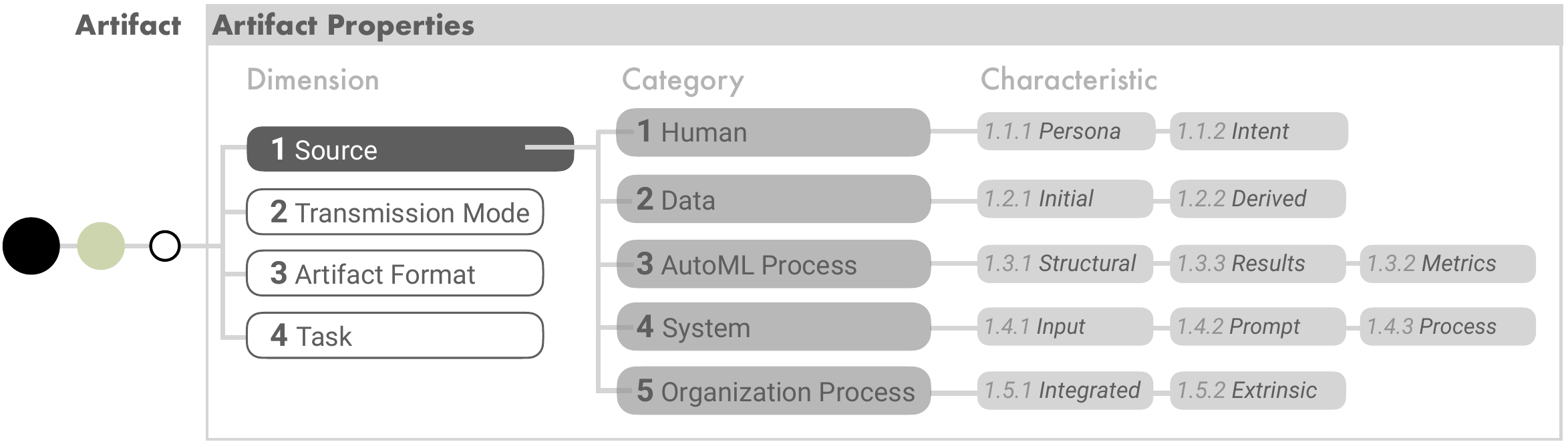}
    \caption{The \textbf{Source} dimension of an artifact and the categories and characteristics it encompasses}
    \label{fig:source_tax}
\end{figure}
\noindent\Dim{Dimension 1: Source} (\textit{\textbf{``What generated the artifact?''}}) Identifying the artifact's source helps provide context and a sense of provenance of how the decisions were made throughout an AutoML process. In fully automated data science processes, these artifacts are generated by computational processes, which we refer to as `the machine', without human interventions. However, as full automation is both challenging to achieve and not always desirable, in reality, artifacts can have a variety of sources. For example, a visual analytics mixed-initiative system that operates on top of an AutoML pipeline. In such a system, an analyst can arrive at a set of insights through a combination of automated decisions made by a back-end model and human inputs provided through the interface made along the way~\cite{Heer:agency:2019,Sperrle:HCE_STAR:2021,Sperrle:EuroHILML:2020}. At a high level, artifacts can have human or machine sources. However, in our taxonomy development process, we were also able to define an additional layer of granularity to artifact sources. Human artifacts can be sourced from individual or organizational processes. Machine artifacts can be sourced from the AutoML processes and the overall software infrastructure (or system) that orchestrates the automated data science processes. Finally, we separate data as its own unique source as it cross-cuts both human and machine sources. These more granular source delineations are categories in our taxonomy that have additional characteristics. While we found that many artifacts generally have distinct sources, some can have multiple sources. For example, many artifacts concerning data augmentation can be sourced from a combination of human intents and derivations from the initial dataset. Sources of human input can also result from prompts by the system that explicitly seek user feedback.

\begin{itemize}
    \item \Cat{Category 1.1: Human} We found that humans act as sources to AutoML pipelines primarily by providing inputs in the form of goals and requirements, specifications~\cite{hong2020evaluating,gijsber:automl_benchmark:2019}, and interactions with a system~\cite{hong2020evaluating,Cashman:EMA:2019,sacha:wywc:2017}. While human can refer to one or multiple individuals providing input, we prefer the more narrow interpretation of a single human providing input to, or interacting with an AutoML pipeline. As will become clear, `organizational processes' is a better source designation to describe multiple humans working together. Amongst individual human sources,  we found two characteristics that added important context: persona and intent. 
    
   We found that artifact types can differ based upon the\Char{persona (c1.1.1)}~\cite{sant:automl:2021,wang:atmseer:2018,Crisan:AutoML:2021} of the individual carrying out the analysis. AutoML systems can be leveraged by individuals not trained in data science or machine learning. We posit the nature of those inputs and the affordances they use to supply those inputs will be different than those with more area expertise. For example, individuals trained in data science of machine learning might produce more codebase artifacts through their use of notebooks~\cite{Wang_April:Notebook:2019b}, while other personas may rely more on no-code solutions, and their inputs are more likely captured through interface widgets or other types of semantic interactions~\cite{Gehrmann:SemanticInference:2020,Endert:2012}. 
    
    \vspace{1.5mm}
    Another important characteristic of human source artifacts is the \Char{intent (c1.1.2)} of the individual. These artifacts can appear as user preference models, analysis types, or even model tasks (the analyst chooses a model optimized for a specific task). The HCI, Vis, and ML communities have used different terminologies to define what a person wishes to do in an analysis process. Tasks is a common term used in all three communities (i.e.~\cite{sant:automl:2021,Brehmer:Typology:2013,Wong:NeuralAutoML:2018} ) and these can be tied to goals~\cite{Lam:GoalstoTasks:2018} or preferences. Recently, visualization researchers have begun using intent as a general way to capture this spectrum, from an individual's tasks to their goals~\cite{gadhave:vis_intnet:2020,setlur:intent:2019}. We opted to use this terminology because it aligned well with the diversity of artifacts our analysis captured. 
    
    \vspace{1.5mm}
    \item \Cat{Category 1.2: Data} Data are perhaps the most obvious artifact of an AutoML process and one that needs the least explanation. In our taxonomy, the primary characteristics of data differentiate whether it is an initial input or whether it is derived from the AutoML process.
    
    \vspace{1mm}
    \Char{Initial (c1.2.1)} datasets are sometimes also referred to as raw data. We refrain from using the word `raw' largely because no dataset truly exists in such a state~\cite{Davis:rawdata:2013,Gitelman:rawdata:2013}. Instead, we use the term `initial' dataset, in lieu of the `raw' terminology. Furthermore, the terminology of 1initial' acknowledges that a  dataset may be further transformed or augmented either by a human or an AutoML processes before a machine learning model is applied.
    
    \vspace{1mm}
    In contrast, \Char{derived (c1.2.2)} datasets result when transformations are applied to the initial data. These transformations can result from data cleaning or wrangling operations~\cite{kasica:tablescraps:2021} (including feature encoding~\cite{yao:humanout:2019,Yang:AutoML:2020,sant:automl:2021}, the derivation of new features~\cite{wang:DS_automation:2021,Zimmer:Auto-Pytorch:2021}, or creating a new representations via data or feature embedding~\cite{Yang:AutoML:2020}). The resulting derived datasets are generated by the AutoML processes and changes in their compositions can be useful to understand how processes arrived at its final set of results~\cite{Celik:Adaption:2021}.
    
    \vspace{1.5mm}
    \item \Cat{Category 1.3: AutoML Process} Different levels of automation directly influence how many and what kinds of artifacts are generated by an AutoML process. Given that AutoML can theoretically range from hyperparameter tuning to a full end-to-end data science pipeline~\cite{Zoller:BenchmarkAS:2021,sant:automl:2021,Heffetz:deepline:2020}, the spectrum of possible artifacts stem from AutoML processes can be very broad. However, we identified three characteristics of artifacts that span this spectrum: structure, metrics, and results. 
    
    \Char{Structural (c1.3.1)} characteristics of artifacts describe a component of an AutoML pipeline, such as a machine learning model, or an end-to-end pipeline of steps that also encompasses data preparation, feature engineering, and reporting~\cite{wang:vega:2020,Zoller:BenchmarkAS:2021,sant:automl:2021,Yang:AutoML:2020,Heffetz:deepline:2020,feurer:autosklearn:2020}. We additionally extended the definition of structural characteristics to include algorithmic artifacts that constitute training or tuning a specific component~\cite{golovin:googleVizer:2017}, the architecture or more complex models like neural networks~\cite{Haifeng:autokeras:2019}, or pipeline topology~\cite{Yang:AutoML:2020}, configuration space~\cite{feurer:autosklearn:2020,Zoller:BenchmarkAS:2021}, or search space~\cite{wang:atmseer:2018,pipelineprofiler2020,nikitin:automated:2021}. Lastly, we include a model's tasks as part of its structural characteristics, as they play an important role in understanding what the model is intended to do while adding context to architecture. Structural characteristics often take the form of specifications supplied by the end-users or are automatically generated by the AutoML processes. For example, we consider the final architecture or fit of a model to be an automatically generated artifact with structural characteristics resulting from an algorithmic process.
    
    \vspace{1mm}
    \Char{Metrics (c1.3.2)} and \Char{results (c1.3.3)} are two complementary characteristics and perhaps the most widely scrutinized aspects of AutoML process artifacts. Metrics refers to measures that describe the model training, validation, and testing performance. These can take various forms depending on the type of model used and the task it is intended to solve. However, basic measures such as overall or average accuracy tend to be the most commonly reported. Metrics are intimately tied to the result of a component or pipeline applied to a data set. Again, the precise nature of this result depends upon the model task. Two commonly used types are classification and clustering tasks; however, more advanced models enable a more complex set of tasks such as document summarizing, text or image generation, among others. 
    
    \vspace{1.5mm}
    \item \Cat{Category 1.4: System} AutoML processes sit within a larger software ecosystem that orchestrates and carries out the computational instructions of its different components (i.e., data cleaning, feature engineering, or machine learning steps). Artifacts tend to be generated by a system, and we identified three characteristics of such artifacts: inputs, prompts, and processes. 
    
    Characteristics of these artifacts concerned the ways that they were either provided or generated by the system. Some artifacts operate as  \Char{inputs (c1.4.1)}, which can come from human processes or result from data or other types of artifacts transferred between an AutoML process and the computational layer of a system. These can include configuration files for the computational environment~\cite{Cambronero:Software_AutoML:2021}, computational budgets~\cite{wang:atmseer:2018}, or source code~\cite{wang:DS_automation:2021,Cambronero:Software_AutoML:2021}.
    Artifacts are also generated as a result of the system presenting a \Char{prompt (c1.4.2)} to an individual for some input, or through an automatic \Char{process (c1.4.3)}. Alerting mechanisms can be a common way to prompt an individual for some action; this action produces an artifact that can trigger a change to the AutoML pipeline. For example, alerting an individual to a high correlation between two variables in their input dataset can lead them to remove a feature from a model. The alert is generated by an automated process that carries out the correlation checking and is itself an artifact, but the choice the user makes (whether to remove the feature from the model or not) results from the prompt itself; the artifact is a user's choice and has the characteristic of being generated by a prompt.
    
    \vspace{1.5mm}
    \item \Cat{Category 1.5: Organizational Process} AutoML technology is used in conjunction with existing business and organization practices~\cite{Crisan:AutoML:2021}. These processes generate artifacts that can act as input and integrate directly into AutoML processes while others exert an extrinsic influence but do not provide any direct input. 
    
    \vspace{1mm}
    Organizational artifacts that have an \Char{integrated (c1.5.1)} characteristic when they directly influence how AutoML pipelines are trained, evaluated, and finally used in decision making. For example, the data schemas that define the structure of the data are influenced by business practices. However, schemas influence the type of data that is collected, how it is stored and accessed, which can be used or limit what is achievable in an AutoML  process~\cite{dellermann:future_collab:2021}.  Other artifacts that constitute integrated organizations process include data augmentations, through contextual augments (i.e. human supplied semantic annotations~\cite{estevez:semantic_annots:2019} or ontologies~\cite{Cashman:EMA:2019}) and benchmark datasets~\cite{Zoller:BenchmarkAS:2021,Haifeng:autokeras:2019,zhang:autocp:2020,Olson:tpot:2016}. We found that these artifacts closely reflected how organizations carry out their practices, unlike a machine learning model whose underlying mathematical specifications are largely agnostic to organizational practices. 
    
    \vspace{1mm}
    Although not the focus of our research (Section~\ref{problem-statement}), we also made space of \Char{extrinsic (c1.5.2)} organizational processes, which include legal procedures or practices within the organization that dictate the use and limitations of AutoML technology. These artifacts are not directly integrated into the processes of specifying, developing, or training aspects of an AutoML pipeline, as, for example, data augmentations are. They are a step removed from the AutoML processes, even though they add relevant contextual information; hence, we classified these artifacts as extrinsic.  
\end{itemize}


\begin{figure}[h!]
    \centering
    \includegraphics[width=0.8\columnwidth]{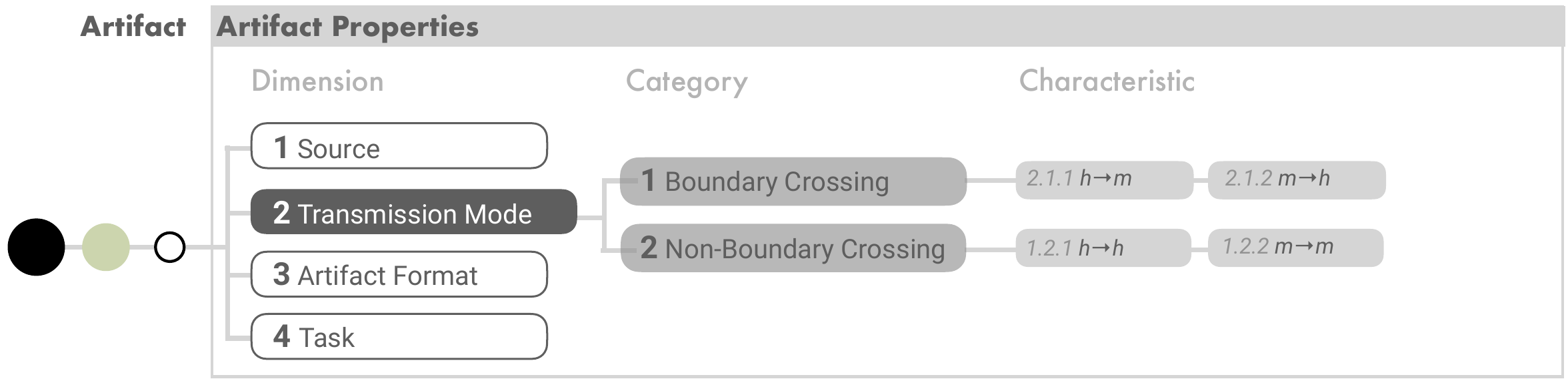}
    \caption{The \textbf{Transmission} dimension of an artifact and the categories and characteristics it encompasses}
    \label{fig:tm_tax}
\end{figure}
\noindent\Dim{D2: Transmission Mode}(\textbf{\textit{``Does it cross the boundaries between human and AutoML processes?''}}) Our work emphasizes artifacts that cross boundaries between machine and human processes. However, there still exist artifacts that do not cross this boundary but that are important for transparently describing an AutoML processes. We include space for both in our taxonomy.

\vspace{1.5mm}
\begin{itemize}
    \item \Cat{Category 2.1: Boundary Crossing} artifacts are passed between the human and AutoML processes, and vice-versa, which we interpret as characteristics of an artifact. An artifact that goes from a human to AutoML processes, \Char{h$\rightarrow$m (c2.1.1); human-to-machine}, serves as input to the AutoML. This input includes artifact sources stemming from the data they provide, specifications and configurations, or even the imposition of integrated or extrinsic processes, such as requirements documentation~\cite{wang:DS_automation:2021,alletto:randomnet:2020,Cambronero:Software_AutoML:2021} which can dictate what the AutoML process should do. We also consider actions that govern an AutoML process and determine how it proceeds, for example, whether a specific model can be deployed, to also possess h$\rightarrow$m characteristics, although none of the research papers we reviewed in building our taxonomy contained such an explicit artifact. AutoML processes can automatically generate outputs in the form of reports or alerts intended for human consumption. These artifacts have an \Char{m$\rightarrow$h (c2.1.2; machine-to-human)} characteristics. Increasing AutoML processes produce, or are expected to produce,  explanations for their outputs in the form of reports~\cite{wang:DS_automation:2021} or automatically generated model cards~\cite{Mitchell:model_cards:2019}. Automatic methods for detecting anomalies in the data or model~\cite{elshawi2019automated}, especially the presence of concept drift~\cite{Celik:Adaption:2021}, are initiated by the AutoML system. These automated methods generate artifacts to present to a user in the form of alerts, automatically generated reports, or dashboards.
    
    \vspace{1mm}
    \item \Cat{Category 2.2: Non-Boundary Crossing} We limited the number of non-boundary crossing items we included in our taxonomy; a full survey of such artifacts could likely fill one or several research papers on their own. Moreover, other research literature has examined these such artifacts ranging from knowledge management~\cite{Kreiner:tacit_artifact:2002} to APIs and other considerations of computational infrastructure. To be able to connect our taxonomy to such artifacts we have included  \Char{h$\rightarrow$h (c2.2.1; human-to-human)} and \Char{m$\rightarrow$m (c2.2.1; machine-to-machine)} characteristics for non-boundary crossing artifacts. 
\end{itemize}

\begin{figure}[h!]
    \centering
    \includegraphics[width=0.8\columnwidth]{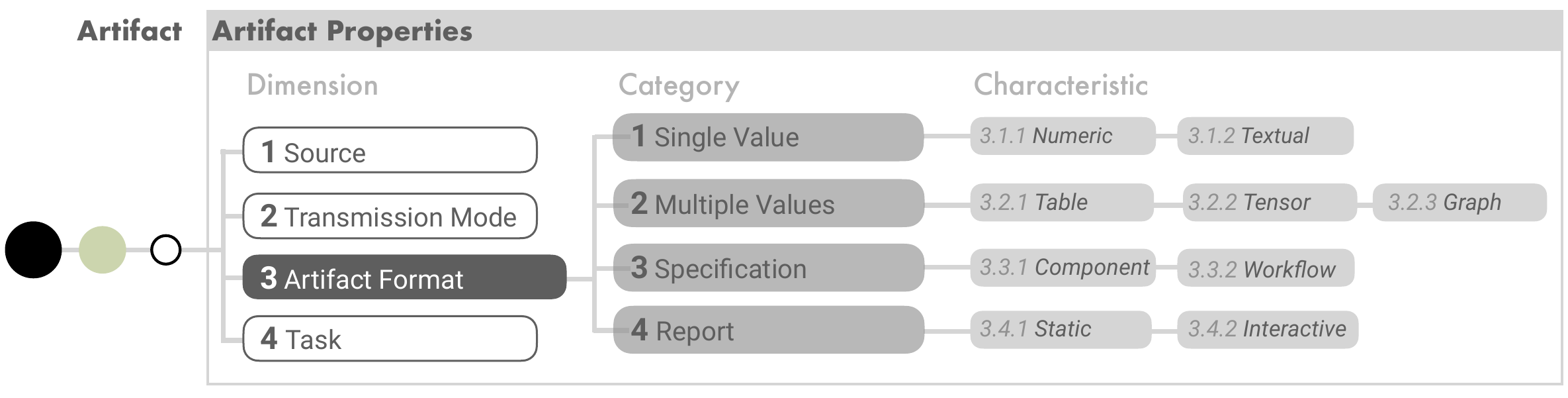}
    \caption{The \textbf{Artifact Format} dimension of an artifact and the categories and characteristics it encompasses}
    \label{fig:af_tax}
\end{figure}

\noindent\Dim{D3: Artifact Format}(\textbf{\textit{``What shape does the artifact take?''}}) As a practical consideration in our taxonomy is the different forms that artifacts take. Our observation is that different systems produce artifacts in different formats, although current systems place an emphasis on primarily numerical artifacts. Systems that visualize AutoML process, such as PipelineProfiler~\cite{pipelineprofiler2020}, ATMSeer~\cite{wang:atmseer:2018}, or AutoVizAI~\cite{Weidele:autovizai:2020}, are informed by the format these artifacts take when consider what to visualize and how. As we take a more expansive look at AutoML processes and the hand-off between human and machine processes, and as AutoML systems expand to greater levels of automation, we argue that the diversity of artifact formats grows. It is important to acknowledge this artifact format diversity in creating future AutoML systems or tools that aim to surface these artifacts of a diversity of data science and non-data science personas. Moreover, these artifacts can constitute both inputs to, or outputs of, and AutoML processes. In our analysis, we identified four categories of these artifact formats : single values, multiple values, specifications, and reports.  

\begin{itemize}
    \item \Cat{Category 3.1: Single Value} It is common for AutoML artifacts to be a single value, for example, a single summary statistic or a text alert informing an individual of some result. We found that either \Char{numeric (c3.1.1)} or \Char{textual (c3.1.2)} characteristics were common for such artifacts. Annotations on the data~\cite{estevez:semantic_annots:2019,Renan:ml_prov:2019} or AutoML pipeline~\cite{Spinner:explainer:2020} and feedback from individuals also tend to come in the form of comments and thus are also artifacts with textual characteristics. 
    
    \vspace{1mm}
    \item \Cat{Category 3.2: Multiple Value} A natural extension of single value artifacts are those that contain multiple values. We delineate \Char{table~(c3.2.1)} and \Char{tensor~(c3.2.2)} characteristics for the shape of these data. In the machine learning literature, tensor strongly implies an N-dimensional numeric array, which can range form a single scalar value (N=0) to a vector (N=1), and a finally a multi-dimensional array (N>1). Because the term tensor has such a strong numeric connotation, we include `table' terminology to allow for artifacts mixed types of data (numeric, ordinal, categorical);  a table with one column is just a list. We did consider whether single values should simply be considered special cases of tables and tensors, but, because we also wanted to emphasize data with similar or mixed variable (column, attribute) types we opted to separate single and multiple values in order to make this characterization easier to identify.
    \\
    We also considered a \Char{graph (c3.2.3)} to constitute multiple data types. The word `graph' is overloaded and can mean either a visualization or a type of data structure. Here, we use the term to mean a graph data structure that constitutes nodes and edges. Moreover, we use graph as a general term to encompass both network and tree structures, again with these being special cases on graphs; a tree is a graph with a hierarchical, directed, and acyclic structure.  Behavioral graphs, interaction logs, or interaction sequences~\cite{battle:eva:2019,hong2020evaluating,Cashman:EMA:2019,stitz:kx_pearls:2020} are common examples of artifacts with graph characteristics that we found. Initial datasets are also artifacts that can have a graph characteristics, for example, ontologies or knowledge graphs~\cite{von_Rueden:informedml:2021,agarwal:knowledge:2021}.
    
    \vspace{1mm}
    \item \Cat{Category 3.3: Specification}
    
    As AutoML processes grow in complexity from focusing on model training and expanding to multiple stages of a data science processes. As a result an AutoML process can be described as a set of interchangeable components~\cite{Yang:AutoML:2020,wang:vega:2020} (also referred to as an `ML primitive'~\cite{Heffetz:deepline:2020,pipelineprofiler2020}) that are pieced together into a final workflow configuration. We use an expanded definition of an AutoML component here to include not such the machine learning model and it's dependencies (feature engineering, data preparation), but also to include reporting, task formulation, and other such stages~\cite{sant:automl:2021}. Depending on the level of automation, specifications can represent individual components or entire workflows. These specifications typically take the form of configuration files or source code, however, others have explored a broader space of specifications that includes equations and logic rules~\cite{von_Rueden:informedml:2021}\footnote{Although the authors consider these to be knowledge representations, they would fit our definition of a specification.}.
    
    Depending upon the level of automation, an individual \Char{component~(c3.3.1)}  can either be specified entirely by a human or the specification can be automatically generated by a learning processes. They type of specification will vary depending on the component that is being specified. For example, specifications for a machine learning model component will differ from specifications for a feature generation component which will also be different than a data visualization component. While many of these processes were formally the manual work of data scientists~\cite{Zoller:BenchmarkAS:2021,wang:DS_automation:2021} these processes are becoming increasingly automated~\cite{sant:automl:2021}.
    \\
     Currently, many AutoML systems and toolkits arguably focus primarily on the machine learning model component and the optimization of it's structural specifications~\cite{Yang:OBOE:2019,Olson:tpot:2016,Thorton:AutoWeka:2013}, or in the case of deep learning it's architectural specifications~\cite{Haifeng:autokeras:2019,Zimmer:Auto-Pytorch:2021}. However, as more sophisticated end-to-end AutoML system arise, an individual need not specify many of the components as these can be derived automatically by search a AutoML \Char{workflow~(c3.3.2)}  design. In these cases, an individual may only need to specify the preliminary configurations of the search space~\cite{Zoller:BenchmarkAS:2021,Heffetz:deepline:2020,wang:atmseer:2018,wang:vega:2020,alletto:randomnet:2020,yao:humanout:2019,feurer:autosklearn:2020,pmlr-v64-salvador_adapting_2016}. 
    
    \vspace{1mm}
    \item \Cat{Category 3.4: Report}  Currently, many reporting tasks are done by humans and are not regularly considered part of an AutoML processes, but we see this changing in the future.  Over time, AutoML processes will be automatically generating explainers~\cite{Spinner:explainer:2020}, reports~\cite{wang:DS_automation:2021,Hong:interpret:2020}, and data visualizations~\cite{lee:lux:2021}. These will exist along side human reporting artifacts around the largely analytic goals of an AutoML processes, decision making provenance, and finally, dashboards that report key performance indicators~\cite{wang:DS_automation:2021}. These final output of these reporting artifacts depend upon how they are specified either by individuals or by the AutoML processes. We find that they general have two characteristics, \Char{static~(c3.4.1)}, for example PDFs or presentations, or \Char{interactive~(c3.4.2)} dashboards.
\end{itemize}

\begin{figure}[h!]
    \centering
    \includegraphics[width=0.8\columnwidth]{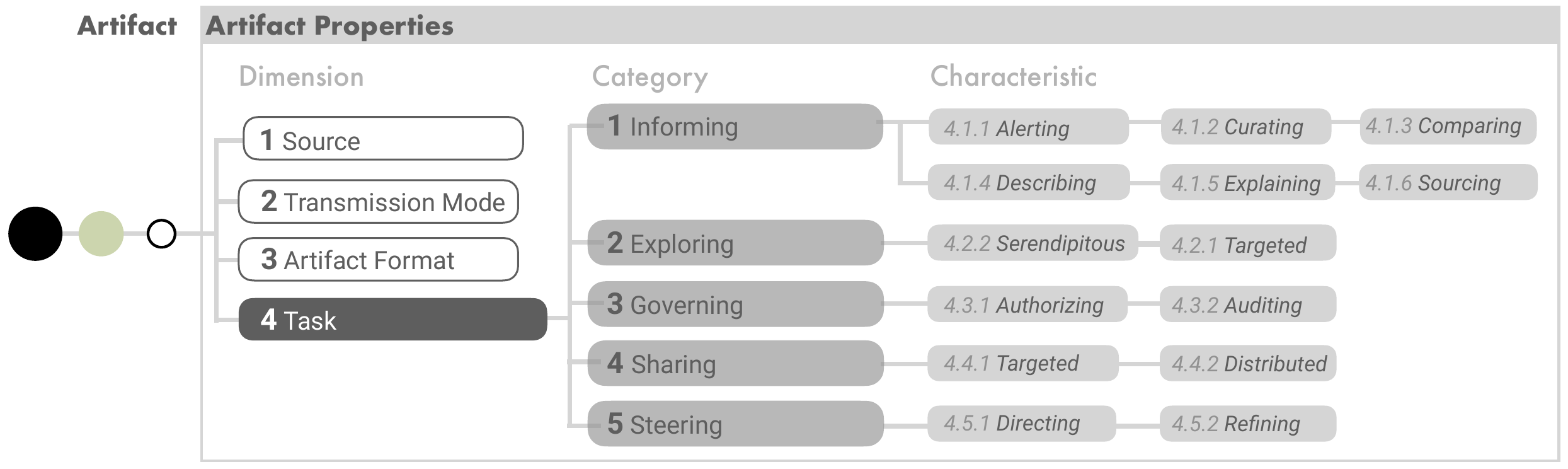}
    \caption{The \textbf{Task} dimension of an artifact and the categories and characteristics it encompasses}
    \label{fig:task_tax}
\end{figure}
\noindent\Dim{D4: Task}(\textbf{\textit{``What is its intended purpose?''}}). Human analytic tasks and machine learning model tasks have complementary, if not overlapping, objectives~\cite{Sperrle:HCE_STAR:2021}. The tasks that either a human or machine wishes to accomplish effects the characteristics of the artifacts that are hand-off between them, especially if this tasks results in a artifact the crosses the boundary between human and machine processes. In our analysis, we have applied our own judgement to establish the types of tasks that different AutoML artifacts are intended to support. However, we also previous task taxonomies specific of AutoML and machine learning~\cite{dellermann:future_collab:2021,tateman:repo_taxonomy:2018,von_Rueden:informedml:2021}, typologies of visual analysis~\cite{Brehmer:Typology:2013,Lam:GoalstoTasks:2018}, and other classification systems~\cite{Sacha:Vis2ML_ontology:2019,sant:automl:2021,Renan:ml_prov:2019,Sperrle:HCE_STAR:2021} reported across various disciplines in formulating our taxonomy of the types of task that artifacts can support. Again, end-to-end AutoML systems are an evolving technology and current implementations of these systems vary with respect to what tasks they support. As such, we remind the reader that these tasks are an amalgamation of tasks we've identified in real and theoretical AutoML systems. 
\begin{itemize}
    \item \Cat{Category 4.1: Informing} We believe that the most common task of artifacts is currently to inform. The intended audience for this information can vary depending the whether the artifact is boundary crossing or not.  We identified six characteristics of these artifacts, which we in alphabetical order. \Char{Alerting~(c4.1.1)} characterises are associated with artifacts that arise from data quality or model quality alerts~\cite{Celik:Adaption:2021} in the form when they are take also take on $m\rightarrow h$ characteristics. We speculate that future systems may even be able to automatically incorporate alerts in Te form of textual feedback from $h\rightarrow m$, for example, if a human evaluate spots an error or omission in the AutoML processes and seeks to alert the algorithm. Moreover, we also speculate that there exists informal $h\rightarrow h$ artifacts, in the form of comments, annotations, or other means that initiate corrective mechanisms in a model following a review of the results or decisions.  \Char{Curating~(c4.1.2)} was a characteristic common of shared or saved insights, which we found to be a common mode of sharing key findings from exploratory visual analyses~\cite{wongsuphasawat:voayger:2016}. Increasingly, AutoML system can produce their own set of curated insights that are presented in a rank-order for individuals to consider.  Artifacts with \Char{comparing~(c4.1.3)} characteristics were in many cases focused on comparing different states of the individual workflow components; often these were also data visualizations. For example, an individual may wish to conduct a sensitivity analysis on a threshold for a classification model. Alternatively, they may also wish to compare a component over time to see how it evolves. Comparison tasks have been extensively studied in visualization research, and there are complimentary taxonomies~\cite{Gliecher:Comparison_2:2018,Gliecher:comparison_taxonomy:2011} here that can add even more context to comparison artifacts from AutoML processes.  Artifacts with \Char{describing~(c4.1.4)} characteristics described the state of an AutoML component or process. These include specifications, requirements, or regulatory documents that indicated either what an individual component was and  what the sequence or configuration the AutoML components~\cite{wang:DS_automation:2021,Cambronero:Software_AutoML:2021}. Moreover, we argue that descriptive artifacts can also take the form of statistical analysis that provide a summary of the data, results, or other properties of the AutoML processes~\cite{sant:automl:2021,Cashman:EMA:2019}. We chose to differentiate these descriptive artifacts from those whose primary purpose is intended to be \Char{explanatory~(c4.1.5)}. Explanations focusing on the modeling components are increasingly important for transparency and are being automatically produced by these components~\cite{wang:DS_automation:2021,Sperrle:HCE_STAR:2021,Brent:Explain:2019,Spinner:explainer:2020}. These artifacts can take the form of reports, but also the outputs of techniques like SHAP, LIME, and the like~\cite{Harmanpreet:XAItechniques:2020}. While these explanations tend to focus on black-box models, we also extend our definition to the self-explanatory characteristics of so-called ``white-box'' models~\cite{ocatavio:black-v-white:2019,barrero:xai:2020}. The final characteristics of artifacts with an informing intent is  \Char{sourcing~(c.4.1.5)}. Sourcing artifacts are those focusing on the lineage and analysis history of an AutoML processes. Even with full automation, human oversight, for example by interacting with other artifacts, can trigger a refinement or retraining of an AutoML processes' configuration~\cite{elshawi2019automated,Sperrle:EuroHILML:2020,Liu:Paths:2020}. Through each iteration, either human driven or triggered by automated process~\cite{Celik:Adaption:2021}. Provenance processes can also capture an individuals interactions with different components, for example data visualizations~\cite{stitz:kx_pearls:2020,Cutler:Trrack:2020}, can could influence an machine learning component through semantic interactions~\cite{Gehrmann:SemanticInference:2020,Endert:2012,Brown:2012}. While sourcing characteristics can also be considered descriptive, we argue that, like explanations, they have a more specific and active role beyond simply describing the state of the systems or some component and thus should be considered separately. 
    
    \vspace{1mm}
    \item \Cat{Category 4.2: Exploring} Artifacts can be generated as an individual, or even an automated processes, explores the data, model, or pipeline configuration space. We differentiate between exploratory artifacts that are \Char{targeted~\c(4.4.1)} for some specific aim (for example, hypothesis verification) and those that arise through purely \Char{serendipitous~\c(4.4.2)} discovery. WE consider many artifacts that arise from an exploratory visual analysis processes~\cite{battle:eva:2019,stitz:kx_pearls:2020,Cutler:Trrack:2020}, which can include bookmarked or saved insights that are curated~\cite{wongsuphasawat:voayger:2016}, to be exploratory artifacts with serendipitous characteristics.
    
    \vspace{1mm}
    \item \Cat{Category 4.3: Governing} Governance processes were not widely considered in the AutoML literature, although they appear in more recent work~\cite{wang:DS_automation:2021,Crisan:AutoML:2021}. These processes will only grow in importance over time as legal requirements and organizational practices change to regulate ML/AI and AutoML technology more generally~\cite{Winfield:ethics_transparency:2018,zhangB:ethics:2021}. From the research that does exist, we propose two characteristics of governance artifacts: authorizing and auditing. \Char{Authorizing~(c4.3.1)} concerns enabling an AutoML system or even an individual analyst to to execute some component, again, who performs the work depends on the level of automation inherent in the system.  We found some evidence for such possible artifacts in~\cite{Crisan:AutoML:2021} and~\cite{wang:DS_automation:2021} in their description of a desire for oversight, for example having a data scientist authorize the deployment of a model created by someone else in the organization, or having automated rules that enforce, for example, anti-bias rules. Complementary to these artifacts are those that enable \Char{auditing~(c4.3.2)}, for example, decision optimization forensics reports~\cite{wang:DS_automation:2021}. Artifacts with auditing characteristics rely on a variety of artifacts, especially those that are intended to inform. 
    
    \vspace{1mm}
    \item \Cat{Category 4.4: Sharing} Even when humans do not aim to intervene at all in AutoML processes, the results of these process are integrated into other human and organizational processes to share knowledge. Artifacts that are intended to be \Char{targeted~(c4.4.1)} to specific personas have a specific purpose. We consider computational notebooks~\cite{zhang:ds_collab:2020} to be artifacts possessing such characteristics, as they are intended to be be shared with other technical data workers or simply with the data scientist themselves.  Artifacts with \Char{distribution~(c4.4.2)} characteristics are those that aim to have a wide audience. These can include presentations or reports, many of which rely on the compilation and analysis of other AutoML artifacts.
    
    \vspace{1mm}
    \item \Cat{Category 4.5: Steering} Artifacts can act to orient or intervene within an AutoML process. Artifacts with \Char{directing~\c(4.5.1)} act to initialize or orient the AutoML processes in a specific trajectory. These artifacts can include a human setting initial goals, providing training data, or providing an initial configurations for components~\cite{yao:humanout:2019}. They can also include algorithmic processes that define the final configuration of AutoML components. In contrast, artifacts with ~\Char{refining~(c4.5.2)} characteristics are subtler than directing as they largely build off of the existing model structure. Fine-tuning operations, example Transfer Learning~\cite{barrero:xai:2020,Pan:transfer_learning_survey:2010,Zhuang:transfer_learning_survey:2021}, produce artifacts for refining an existing neural architecture~\cite{Wong:NeuralAutoML:2018}, rather than conducting an expensive neural architecture search~\cite{alletto:randomnet:2020,neto:nasirt:2020,Haifeng:autokeras:2019,wang:vega:2020,Wong:NeuralAutoML:2018,Zoph:NAS:2017}. We consider the former to be  a refinement, while the latter uses an initial design space configuration to direct the overall neural architecture search process.
\end{itemize}

\end{document}